\documentclass[]{mnras}
\usepackage{epsfig}
\usepackage{graphicx}
\usepackage{mathtools}
\usepackage{pgf}
\usepackage{float}
\def\gsim{\vcenter{\hbox{$>$}\offinterlineskip\hbox{$\sim$}}}

\title[M\,33 monitoring. VI]{The UK Infrared Telescope M\,33 monitoring
project -- VI. Feedback from dusty stellar winds across the galactic disc}
\author[Javadi et al.]{\parbox{17cm}{\vspace{-3mm}Atefeh Javadi$^{1}$,
                                     Jacco Th.\ van Loon$^{2}$,
                                     Mina Alizadeh$^{1}$,
                                     Seyed Azim Hashemi$^{3}$
                                 and Elham Saremi$^{4,1}$\vspace{3mm}}\\
$^{1}$School of Astronomy, Institute for Research in Fundamental Sciences
      (IPM), P.O.\ Box 19395-5531, Tehran, Iran\\
$^{2}$Lennard-Jones Laboratories, Keele University, ST5 5BG, UK\\
$^{3}$Department of Physics, Sharif University of Technology, Tehran,
      84156-83111, Iran\\
$^{4}$School of Physics, Birjand University, Birjand, P.O.\ Box 97175-615,
      Iran}
\date{Resubmitted: 6 June 2018; in its original form 15 February 2018}
\pagerange{\pageref{firstpage}--\pageref{lastpage}}
\pubyear{2018}
\begin{document}
\maketitle
\label{firstpage}
\begin{abstract}
We have conducted a near-infrared monitoring campaign at the UK InfraRed
Telescope (UKIRT), of the Local Group spiral galaxy M\,33 (Triangulum).
In this sixth paper of the series, we measure the
 dust and gas mass-loss rates
by the pulsating Asymptotic Giant Branch (AGB) stars and red supergiants
(RSGs) across the stellar disc of M\,33.
We combined our time-averaged near-IR photometry with the multi-epoch mid-IR
photometry obtained with the {\it Spitzer} Space Telescope, and employed a
combination of spectral energy distribution modelling and scaling relations.
We found that the mass-loss rate is approximately proportional to
luminosity (birth mass), with additional weaker dependence on pulsation period
and/or amplitude (reflecting stellar evolution). As a population, AGB stars
contribute most to the mass return into the interstellar medium (ISM).
Super-AGB stars also reach very high mass-loss rates, in excess
of $10^{-4}$ M$_\odot$ yr$^{-1}$. The mass loss of RSGs appears to be subject to
different modes, with rates well below, around, and well above the nuclear
burning timescale. The timescale for the dominant mass loss phase is
$\sim0.6$--$2\times10^5$ yr, shorter than the thermal-pulsing AGB or RSG
phases. The rate at which stars return mass to the ISM, $\sim0.1$ M$_\odot$
yr$^{-1}$ is about four times lower than the star formation rate, which would
deplete the current ISM mass within about a Gyr, thus requiring additional,
external gas supplies to sustain the long-term future of star formation in
M\,33.
\end{abstract}
\section{Introduction}

Mass loss on the Asymptotic Giant Branch (AGB) truncates these stars'
evolution, leaving behind a carbon--oxygen white dwarf and possibly preventing
the electron-capture induced explosion of the oxygen--neon--magnesium cores of
massive (super-)AGB stars.
While more massive stars
(with birth masses $\gsim 8$ M$_\odot$)
are incapable of avoiding core collapse, mass loss
during the red supergiant (RSG) phase can severely deplete the mantle of the
star and even force a return to the blue
(Georgy 2012; Georgy et al.\ 2012).
Besides the dramatic effect on stellar evolution, mass loss determines the
stellar remnant distribution.
Theoretical models (Dwek 1998) show that stellar mass loss replenishes
the gas in the interstellar medium (ISM), and helps sustain star formation.
This material is enriched in nitrogen, helium, carbon and/or slow neutron
capture elements depending on the stellar mass (Ventura et al.\ 2018). These
cool stars also produce dust grains, that play an important role in the
temperature regulation of the ISM, chemistry, and the formation of planets.
Dust formed by RSGs may not survive the ensuing supernova (Laki\'cevi\'c
et al.\ 2015; Temim et al.\ 2015) and in any case grains are heavily processed
in the ISM (Jones et al.\ 2014).
The feedback occurs on timescales as rapid as $10^7$ yr for RSGs, affecting
the molecular cloud environment in which these stars had been born, to as long
as $10^{10}$ yr for $\sim1$ M$_\odot$ AGB stars, i.e.\ far from their
birthplaces
 (Ekstr\"om et al.\ 2012).
Stellar mass loss is thus a critical ingredient in driving galaxy evolution.

Fortunately, AGB stars and RSGs are relatively easy to detect, as they become
not only very luminous ($\sim10^{3.5-5.5}$ L$_\odot$) but also very red,
and thus stand out at infrared (IR) wavelengths above other types of
stars within galaxies (Davidge 2000, 2018).
The tenuous mantles of these inflated stars become easily excited
leading to radial oscillations,
and during the heaviest mass loss phases they pulsate in the fundamental mode
(AGB stars) or first overtone (RSGs) (e.g., Whitelock, Feast \& Catchpole
1991; Wood 2000). These pulsations, on timescales of
 $\sim10^2$--$10^3$
d, cause large variations in brightness that can exceed a magnitude even in
the IR. IR surveys for long period variables (LPVs) are therefore an effective
route to identifying the population of stars undergoing the heaviest and
dustiest mass loss. We have embarked on a programme to do exactly this, in the
Local Group spiral galaxy M\,33 (Messier 1771) otherwise known as Triangulum
(Hodierna 1654).

The proximity of M\,33 (distance modulus $\mu=24.9$ mag -- Bonanos et al.\
2006), modest size ($\sim1^\circ$) and favourable orientation angle
(56--$57^\circ$ -- Zaritsky, Elston \& Hill 1989; Deul \& van der Hulst 1987)
render M\,33 an exquisite target to study stellar mass loss, stellar
evolution, and feedback processes across a galaxy not all too dissimilar from
our own Milky Way -- which is mostly obscured from our view by the dust in the
Galactic plane (van Loon et al.\ 2003; Benjamin et al.\ 2005;
Ishihara et al.\ 2011) and plagued by distance ambiguities.
Populations of AGB stars and RSGs have been identified in M\,33 (Cioni et al.\
2008; Drout, Massey \& Meynet 2012), among which many are dusty LPVs (McQuinn
et al.\ 2007; Thompson et al.\ 2009; Javadi, van Loon \& Mirtorabi 2011a).

The main objectives of our project are described in Javadi, van Loon \&
Mirtorabi (2011c): to construct the mass function of LPVs and derive from this
the star formation history (SFH) in M\,33; to correlate spatial distributions
of the LPVs of different mass with galactic structures (spheroid, disc and
spiral arm components); to measure the rate at which mass is lost and dust is
produced and fed into the ISM; to establish correlations between the mass-loss
and dust-production rate, luminosity, and amplitude of an LPV; and to compare
the {\it in situ} gas and dust replenishment with the ISM mass and star
formation rate (SFR). Paper I in the series presented the photometric
catalogue of stars in the inner square kpc (Javadi et al.\ 2011a), with Paper
II presenting the galactic structure and SFH (Javadi, van Loon \& Mirtorabi
2011b), and Paper III presenting the mass-loss mechanism and dust production
rate (Javadi et al.\ 2013). Paper IV presented the photometric catalogue of
stars in a nearly square degree area covering much of the M\,33 optical disc
(Javadi et al.\ 2015), with Paper V presenting the SFH across the enlarged
survey area (Javadi et al.\ 2017). This is Paper VI, presenting the analysis
of the gas and dust mass return to the M\,33 disc.

\begin{figure}
\centerline{\psfig{figure=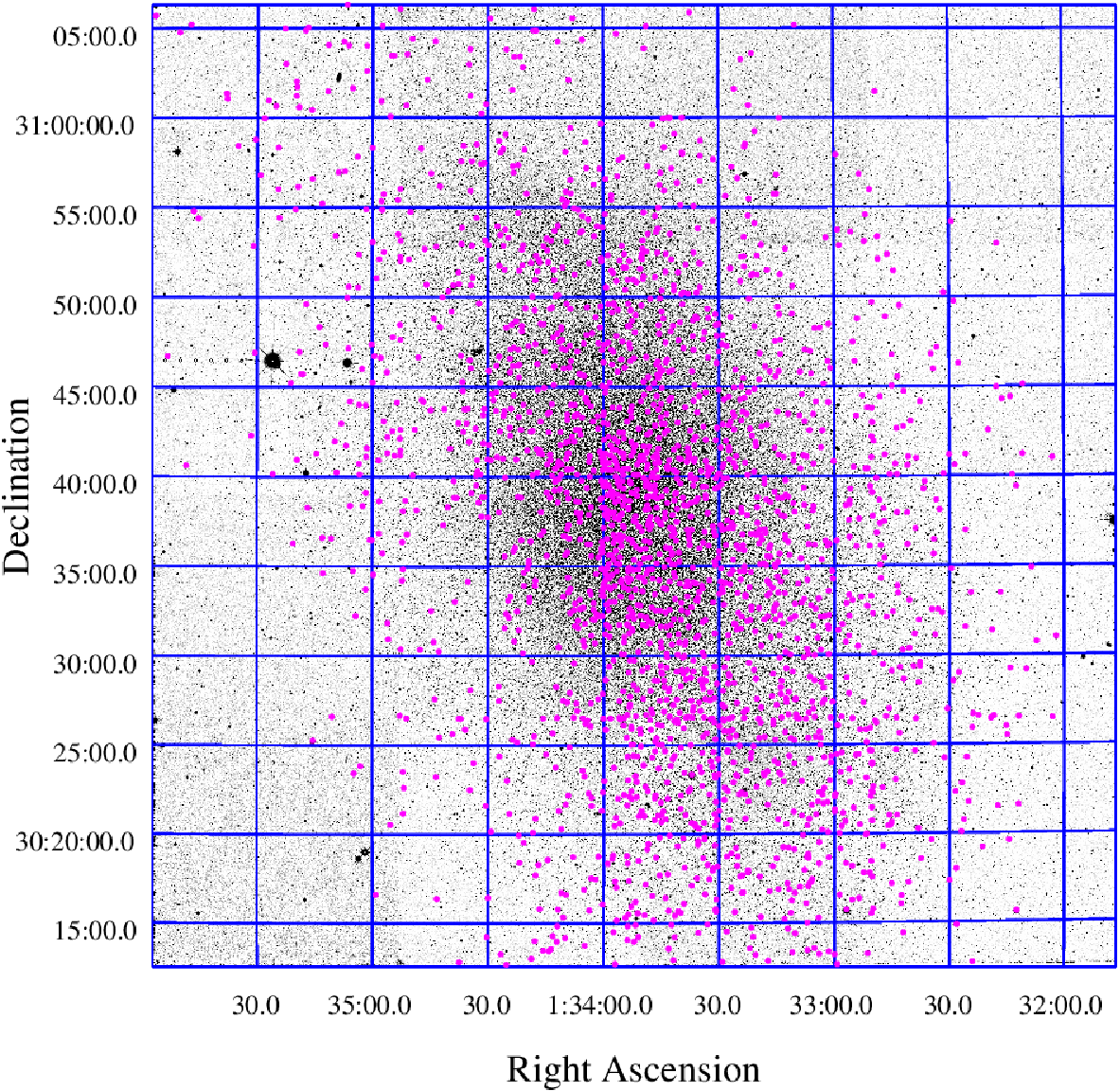,width=84mm}}
\caption[]{Mosaic image of M\,33 with overlain in magenta those WFCAM
variables that are also identified with {\it Spitzer}.}
\end{figure}

\section{Data}

To derive the mass-loss rates of evolved stars we make use of two data sets;
our own near-IR data in the J, H and K$_{\rm s}$ bands (Javadi et al.\ 2015)
and archival mid-IR {\it Spitzer} data at 3.6, 4.5 and 8 $\mu$m (McQuinn et
al.\ 2007). Below we describe each of these data sets in detail.

The photometry has not been corrected for interstellar reddening. Due to
the high Galactic latitude ($-31^\circ$) and favourable orientation, the total
(foreground+internal) reddening to stars within M\,33 is typically low:
$E(B-V)\sim0.10$ mag (Freedman, Wilson \& Madore 1991) or 0.16 mag (Massey et
al.\ 1995). For a standard ($R_V=A_V/E(B-V)=3.1$) reddening law, this
translates into a visual extinction of $A_V\sim0.3$--0.5 mag but only
$A_{K{\rm s}}<0.05$ mag and $(J-K_{\rm s})<0.08$ mag. Indeed, comparison between
the observed colours and theoretical isochrones (e.g., papers III--V) suggests
this is the case for the bulk of the stars. At mid-IR wavelengths the
attenuation diminishes further still (see van Loon et al.\ 2003). Compared to
the effects of circumstellar reddening when mass loss starts to matter, and
the uncertainties in the photometry and scaling relations that we shall derive
later on, interstellar reddening is generally negligible. Occasionally,
though, stars may experience more severe interstellar reddening. These cases
are difficult to ascertain individually, but we return to this issue when
interpreting the inferred mass-loss rates of the fainter, non-variable stars.

\subsection{Near-IR data}

In paper IV we described the
observations and
method employed to identify large-amplitude LPVs across the galactic disc of
M\,33 with the WFCAM imager on the United Kingdom IR Telescope (UKIRT) on
Mauna Kea, Hawai'i. The area that was monitored covers almost a square degree
($53^\prime\times53^\prime$) encompassing most of the star-forming disc of M\,33.

We refer the reader to paper IV for all details pertaining to the data
and variability search, but briefly summarise here the salient points.
The observations were done
at 5--8 epochs
between 2005--2007 in the K$_{\rm s}$ band (2.2 $\mu$m),
reaching a depth of $K_{\rm s}\approx 19$ mag,
with occasional observations in the J and H bands (1.28 and 1.68 $\mu$m,
respectively) for the purpose of obtaining colour information. The photometric
catalogue comprises 403\,734 stars, among which 4643 stars were identified as
LPV AGB stars, super-AGB stars and RSGs.

\subsection{Mid-IR data}

Five epochs of {\it Spitzer} Space Telescope imagery at 3.6, 4.5 and 8 $\mu$m
were analysed by McQuinn et al.\ (2007) to identify variable stars using a
similar method to ours.
We refer to McQuinn et al.\ (2007) for all details pertaining to the data
and variability search, but briefly summarise here the salient points. The
observations were done between 2004--2006, thus substantially overlapping with
our WFCAM survey. The photometry typically reaches $[3.6]\approx 17$ mag and
$[8]\approx 14$ mag except in the most crowded regions.
The {\it Spitzer} images covered nearly a square degree, (only) slightly
larger than our WFCAM survey; out of 40\,571 {\it Spitzer} sources, only 2868
stars fall outside the WFCAM monitoring coverage. Among the stars that {\it
Spitzer} detected within the region in common with our survey, 36\,411 are
also in our photometric catalogue, down to a little below the RGB tip
(paper IV).

Of the stars in common, 985 stars were identified as variables in both surveys
(Fig.\ 1), but two were saturated and therefore excluded from further
analysis. This means that 3658 of the WFCAM variable stars were not identified
as variables in the {\it Spitzer} survey, which is mainly because of the
limitation of {\it Spitzer} in detecting the fainter, less dusty variable red
giants
(see figures 31 \& 32 in paper IV).
On the other hand, the {\it Spitzer} survey identified 2923 variables,
suggesting a one-third completeness level of the WFCAM variable star survey --
this agrees with our internal assessment from a comparison between the WFCAM
and UIST data on the central square kpc (Paper IV). Generally, both surveys do
well in detecting dusty variable AGB stars (and RSGs); this is crucial to
estimate mass-loss rates based on IR photometric data.

\subsection{Combined dataset}

The near-IR and [3.6] photometric completeness for the intermediate-mass
and high-mass AGB stars and RSGs is near 100 per cent (paper IV),
except for the very reddest carbon stars that can have $K_{\rm s}>20$ mag at
the distance of M\,33 (van Loon et al.\ 1997, 2006, 2008; Gruendl et al.\
2008;
Matsuura et al. 2009).
The latter are extremely rare, but depending on the exact number and (very
high) mass-loss rates they could potentially be important for stellar
evolution and the injection of carbon grains into the ISM.
We shall come back to this possibility in the discussion of our results.
OH/IR stars can also become very red (Wood et al.\ 1992; van Loon et al.\
2001a), but they remain relatively bright in the K$_{\rm s}$ band. For stars
with $12<K_{\rm s}<18$ mag and $(J-K_{\rm s})>2.5$ mag, we detected 1763 stars
among which we identified 1042 variables. This implies a completeness level
for the dustiest evolved stars of $\sim59$ per cent.

Figure 2 shows the mid-IR colour--magnitude diagrams (CMDs) for the stars that
are identified in both surveys, overlain with Padova isochrones (Marigo et
al.\
2017).
There is a generally excellent agreement between data and models, except for
very bright sources with K$_{\rm s}<13$ mag, and the brightest 8-$\mu$m sources.
As discussed in paper IV, stars with K$_{\rm s}<11$ mag and (often) redder
H--K$_{\rm s}$ than J--K$_{\rm s}$ are foreground stars, and possibly saturated.
Furthermore, a useful number of stars (4694 stars) are detected at 8 $\mu$m
because of excess emission from circumstellar dust, which allows direct
derivation of their mass-loss rates. These stars form a branch of increasing
mid-IR brightness with increasing near-/mid-IR colour. The brightest mid-IR
objects, [8] $<10$ mag, with very red colours, $([3.6]-[8])\sim5$ mag,
however, deviate from this main branch. These were not (in general) identified
by us as variable stars, and could be non-stellar in nature -- for instance
background galaxies or compact H\,{\sc ii} regions within M\,33
-- or young stellar objects (Meixner et al.\ 2006; Bolatto et al.\ 2007;
Gruendl \& Chu 2009; Woods et al.\ 2011; Jones et al.\ 2017). Indeed, the
Marigo et al.\ (2017) models cover the range of colours of the LPVs and do not
suggest there are (m)any redder sources we should have seen.

\begin{figure}
\centerline{\psfig{figure=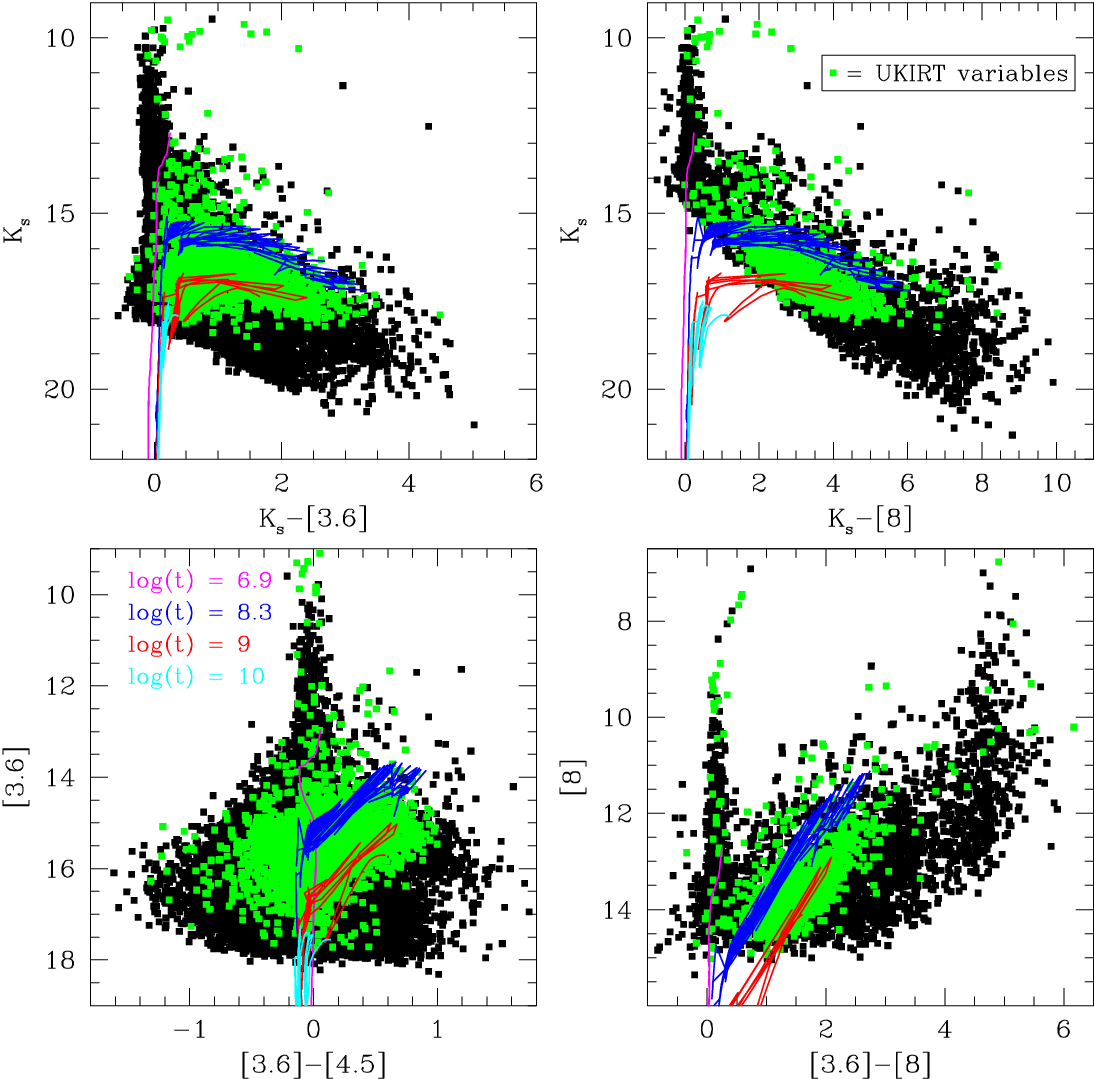,width=84mm}}
\caption[]{Near- and mid-IR colour--magnitude diagrams, overlain with
theoretical isochrones from Marigo et al.\
(2017)
for different ages. UKIRT (WFCAM) variables are highlighted in green.}
\end{figure}

\section{Methods}

We derive the mass-loss rates of the red giant stars in a two-step approach.
First, we model the spectral energy distributions (SEDs) of our near-IR
variables for which mid-IR counterparts have been identified. Then we use
these results to construct relations between the dust optical depth ($\tau$,
here defined at a wavelength of 2.2 $\mu$m)
and bolometric corrections (BCs) on the one hand, and near-IR colours on the
other. Finally, we apply these
near-IR scaling
relations to other red giants stars that were not modelled directly, to derive
their mass-loss rates too. This approach is applicable even for non-variable
stars, although in that case it is less certain that reddening is caused by
{\it circumstellar} dust. All mass-loss rates are expressed in units of
M$_\odot$ yr$^{-1}$ if not already stated so explicitly.

Because a sub-set of AGB stars, carbon stars have a different type of
circumstellar dust, we must try to identify which stars are likely to be
carbon stars. In the absence of spectroscopic confirmation for most of these,
and the limited constraints we have from photometry, we resort to making use
of theoretical expectations. Correcting the observed colours for the effect of
circumstellar dust, we obtain an intrinsic K-band brightness. Using stellar
evolution models (Marigo et al.\ 2008; 2017) we convert this into a birth
mass, given that these are highly evolved stars that will not evolve much in
luminosity. Doing this for the intrinsic brightness renders the method
relatively insensitive to the adopted mass-loss recipe in the models, though
ultimately the goal of this work is to improve those mass-loss prescriptions.
For slightly sub-solar metallicity, as is typical for the intermediate-age
population across much of the disc of M\,33, the mass range for AGB stars to
become carbon stars spans $\sim1.5$--4 M$_\odot$. We thus classify stars for
which the estimated birth mass falls within that range, as carbon stars.
Realising that there is some uncertainty surrounding this classification, we
check against spectroscopically determined stellar types and consider how our
results are affected by potential misclassifications. For more details about
the comparisons with the models, we refer the reader to papers II and V.

\subsection{Modelling the spectral energy distribution}

To model SEDs of WFCAM variables we used the publicly available dust radiative
transfer code {\sc dusty} (Ivezi\'c \& Elitzur 1997). Only variables with at
least two measurements in near-IR bands (K$_{\rm s}$ and J and/ or H) and two
mid-IR bands (3.6, 4.5 and/or 8 $\mu$m) were modelled. We modelled 294 stars,
picked randomly from among the 2185 stars that satisfy the above criteria.
In addition, we modelled the SEDs of the remaining 24-$\mu$m sources from
Montiel et al.\ (2015) -- because they could potentially contribute a large
fraction to the total dust and mass budget -- and discuss the results in a
separate section.

{\sc dusty} calculates the radiation transport in a dusty envelope. We fixed
the input temperatures of the star and of the dust at the inner edge of the
circumstellar envelope, at 3000 and 900 K, respectively. The density structure
is assumed to follow from the analytical approximation for radiatively driven
winds
(Ivezi\'c, Nenkova \& Elitzur 1999). This obviates the need to assume or
measure the outflow velocity, as it is implicit in the relation between
luminosity, optical depth, gas-to-dust mass ratio and mass-loss rate (see
below).
In some cases no acceptable match to the observed SED could be accomplished,
and for these cases we changed the stellar
temperature or the dust temperature at the inner edge.
While molecular bands affect in particular the optical spectrum of the
underlying star, we do not use optical photometry to constrain the SED and the
difference between using (unconstrained) template spectra and black-bodies for
the derived values of $\tau$ and $L$ is within the margins of error resulting
from assumptions about the circumstellar envelope.
We used amorphous carbon dust (Hanner 1988) and a small amount
(15 per cent)
of silicon carbide (P\'egouri\'e 1988) for carbon stars, and astronomical
silicates (Draine \& Lee 1988) for M-type stars. Any misclassification of
carbon stars or M-type stars will typically cause an order of magnitude
difference in the estimated mass-loss rate, and henceforth we consider the
effect of potentially having been too generous in classifying stars as carbon
stars.

We tried different values for $\tau$, and for each of these the luminosity
($L$) was scaled until an acceptable match was obtained, on visual inspection.
This scaling factor only depends on the measured flux and the distance to
M\,33, to which we assumed all stars belong. The BC value for any waveband is
then determined from the difference between the bolometric magnitude for that
luminosity and distance, and the measured magnitude in that band.
We assumed different values for the gas-to-dust mass ratio ($\psi$) depending
on the distance from the centre of M\,33, to reflect the metallicity gradient
(Gratier et al.\ 2017): $\psi=200$ within 2.5 kpc from the centre, 250 within
2.5--3.5 kpc, 300 within 3.5--4, 350 within 4--4.5, and 400 beyond 4.5 kpc.
This corresponds, roughly, to a range between solar and that typical of the
Large Magellanic Cloud (LMC), and is consistent with the radial variation of
the interstellar gas-to-dust ratio across M\,33 (Rela\~no et al.\ 2018).
The mass-loss rate then follows from the self-similar description of
radiation-driven winds (see below), determined by the values for $\tau$, $L$
and $\psi$. Thus,
we obtained optical depth and luminosity and hence mass-loss rate ($\dot{M}$)
for 294 stars. Based on the estimated birth mass of the star (see Paper V),
117 are M-type stars and 177 are carbon stars ($M_{\rm birth}\sim1.5$--4
M$_\odot$) (Table 1); examples are presented in figures 3 (carbon stars) and 4
(M-type stars)
and the full set of SED plots are available electronically.

These graphs suggest that even with a small number of data points, when the
free parameters are limited, we still can constrain the model quite well.
While it is not always possible to tell, on the basis of the fit, which
species of dust is present, the preferred dust species often yields better
fits and hardly ever yields worse fits. In some stars the 4.5-$\mu$m datum is
anomalously faint compared to adjacent bands. This may be due to molecular
absorption, which was not included in our SED modelling. The fundamental
ro-vibrational band of CO at 4.6 $\mu$m can be strong, especially in the
extended atmospheres of pulsating red giant stars (cf.\ Nowotny et al.\ 2013).
The 3-$\mu$m C$_2$H$_2$+HCN band is very strong in carbon stars, but it falls
largely outside the IRAC 3.6-$\mu$m bandpass; the 3.8-$\mu$m C$_2$H$_2$ band,
while strong in metal-poor carbon stars (van Loon et al.\ 2006, 2008), is not
expected to be nearly as strong in the carbon stars in the disc of M\,33.

\begin{figure}
\centerline{\psfig{figure=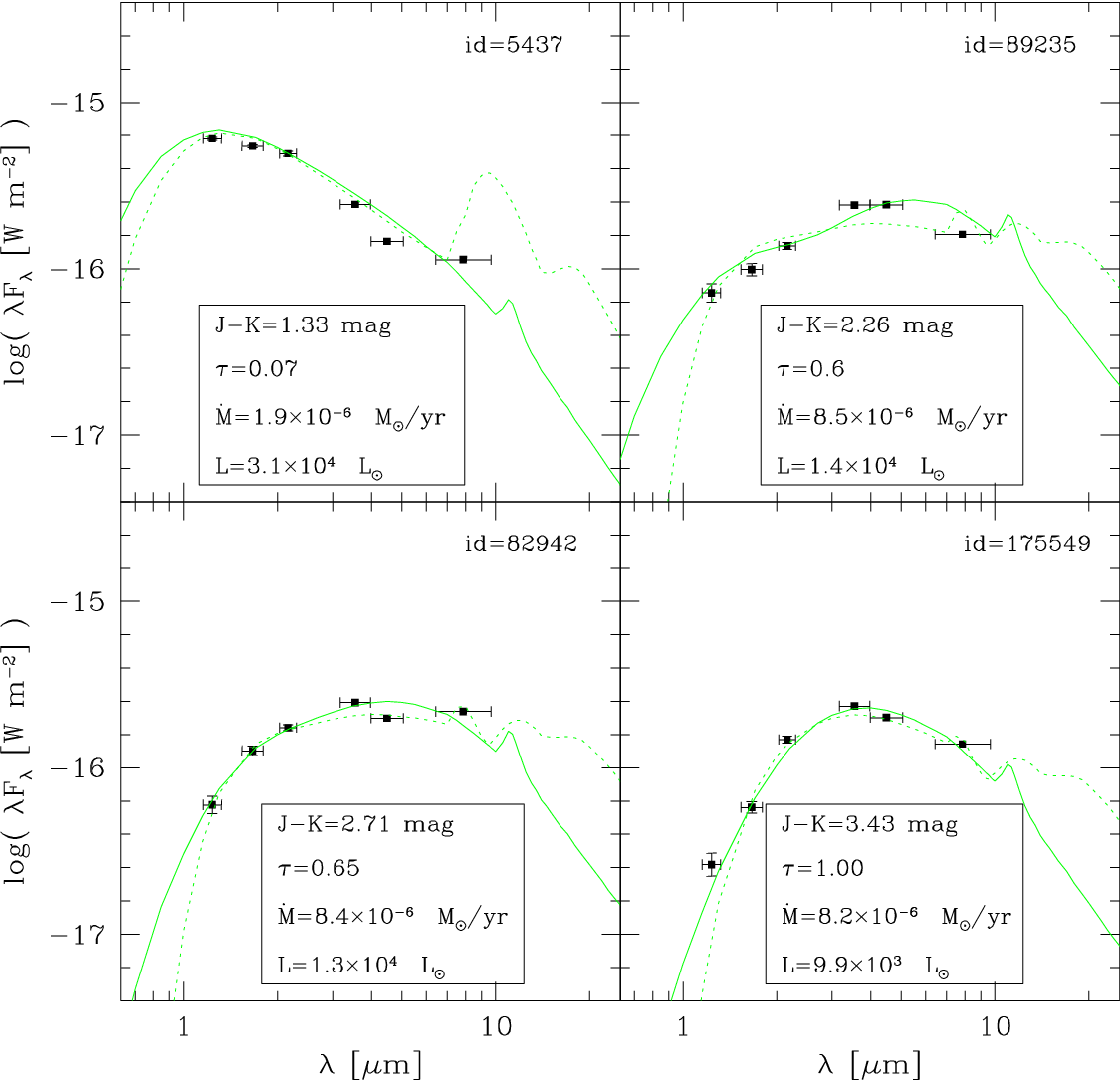,width=84mm}}
\caption[]{Example SEDs of presumed carbon stars. The horizontal "errorbars"
on the data represent the width of the photometric passbands. The best
matching SEDs modelled with {\sc dusty} are shown with solid lines. For
comparison, the best matching fits using silicates are shown with dotted
lines.}
\end{figure}

\begin{figure}
\centerline{\psfig{figure=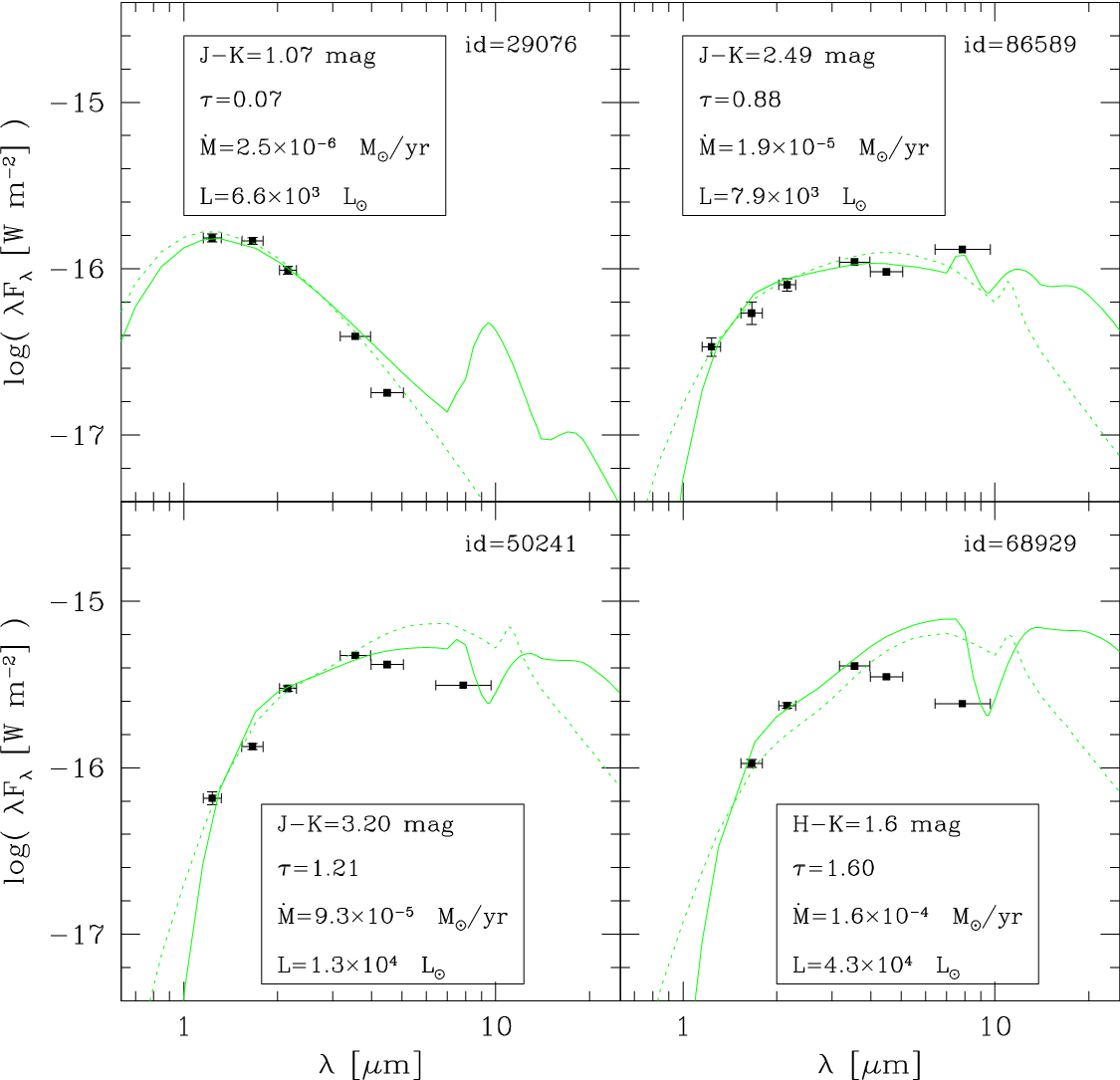,width=84mm}}
\caption[]{Example SEDs of presumed M-type stars. The horizontal "errorbars"
on the data represent the width of the photometric passbands. The best
matching SEDs modelled with {\sc dusty} are shown with solid lines. For
comparison, the best matching fits using amorphous carbon and silicon carbide
are shown with dotted lines.}
\end{figure}

\begin{table}
\caption{UKIRT ID No.\ (Paper IV), optical depth ($\tau$), luminosity ($L$)
and mass-loss rate ($\dot{M}$) determined by modelling the SED with {\sc
dusty}. Note that there is a great deal of uncertainty in the classification
into carbon and M-type stars. (The full table is available electronically.)}
\begin{tabular}{rccc}
\hline\hline
   ID & $\tau$ & $\log L/{\rm L}_\odot$ & $\log\dot{M}$ (M$_\odot$ yr$^{-1}$)  \\
\hline
\multicolumn{4}{l}{\it M-type stars}                                        \\
 1294 & 0.01   & 4.86                  & $-5.53$                            \\
 2492 & 0.15   & 5.01                  & $-4.34$                            \\
20682 & 0.15   & 3.90                  & $-5.31$                            \\
...   & ...    & ...                   & ...                                \\
\multicolumn{4}{l}{\it carbon stars}                                        \\
 5478 & 0.04   & 4.26                  & $-6.44$                            \\
12019 & 0.45   & 4.49                  & $-5.09$                            \\
12715 & 0.62   & 4.53                  & $-4.63$                            \\
...   & ...    & ...                   & ...                                \\
\hline
\end{tabular}
\end{table}

\subsection{Scaling relations}

On the basis of self-similarity of radiatively-driven winds there are scaling
relations (Ivezi\'c \& Elitzur
1995,
1997), such that the combination of $(\tau L^{3/4})/(\psi^{1/2}\dot{M})$ is
approximately constant
(Ivezi\'c et al.\ 1999).
Figure 5 suggests the scatter is $\sim0.2$ dex, due to slight mismatches of
$\tau$ and BC$_{K{\rm s}}$ to the exact shape of the SED. The offset of
$\sim0.7$ dex between M-type and carbon stars is due to the different
opacities of the grains. Carbon stars fitted with silicates end up on the
M-type locus, but the associated mass-loss rate would become higher.

\begin{figure}
\centerline{\psfig{figure=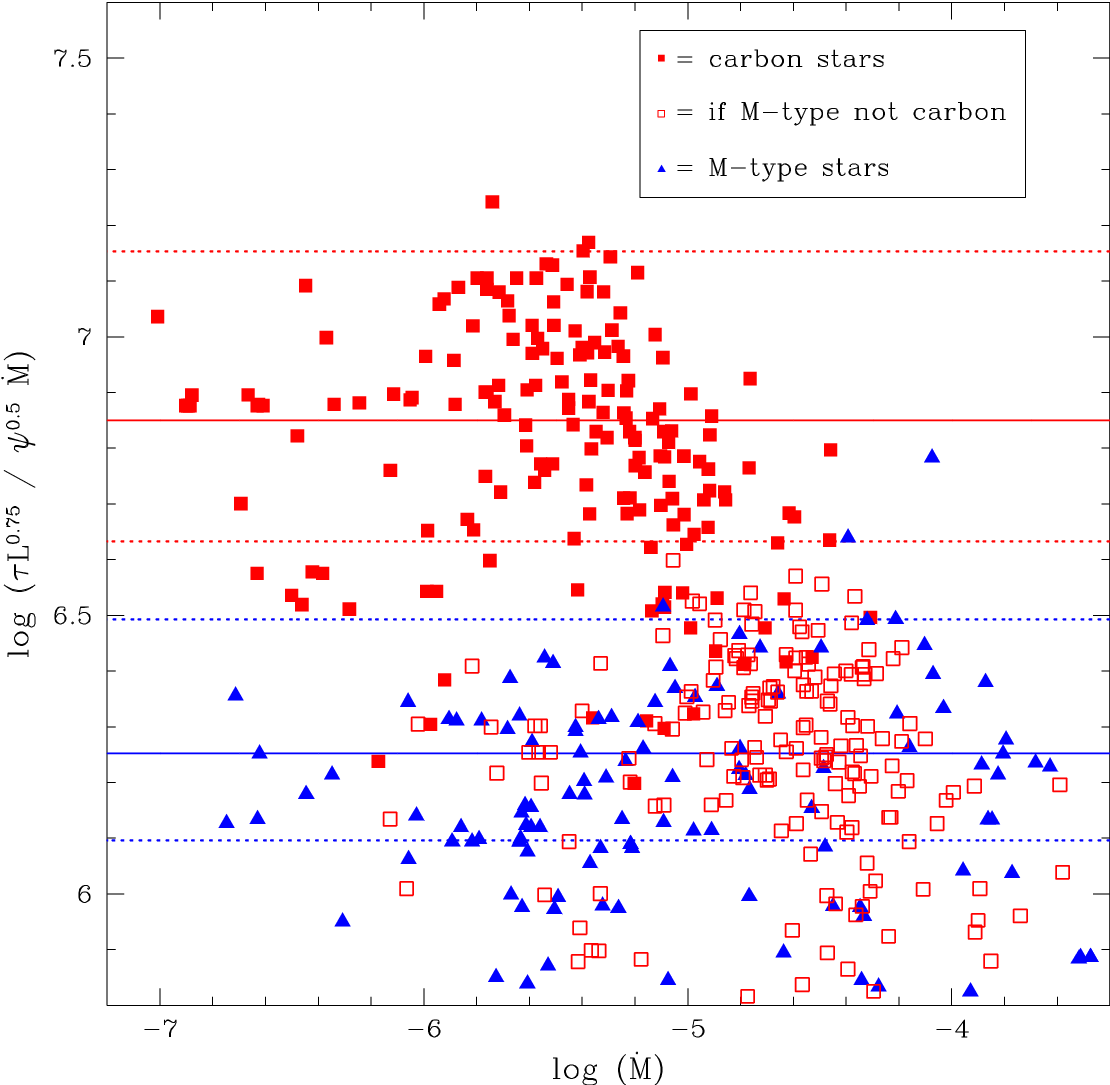,width=84mm}}
\caption[]{Combinations of the optical depth ($\tau$), luminosity ($L$),
gas-to-dust mass ratio ($\psi$) and mass-loss rate ($\dot{M}$); the horizontal
lines show the average values (solid) and $\pm$ standard deviations (dotted).
The open red squares show the results if the carbon stars are presumed to be
oxygen-rich instead.}
\end{figure}

The $\tau$ versus J--K$_{\rm s}$ and H--K$_{\rm s}$ relations show different
dependencies with colour for M--type stars and carbon stars, with less scatter
for J--K$_{\rm s}$ (Fig.\ 6). These relations quantify how, on average, the
dust causes photometric reddening. When J- (and K$_{\rm s}$-) band photometry
is available, but no mid-IR photometry, the optical depth is determined by
applying the relation constructed between $\tau$ and J--$K_{\rm s}$; if only
H- (and K$_{\rm s}$-, but not J-) band photometry is available, the optical
depth is estimated by using the relation constructed for H--K$_{\rm s}$ (Table
2). The relation between $\tau$ and J--K$_{\rm s}$ for both chemical types
tends to bend a little downwards compared to what we determined for the
central region of M\,33 (see Paper III). It is possible that grain properties
depend on the environment in which they form, and this may vary with location
in a galaxy. We stress that the fairly good relations between optical depth
and colours are due partly to the fact that we have used a single opacity
table
(for each of the two main dust types)
in the modelling with {\sc dusty}. This means the optical depth is only
scattered because of different luminosities and some uncertainty in
determining the best fit; the empirical colours are obviously affected by
photometric uncertainties.

\begin{figure}
\vbox{
\centerline{\psfig{figure=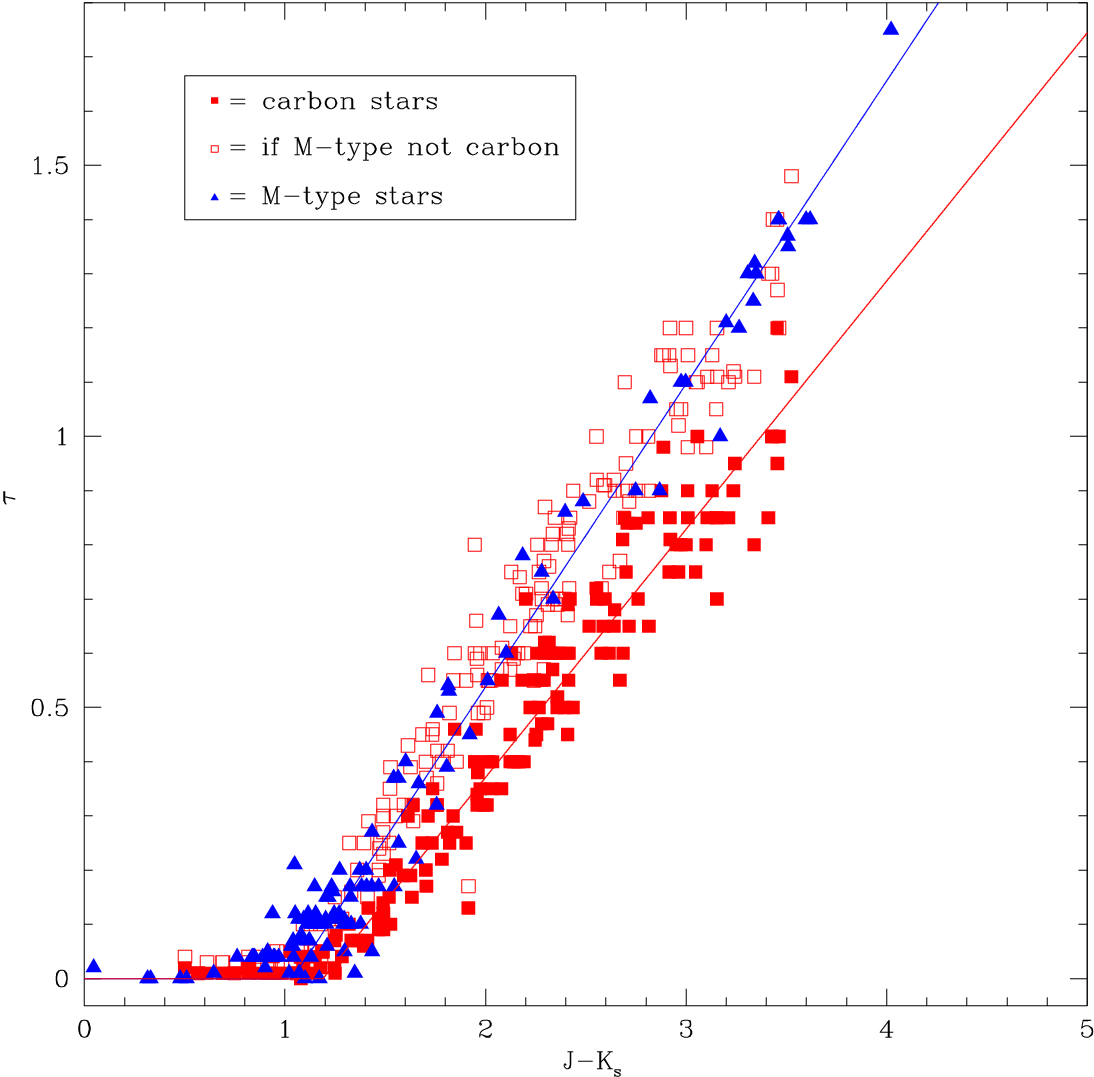,width=84mm}}
\centerline{\psfig{figure=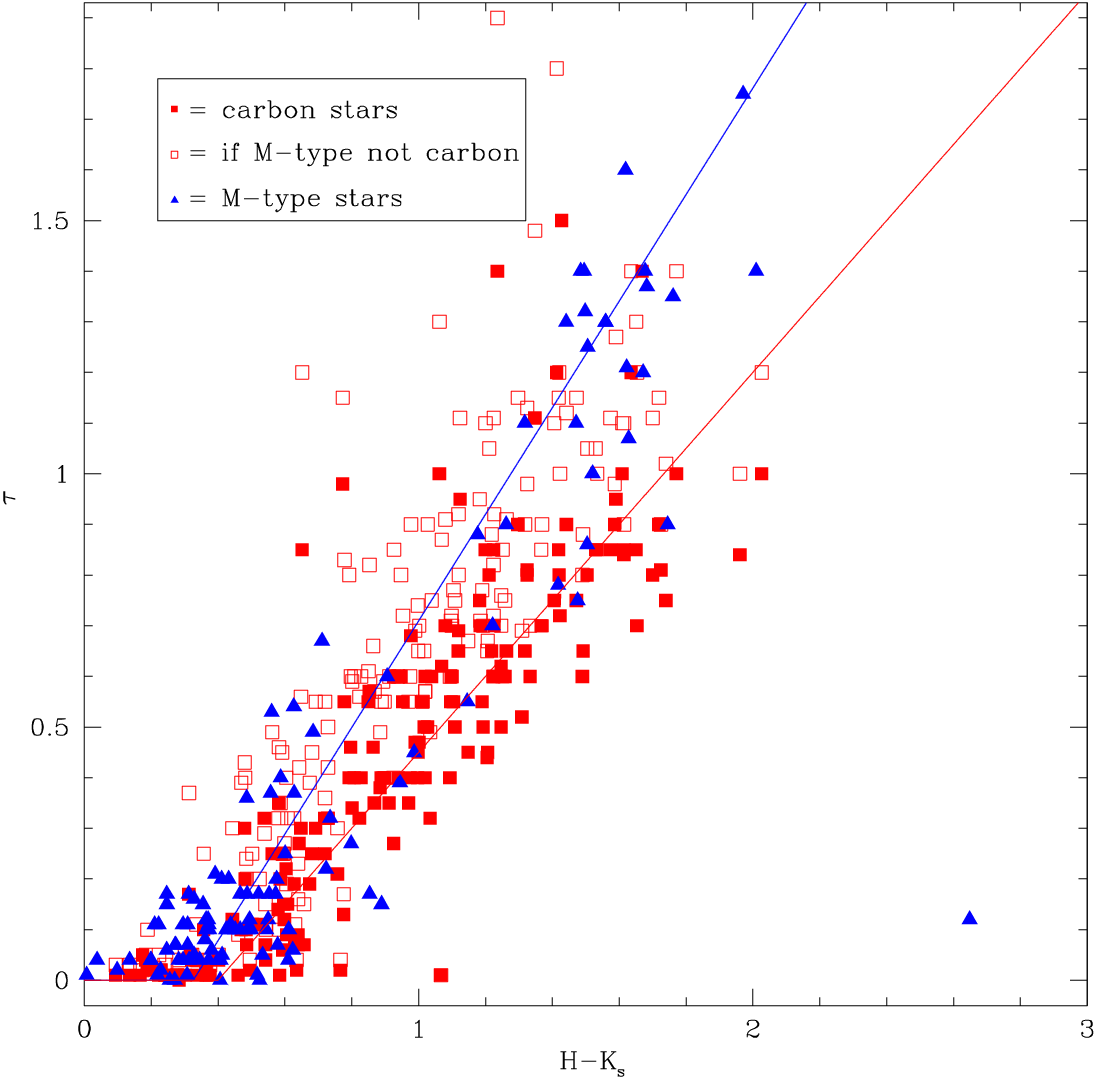,width=84mm}}
}
\caption[]{Relations between optical depth ($\tau$) and near-IR colours (top:
J--K$_{\rm s}$; bottom: H--K$_{\rm s}$) for carbon stars (red squares) and
M-type stars (blue triangles). The open red squares show the results if we
adopt oxygenous dust for the presumed carbon stars. The red and blue lines are
the fitted relations for carbon and M-type stars, respectively.}
\end{figure}

Similarly, the BC to the K-band (BC$_{\rm Ks}$) shows a different dependency on
colour for M-type and carbon stars (Fig.\ 7). Since this relation depends on
the underlying star (its temperature), presumed carbon stars fitted with
silicates end up between carbon stars fitted with carbonaceous dust and M-type
stars fitted with silicates. Eventually, the luminosity is determined for the
stars for which the SED cannot be modelled accurately by applying a
parameterisation of the relation between BC$_{\rm Ks}$ and J--K$_{\rm s}$ if $J$
is available, and H--K$_{\rm s}$ if $H$ (but not $J$) is available (Table 3).

The BCs presented in this paper are in good agreement with the relations
derived for the Magellanic Clouds (Groenewegen \& Sloan 2018). For example,
the peak of the BC$_{\rm Ks}$ versus J--K$_{\rm s}$ derived by Groenewegen \&
Sloan is at $(J-K_{\rm s})\sim2$ mag with $BC_{K{\rm s}}\sim2.6$ mag, and at
$(J-K_{\rm s})\sim4$ mag the derived value is $BC_{K{\rm s}}\sim2$ mag -- both
similar to what we find in M\,33. The shape of the relation is also similar.
The scatter around the mean relation is due to a variety of reasons,
including the effects of geometry of the circumstellar envelope and variations
in metallicity which can alter the photospheric colours and dust properties.
For the dustiest stars, with $(J-K_{\rm s})>5$ mag, the relation might become
flatter (Groenewegen \& Sloan 2018), but this does not affect our analysis as
there are no such red sources in the WFCAM catalogue.

\begin{figure}
\vbox{
\centerline{\psfig{figure=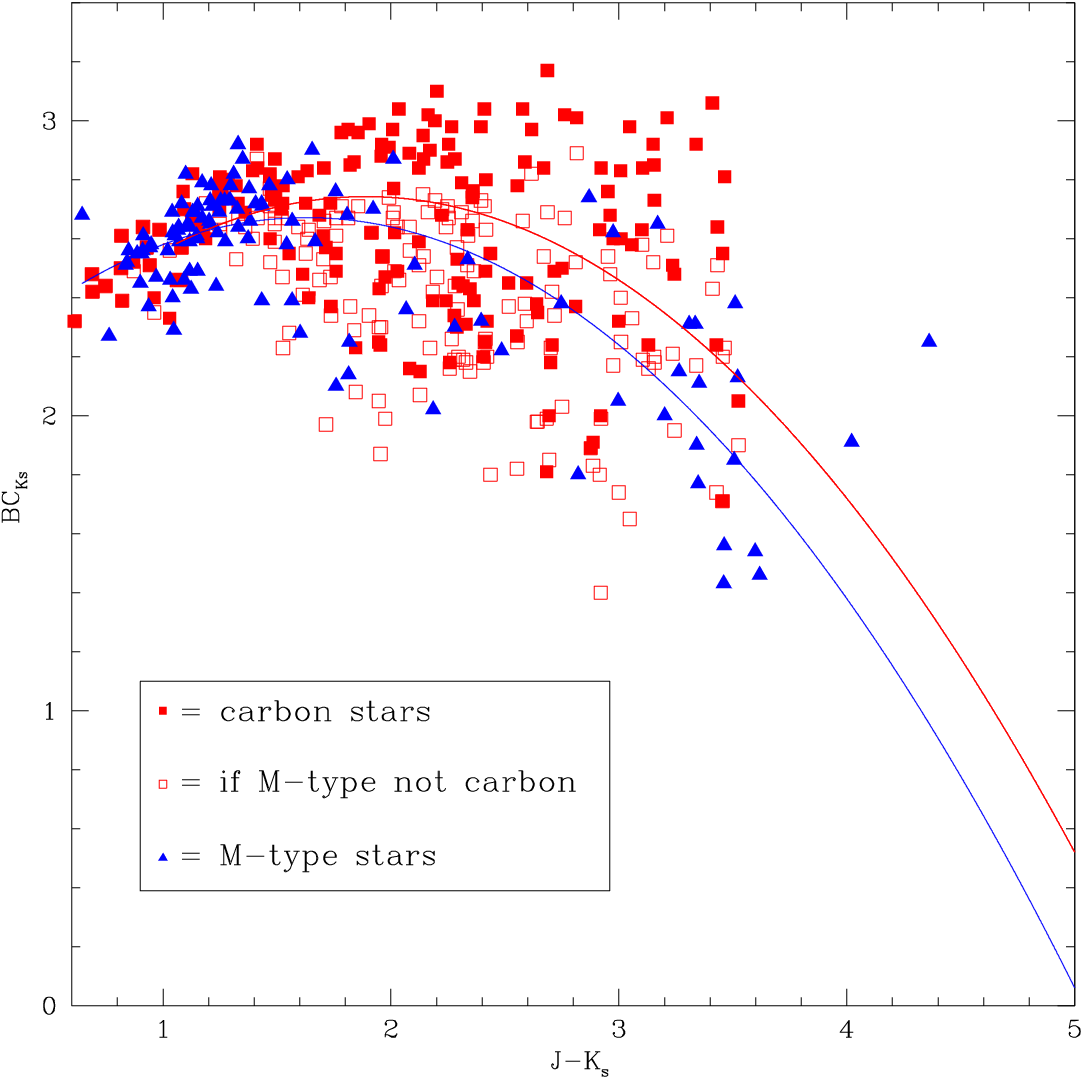,width=84mm}}
\centerline{\psfig{figure=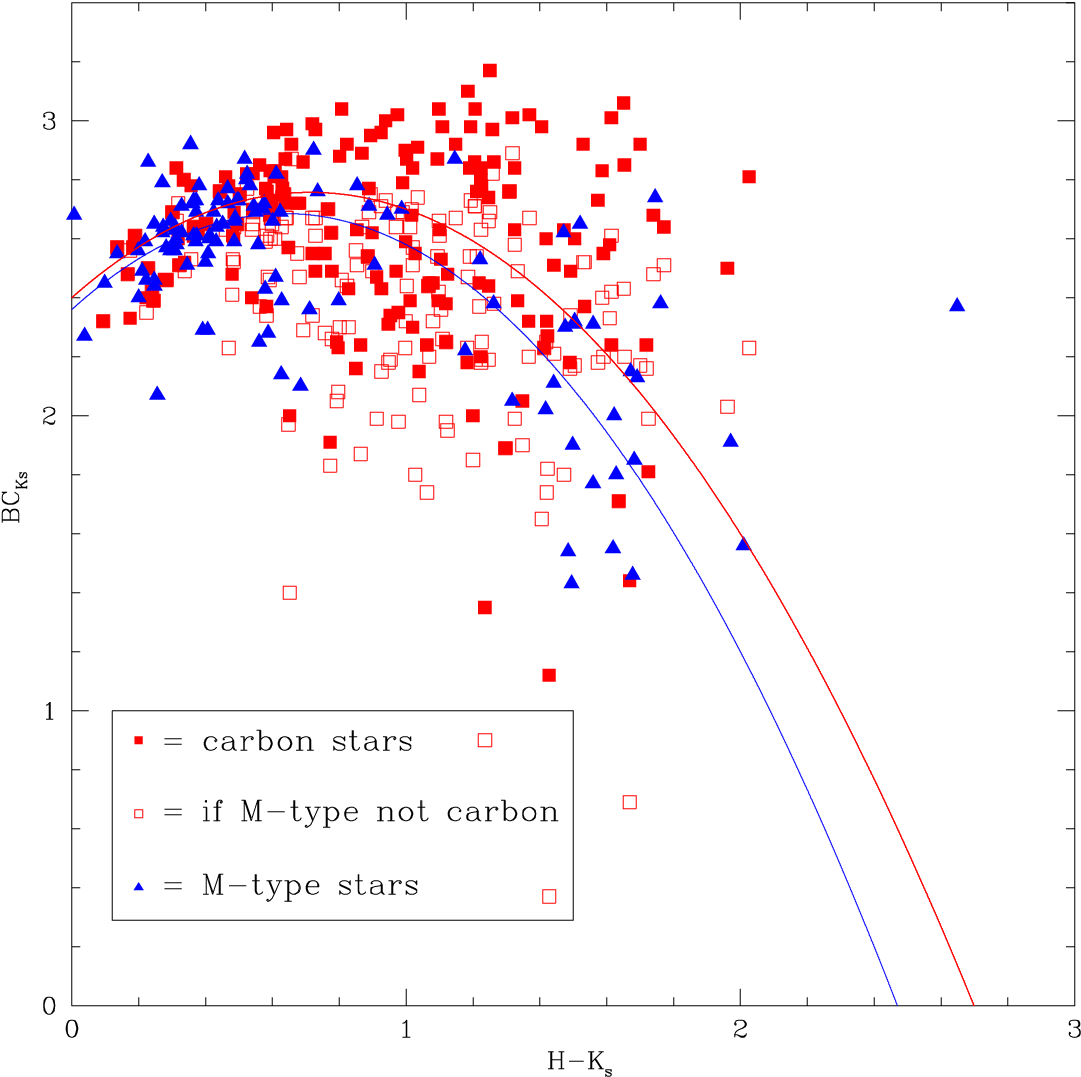,width=84mm}}
}
\caption[]{Relations between bolometric correction to the K$_{\rm s}$ band
(BC$_{\rm Ks}$) and near-IR colours (top: J--K$_{\rm s}$; bottom: H--K$_{\rm s}$)
for carbon stars (red squares) and M-type stars (blue triangles). The open red
squares show the results if we adopt oxygenous dust for the presumed carbon
stars. The red and blue lines are the fitted relations for carbon and M-type
stars, respectively.}
\end{figure}

Finally, when applying these relations to derive the luminosity and mass-loss
of other stars including non-variables, we apply the following criteria:

\begin{itemize}
\item{When stars have no J- or H-band magnitude but are variable, and closer
inspection revealed no suspicion regarding their photometry, we assign a
J-band magnitude equal to a (conservative) detection limit of $J=21$ mag. When
non-variable stars have neither a J- nor H-band magnitude we exclude them from
further analysis.}
\item{As discussed in Paper III, to account for the effect of photometric
errors in our estimate of the total mass return, we do accept negative
mass-loss rates which are a consequence of a negative optical depth for stars
with $(J-K_{\rm s})<1$ mag. This is a robust way to mitigate against an
otherwise unavoidable overestimation of the amount of dust produced by
non-dusty stars. However, in some cases when the J--K$_{\rm s}$ is unreliable
and the star is bright it can cause erroneously high negative mass-loss rates.
While statistically the number of these stars are few, they can have a huge
impact on the total mass return and they are therefore ignored.}
\item{Among the variables, 23 appear to have very large mass. Many of these
stars cannot be RSGs and they were removed from further analysis. We will
discuss these sources individually in the next section.}
\end{itemize}

%
%
\begin{table}
\caption{Parameterisations of relations between optical depth ($\tau$) and
near-IR colour (C) of the form $\tau=a+b\ C$ for $C\geq c$ mag (and $\tau=0$
for $C<c$ mag).}
\begin{tabular}{lccccccc}
\hline\hline
C             & a        & b     & c    & a        & b     & c     \\
& \multicolumn{3}{c}{\it carbon stars} & \multicolumn{3}{c}{\it M-type stars} \\
\hline
J--K$_{\rm s}$ & $-0.545$ & 0.457 & 1.190 & $-0.581$ & 0.559 & 1.036 \\
H--K$_{\rm s}$ & $-0.300$ & 0.750 & 0.400 & $-0.342$ & 1.052 & 0.325 \\
\hline
\end{tabular}
\end{table}

\begin{table}
\caption{Parameterisations of relations between the bolometric correction to
the K$_{\rm s}$ band (BC$_{\rm Ks}$) and near-IR colour (C) of the form
$BC_{K{\rm s}}=a+b\ C+c\ C^2$.}
\begin{tabular}{lcccccccc}
\hline\hline
C   \mbox{ } & & a    & b    & c                      & 
    \mbox{ } &   a    & b    & c                      \\
             & & \multicolumn{3}{c}{\it carbon stars} &
               & \multicolumn{3}{c}{\it M-type stars} \\
\hline
J--K$_{\rm s}$ & & 1.92 & 0.87 & $-0.23$                &
               & 2.06 & 0.75 & $-0.23$                \\
H--K$_{\rm s}$ & & 2.40 & 1.00 & $-0.70$                &
               & 2.36 & 1.02 & $-0.80$                \\
\hline
\end{tabular}
\end{table}

\section{Results}

In the following sub-sections we present the mass-loss rates
and
explore the dependency of the mass-loss rate on stellar parameters, and
present an assessment of the mass return rate into the ISM.
In Appendix A we scrutinise individual sources of a known and/or extreme
nature, in order to assess and enhance the reliability of our sample.

\subsection{Mass-loss rates}

The mass-loss rates and luminosities for selected sources described
 in Appendix A
are shown in figure 8. We have excluded the stars that we believe are not
cool, evolved stars
but also consider the contamination by other types of IR sources.

\begin{figure}
\centerline{\psfig{figure=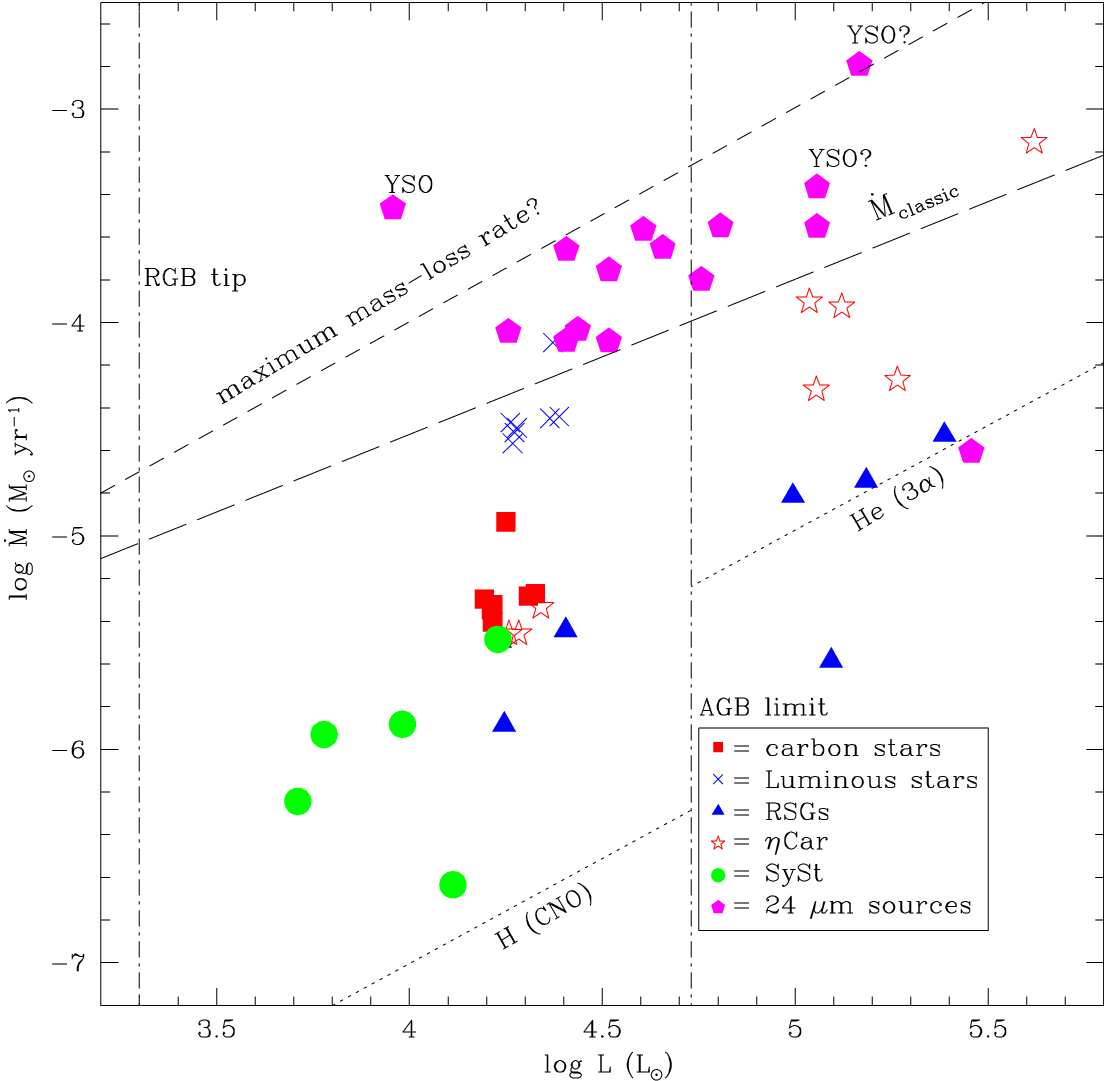,width=84mm}}
\caption[]{Mass-loss rate vs.\ luminosity, for spectroscopically confirmed
carbon stars from Block et al.\ (2007; red squares), 24-$\mu$m sources from
Montiel et al.\ (2015; magenta pentagons) luminous stars from Martin \&
Humphreys (2017; blue crosses), RSGs from Drout et al.\ (2012; blue
triangles), $\eta$\,Carin{\ae} analogues from Khan et al.\ (2013; red stars)
and symbiotic binary stars from Miko{\l}ajewska et al.\ (2017; green points).
The vertical dash--dotted lines mark the tip luminosity of the first ascent
red giant branch (RGB) and the classical limit of the most massive AGB stars
(excluding the effects of Hot Bottom Burning). The dotted lines trace the
mass-consumption rates by shell hydrogen burning (CNO cycle) on the AGB and
core helium burning (triple-$\alpha$ reaction) in RSGs. The dashed lines trace
the limits to the mass-loss rate in dust-driven winds due to single scattering
(classic) and multiple scattering (maximum?).}
\end{figure}

The different types of stars end up where one would expect them: symbiotic
stars are relatively low mass and unevolved, hence their luminosities around
$10^4$ L$_\odot$; carbon stars and some of the IR-luminous stars are further up
the AGB, with most of the known RSGs and $\eta$\,Car candidates reaching
luminosities in excess of $10^5$ L$_\odot$. Likewise, the symbiotic stars have
moderate mass-loss rates ($\sim10^{-6}$ M$_\odot$ yr$^{-1}$), optical carbon
stars and RSGs have slightly higher rates (several $10^{-6}$ M$_\odot$
yr$^{-1}$), and the ``luminous'' sources of IR emission and the most extreme
$\eta$\,Car look-alikes reach the highest rates (several $10^{-5}$ M$_\odot$
yr$^{-1}$). In particular, the 24-$\mu$m sources classified by us as AGB or
RSG are among the most luminous and heaviest mass-losing in the population;
one notable source being VC\,14 -- which is very luminous, and possibly among
the most massive RSGs in M\,33, but losing mass at a rate which is no more
(and no less) than its nuclear burning rate.

The few YSO candidates are also included in the plot;
obviously the {\sc dusty} modelling is not appropriate for YSOs, but it
shows the contamination they cause in a census of mass return. Likewise, the
$\eta$\,Car candidates may include hot(ter) stars for which the {\sc dusty}
modelling would need to be adjusted. Because insufficient information is
available to do this, it is outside the scope of this work.

For our complete sample,
some dependence of mass-loss rate on luminosity is seen (Fig.\ 9); the maximum
mass-loss rate increases with luminosity and the highest mass-loss rates are
generally achieved by the most luminous, most massive large-amplitude variable
stars. This confirms earlier studies in the central region of M\,33 (Paper
III) and in the Magellanic Clouds (van Loon et al.\ 1999b; Srinivasan et al.\
2009). 
While none of the stars in M\,33 attain as high mass-loss rates as given
by the ``maximum mass-loss rate'' ridgeline, especially at the high luminosity
end stars do exceed the single scattering limit. The ridgeline is an extreme
envelope of rates that were measured once in the past (van Loon et al.\
1999b), and given the uncertainties in those data and modelling assumptions it
is possible that the actual limit is somewhat lower -- hence we added a
question mark to it. Note also that among known sources only YSOs (for which
the models would not apply) appear to reach that limit -- see figure 8.
The mass-loss rates for M-type AGB stars and RSGs are similar to those
found in the Solar Neighbourhood (a few $\times10^{-5}$ and $10^{-7}$--$10^{-4}$
M$_\odot$ yr$^{-1}$, respectively; Jura \& Kleinmann 1989). The mass-loss rates
for presumed carbon stars are also in good agreement with those found in the
Milky Way (a few $\times10^{-5}$ M$_\odot$ yr$^{-1}$; Whitelock et al.\ 2006)
and in the Magellanic Clouds ($\sim10^{-5}$ M$_\odot$ yr$^{-1}$; Gullieuszik et
al.\ 2012).

\begin{figure}
\centerline{\psfig{figure=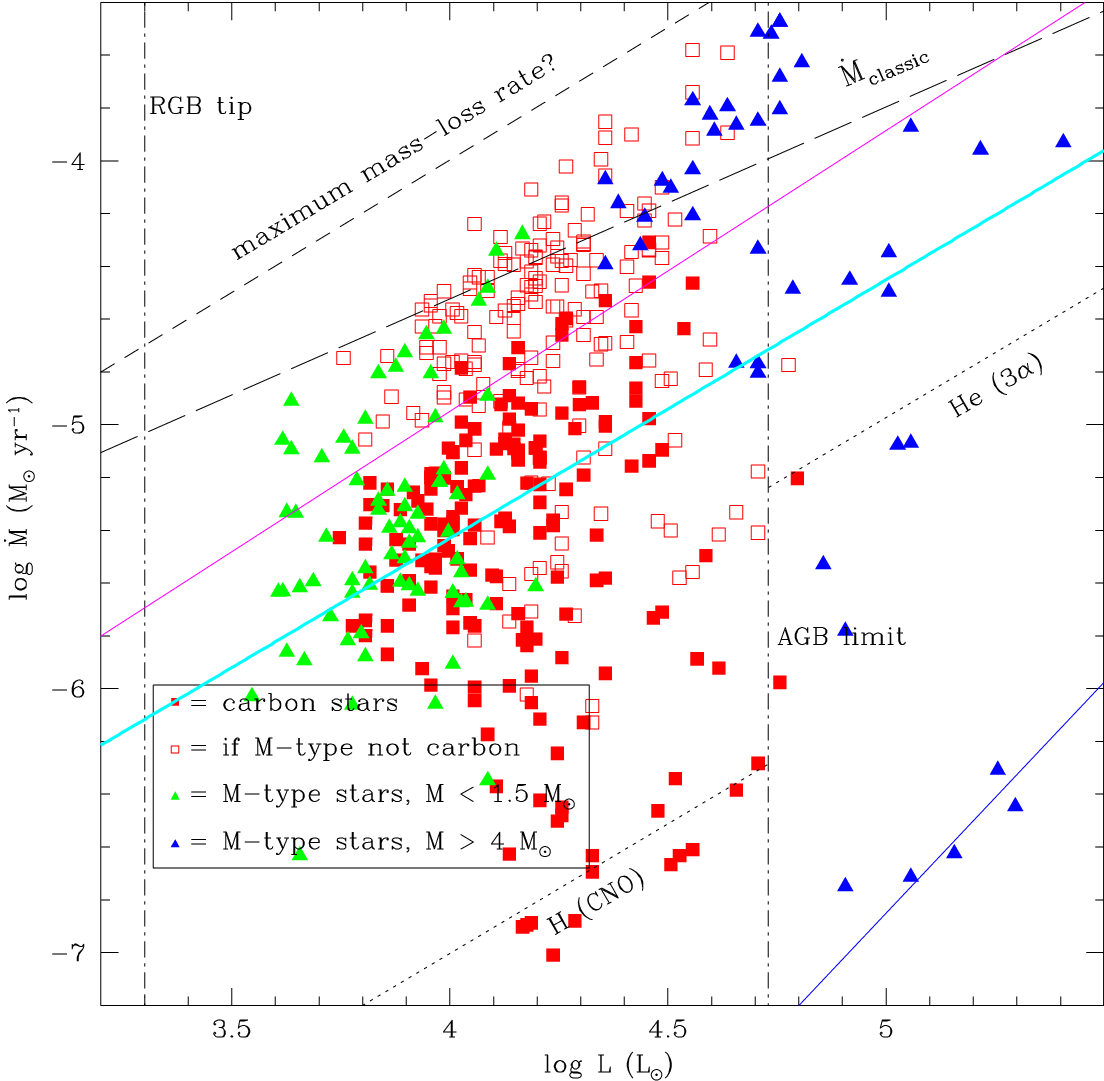,width=84mm}}
\caption[]{Mass-loss rate vs.\ luminosity, modelled with {\sc dusty} for
low-mass AGB stars (green triangles), intermediate-mass carbon stars (red
squares) and massive AGB stars and RSGs (blue triangles). The open red squares
show the results if the presumed carbon stars are presumed to be oxygen-rich
instead. The lines have the same meaning as in figure 8. The magenta line
traces the fit to the mass-loss rate vs.\ luminosity presented in Goldman et
al.\ (2017), whilst the blue line traces the relation found by Verhoelst et
al.\ (2009) for Galactic RSGs.
{\bf The cyan line is the relation given by Eq.\ (1), derived later.}
}
\end{figure}

It is reassuring to see that the RSGs
(certainly stars well above the AGB limit of $\log L/L_\odot=4.73$ --
Wood, Bessell \& Fox (1986))
are generally oxygenous; that the least luminous stars are too, and that the
maximum mass-loss rate increases with luminosity (in fact rather steeply).
Oxygenous stars around -- or slightly fainter than -- the AGB limit with very
high mass-loss rates are probably massive AGB stars, the equivalent of (most
of) the OH/IR stars that are found in the LMC (Wood et al.\ 1992; Marshall et
al.\ 2004; Goldman et al.\ 2017, 2018).

Much of the spread in mass-loss rate at a given luminosity is related to
stellar evolution (van Loon et al.\ 1999b, 2005a; Beasor \& Davies 2016, 2018;
cf.\ Groenewegen \& Sloan 2018). The mass-loss rate increases as the star
approaches its end point of evolution
 (see, e.g., Vassiliadis \& Wood 1993).
The least luminous carbon stars ($\log
L/{\rm L}_\odot<4$)
might not yet be
at the tip of their AGB because M-type AGB stars are found at the same
luminosities but higher mass-loss rates -- the latter
might be
more evolved than the carbon stars; but these carbon stars are not expected to
become M-type again (Marigo et al.\ 2008) so they must still evolve to higher
luminosities before they end their AGB evolution.
Variations in metallicity across M\,33 can also result in overlap between
M-type and carbon stars as the mass range for AGB stars to turn into carbon
stars depends on metallicity (see Marigo et al.\ 2017).
Alternatively, the less luminous carbon stars with lower mass-loss rates might
be in their inter-thermal pulse luminosity dip and/or the more luminous carbon
stars with high mass-loss rates might just be experiencing the aftermath of a
thermal pulse (Olofsson et al.\ 1990; Vassiliadis \& Wood 1993; Mattsson,
H\"ofner \& Herwig 2007). Near the AGB limit we might find relatively massive
carbon stars, which have turned into a carbon star in a final thermal pulse,
which will in fact have decreased the mass-loss rate. An iconic example is
IRAS\,04496$-$6958 in the LMC (van Loon, Zijlstra \& Groenewegen 1999).

Notwithstanding the above, none of the carbon stars seem to reach similar high
mass-loss rates as the most intensely mass-losing oxygenous stars of similar
luminosity.
Because carbon dust is more opaque than silicates are, at the same
mass-loss rates (and luminosities) carbon stars are redder than oxygenous AGB
stars (van Loon et al.\ 2008). The photometric completeness limits could
therefore have resulted in a disproportionate loss of the most extreme carbon
stars from our sample. Alternatively, the optical properties of the grains (be
it the carbon grains or the silicates) may not be accurate (Srinivasan,
Sargent \& Meixner 2011).


The stars with the highest mass-loss rates occupy a continuous sequence that
extends well beyond the classical AGB limit
(Figs.\ 9 and 10).
This strongly suggests that the super-AGB stars, those AGB stars that also
experience nuclear carbon burning (Doherty et al.\ 2017), become as cool and
dust-enshrouded as their less massive siblings; this confirms our earlier
findings in Papers III and VI. Massive AGB stars experience Hot Bottom Burning
(HBB; Iben \& Renzini 1983), which prevents them from becoming carbon stars
and enhances their luminosity above the classical core-mass--luminosity
relation (Boothroyd \& Sackmann 1992). Such stars can have luminosities
exceeding the classical AGB limit, making it difficult to distinguish between
an AGB star in the HBB phase and a super-AGB or RSG.

The empirical mass-loss prescription developed for samples of
OH/IR
stars in the LMC, Galactic Centre and Bulge by Goldman et al.\ (2017) is in
broad
agreement with the trends we see in M\,33. Their mass-loss rate formula is
essentially independent of gas-to-dust ratio, and while it has a dependence on
pulsation period the strongest dependence is with luminosity. In figure 9 we
plot, in magenta, a line we fitted to figure 19 in Goldman et al.\ (2017);
the highest mass-loss rates end up in between the Goldman relation and the van
Loon limit,
whilst somewhat less extreme oxygenous AGB stars and RSGs lie below it.

Since RSGs are typically warmer than AGB stars (van Loon et al.\ 2005a; cf.\
Bonanos et al.\ 2010), the maximum mass-loss rate achieved by RSGs is
comparable to those achieved by the most extreme, cooler and hence less
gravitationally bound AGB stars. As noted in van Loon et al.\ (1999b), RSGs
spend very little time in such extreme phase, or may not always reach it, with
most RSGs losing mass at much more modest rates. Indeed, the relation found by
Verhoelst et al.\ (2009) for Galactic RSGs (blue line in Fig.\ 9), and in good
agreement with the mass-loss rates of RSGs estimated by Groenewegen \& Sloan
(2018), matches the low rates of less evolved RSGs in M\,33.

The direct modelling with {\sc dusty}
has an internal consistency within $\pm0.2$ dex in the mass-loss rate of
oxygenous stars at a given luminosity, and $\pm0.26$ for carbon stars (see
figure 5).
However, the vast number of stars that were identified as variable -- let
alone those that were not --
lack mid-IR photometry and can therefore not be modelled in detail.
While this could mean these stars do not lose mass at high rate, it could
nevertheless lead to an under-estimation of the total mass lost during the
evolution of a star, and to the total mass returned to the ISM. Therefore, in
figure 10 we included all sources detected in our survey (except those
excluded for the reasons explained before), by using the relations we derived
earlier. This would be a grossly optimistic view on mass loss, as especially
the non-variable low-mass stars are more likely to be reddened by interstellar
reddening rather than circumstellar reddening. Indeed, the latter all but
vanish when only the variable sources are considered (right panels in figure
10). Interestingly, most of the RSGs with mass-loss rates $\sim10^{-5}$
M$_\odot$ yr$^{-1}$ are also lost, but the more extreme examples are retained.
On the other hand, extremely dusty carbon stars would have been missed.

\begin{figure*}
\centerline{\vbox{
\hbox{\psfig{figure=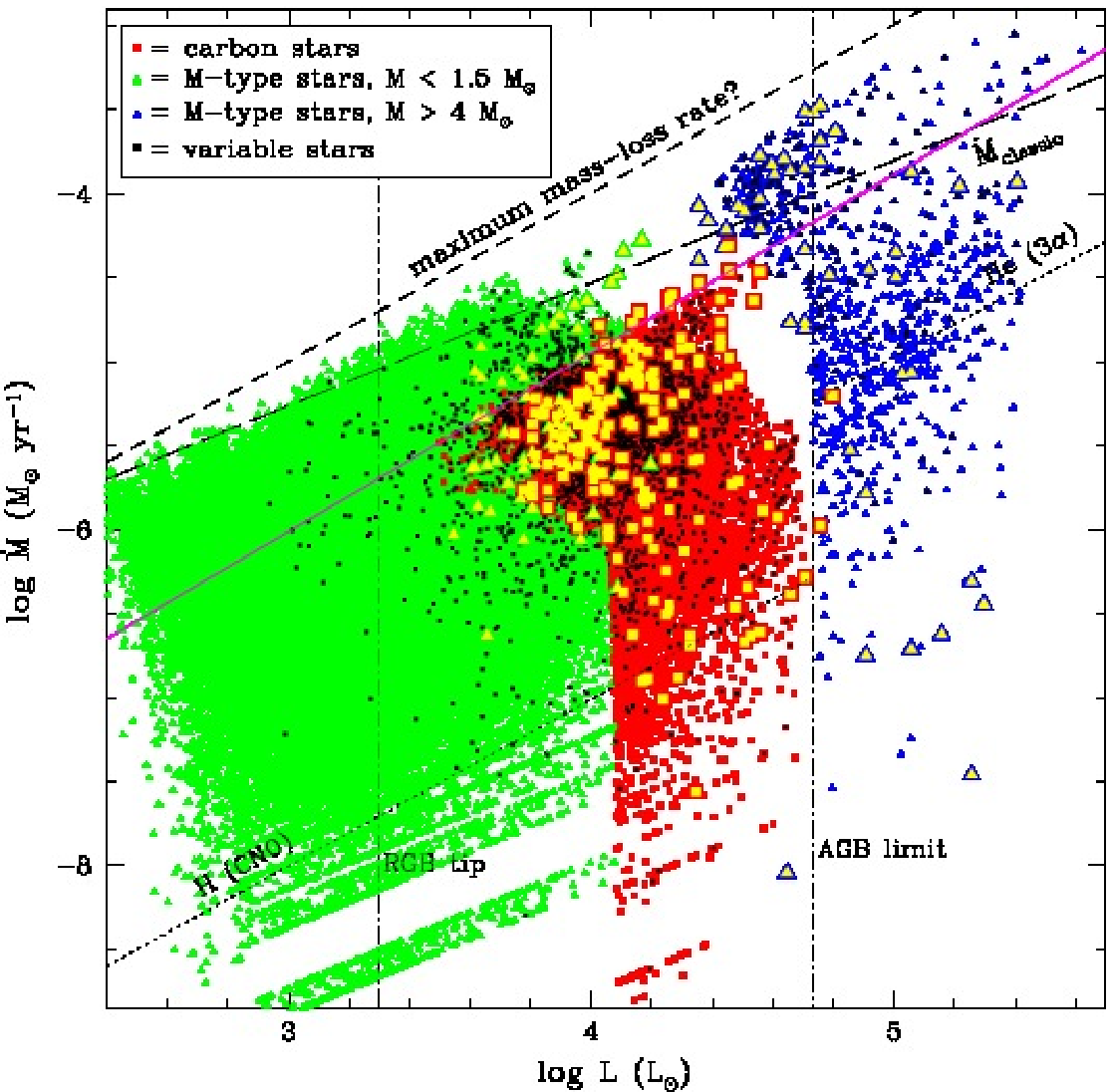,width=88mm}
\psfig{figure=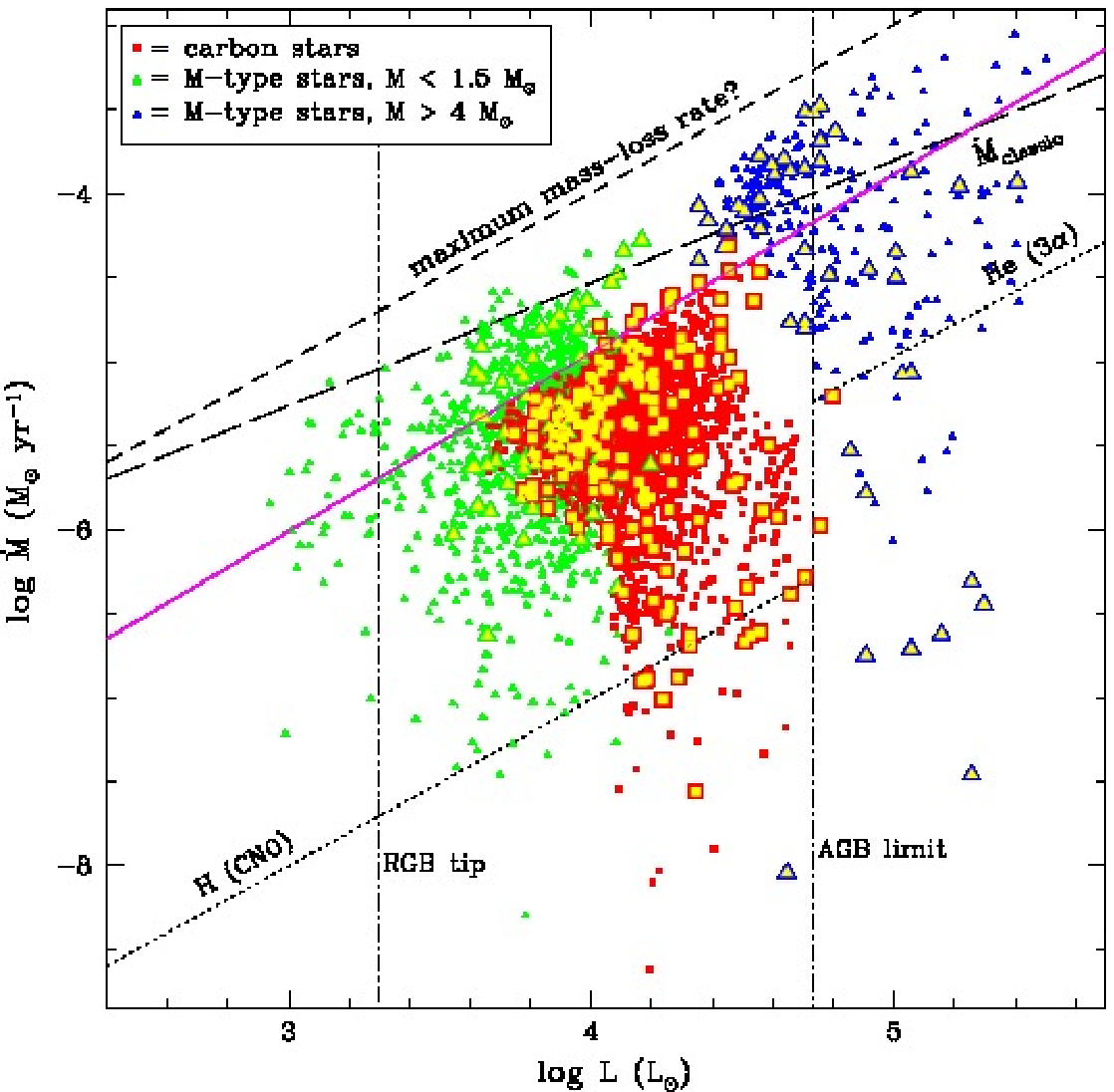,width=88mm}}
\hbox{\psfig{figure=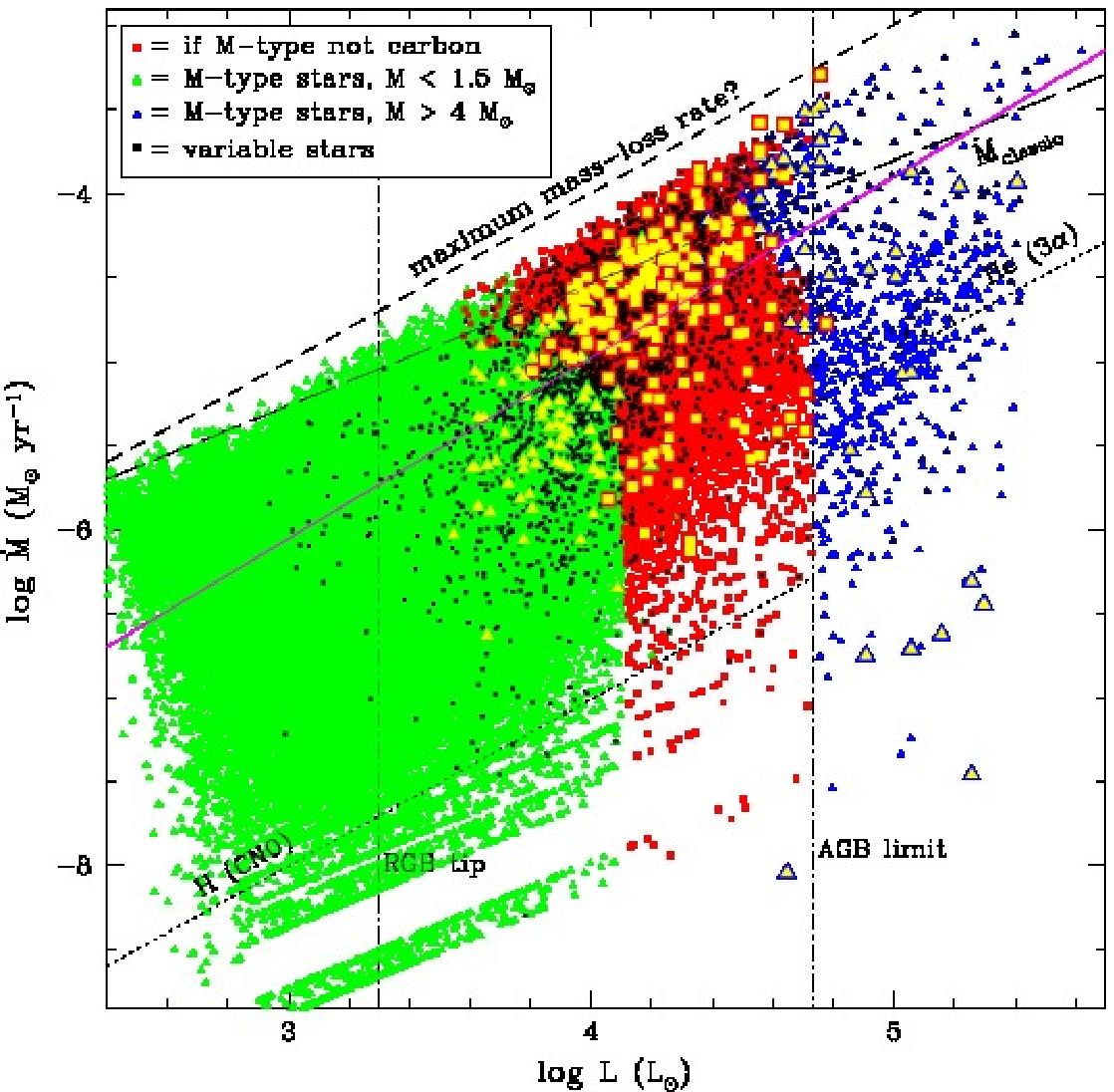,width=88mm}
\psfig{figure=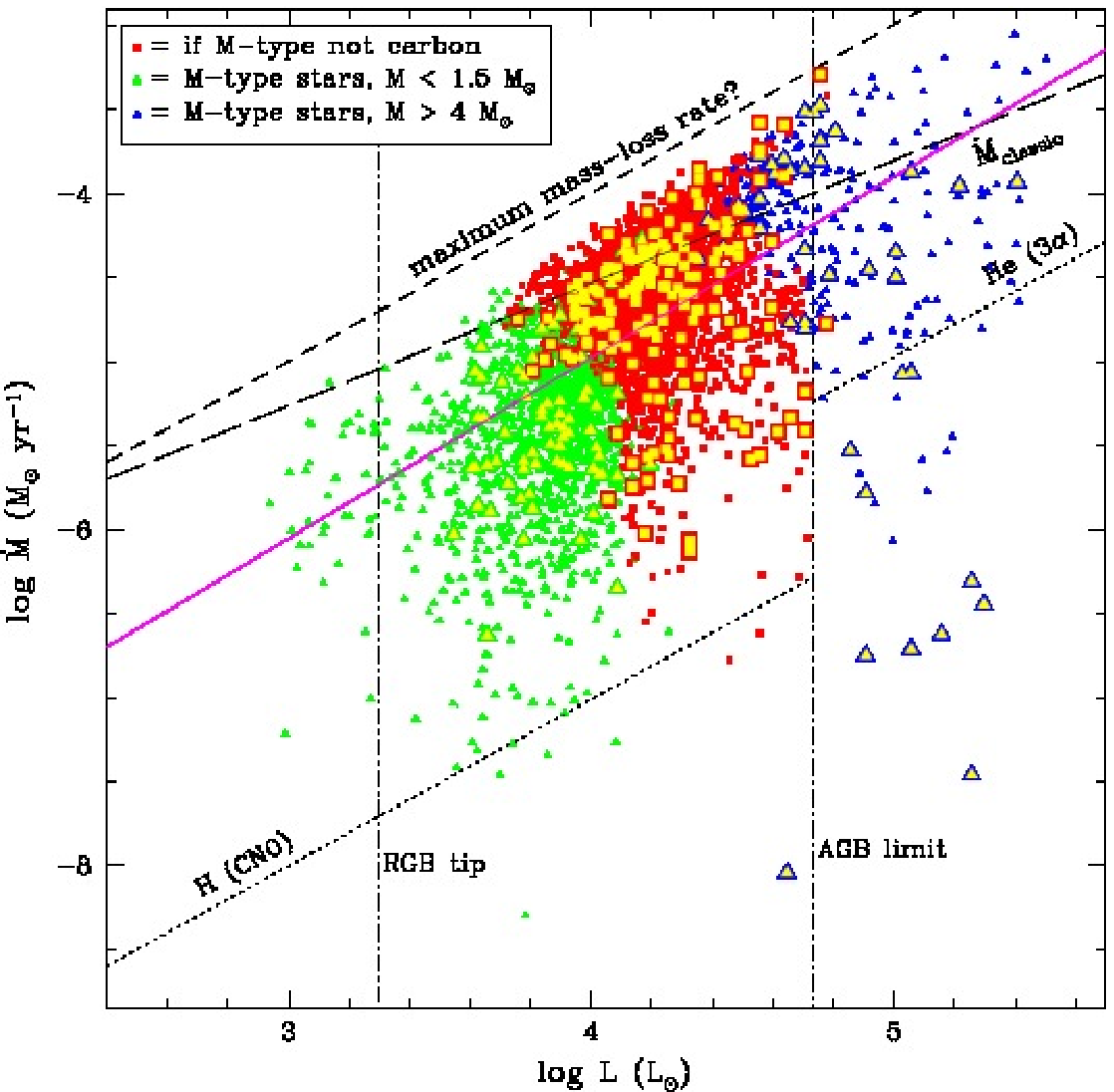,width=88mm}}
}}
\caption[]{Left: mass-loss vs.\ luminosity for all stars including
non-variable stars. The blue and green triangles represent massive luminous
M-type stars and low-mass stars (at lower luminosities), respectively. The red
squares represent presumed carbon stars. Large yellow symbols identify the
stars modelled with {\sc dusty}; other UKIRT (WFCAM) variables are identified
by black squares. The bottom panel shows the results if the carbon stars are
presumed to be oxygen-rich. The lines have the same meaning as in figures 8
and 9. Right: same, but limited to variable stars.}
\end{figure*}

The carbon star population appears offset with respect to the oxygenous
(M-type) stars, creating a blank spot around $\log L/{\rm L}_\odot=4.6$ and
$\log\dot{M}=-5$, and to al esser extent around $\log L/{\rm L}_\odot=4.2$ and
$\log\dot{M}=-4.5$. This is caused by the differences in the model results for
carbon stars vs.\ oxygenous stars due to the (uncertain) optical constants of
the grains, and not all stars will have been classed correctly (as either
carbon or M-type). Indeed, if these stars are treated as M-type stars then
they fill those gaps. While the detection limits could have resulted in the
omission of some of the most extreme carbon stars, this could alleviate the
discrepancy around $\log L/{\rm L}_\odot=4.2$ and $\log\dot{M}=-4.5$ but not at
the higher luminosities and lower mass-loss rates.

\subsection{
Mass-loss rates as a function of stellar parameters}

\subsubsection{Dependency on luminosity}

We have seen that, overall, mass-loss rates increase with luminosity. Or at
least the maximum mass-loss rate that is attained does. At any given
luminosity there is a large spread in mass-loss rate, but our survey is biased
towards the most extreme objects.

A Theil--Sen robust fitting for the M-type stars with optical periods (see
below) yields the following parameterisation of the mass-loss rate in terms of
luminosity:
\begin{equation}
\log\left(\frac{\dot{M}}{{\rm M}_\odot\ {\rm yr}^{-1}}\right)=
(0.98\pm0.18)\times\log\left(\frac{L}{{\rm L}_\odot}\right)-9.35\pm0.05
\end{equation}
This suggests
proportionality between mass-loss rate and luminosity,
and a typical mass-loss rate for a strongly pulsating star of $L=10^4$
L$_\odot$ of $\dot{M}\sim4\times10^{-6}$ M$_\odot$ yr$^{-1}$. Limiting the fit to
those stars of which we had modelled the SEDs yields a very similar result,
with a slope of $1.00\pm0.25$ and a zero point of $-9.42\pm0.15$.
The relation is a good representation of the ``typical'' mass-loss rate,
and in particular follows a sequence of RSGs (Fig.\ 9).

Because much of the remaining scatter is intrinsic, below we explore other
parameters that may contribute to determining the mass-loss rate, and which
have been measured for at least a subset of our sample of stars.

\subsubsection{Dependency on amplitude}

The mass-loss rate increases with increasing K$_{\rm s}$-band amplitude (Fig.\
11). This correlation supports the notion of stellar pulsation driving the
initial development of winds from cool evolved stars (Whitelock, Feast \&
Pottasch 1987).

There is a lot of scatter. This is partly due to the crude estimation of the
amplitudes from the sparsely sampled light-curves. In addition, there are real
variations among stars with the same amplitude (for instance the gravitational
acceleration near the surface as a result of varying mass and radius). When
trying to fit $\log\dot{M}$ as a function of amplitude, $A$ (note that the
amplitude is already a logarithmic quantity) the result is not meaningful and
we do not quote it here.

One may wonder, though, how stars with amplitudes as low as $A<0.5$ mag can
achieve mass-loss rates as high as $10^{-4}$ M$_\odot$ yr$^{-1}$ (Fig.\ 11).
These are the most massive stars in our sample, i.e.\ the most luminous. The
mass-loss rate is more directly related to the absolute amplitude, expressed
in luminosity units (van Loon et al.\ 2008). The K$_{\rm s}$-band amplitude
closely traces the bolometric luminosity variations as this band is near the
peak of the stellar SED, in between the strong attenuation by dust at shorter
wavelengths and its emission at longer wavelengths.

\begin{figure}
\centerline{\psfig{figure=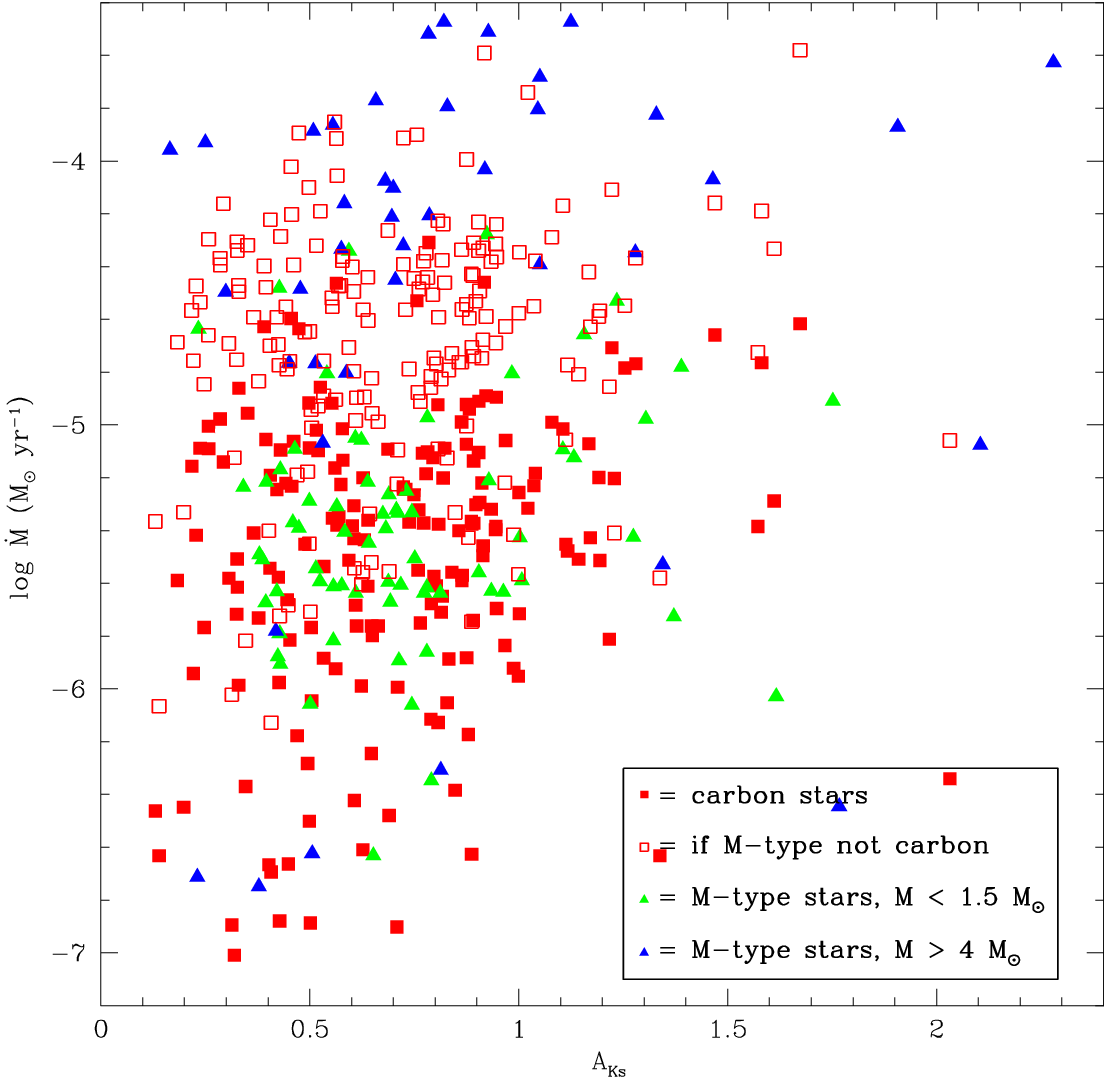,width=84mm}}
\caption[]{Mass-loss rate vs.\ K$_{\rm s}$-band amplitude, for stars modelled
with {\sc dusty} -- the open red squares show the results if the carbon stars
are presumed to be oxygen-rich.
}
\end{figure}



\subsubsection{Dependency on infrared colours}

Often, mass-loss rates are derived from individual infrared colours. However,
these scalings depend on the luminosity as well as the dust properties (van
Loon 2007; Srinivasan et al.\ 2009). We would advocate the same procedure as
we have employed, namely using the colour to estimate the bolometric
correction as well as the optical depth, from which then the mass-loss rate
can be determined.

Indeed, at a given J--K$_{\rm s}$ colour, the spread in luminosity and chemical
type (M-type or carbon) results in a spread in mass-loss rate of $\sim1.5$ dex
(Fig.\ 12). The luminosity dependence is clear from the offset between the
low-mass and high-mass M-type stars. Less luminous stars are smaller, with
more compact envelopes in which the optical depth can grow more easily. Hence,
for a given mass-loss rate, less luminous stars are redder (van Loon et al.\
1999b). Carbon stars, on the other hand, become reddened easily because
carbonaceous dust is relatively opaque compared to the more transparent
silicates found in M-type stars. This means that -- compared to M-type stars
of similar colour -- their mass-loss rate is relatively low.

\begin{figure}
\centerline{\psfig{figure=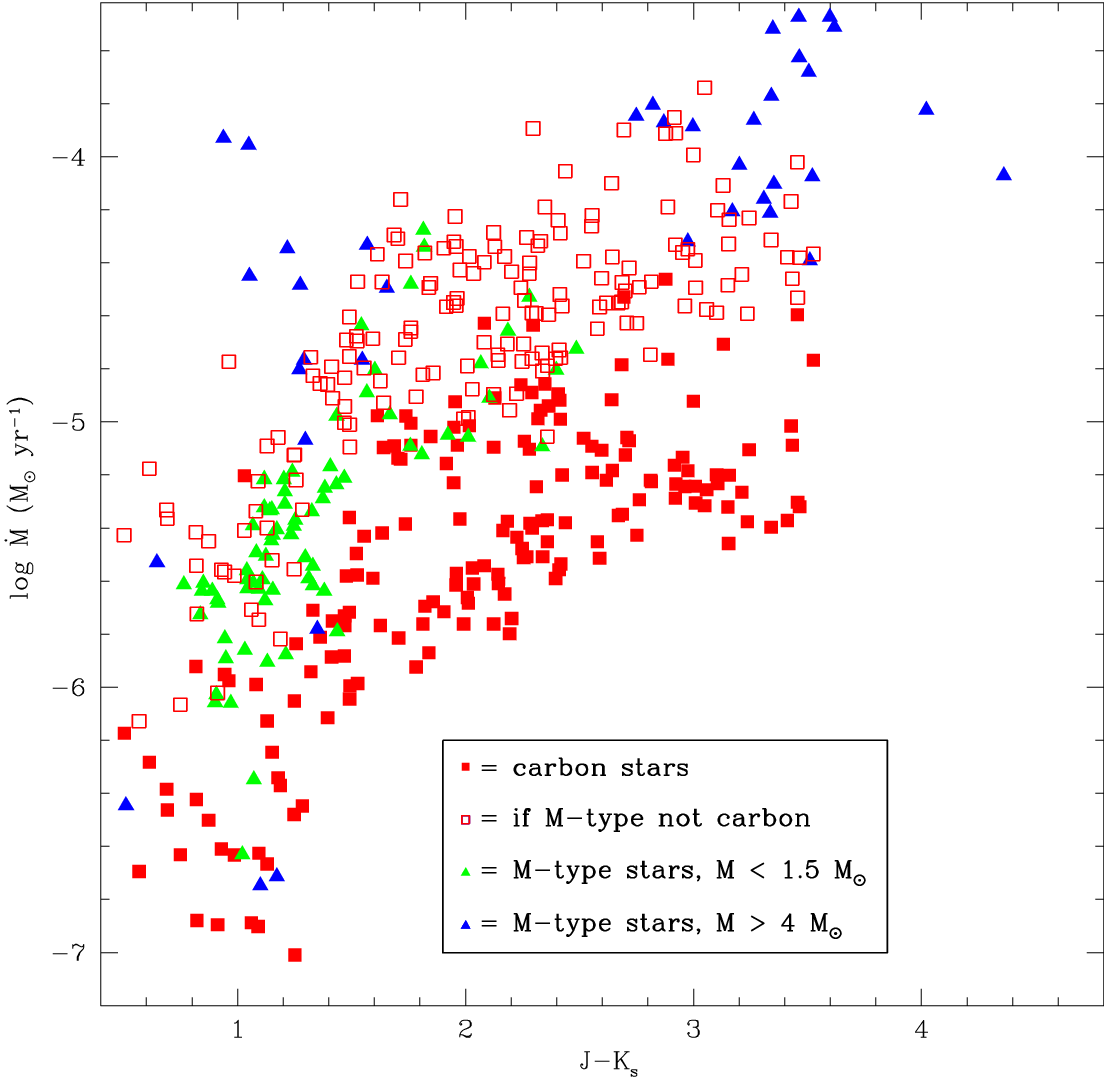,width=84mm}}
\caption[]{Mass-loss rate vs.\ J--K$_{\rm s}$ colour, for stars modelled with
{\sc dusty} -- the open red squares show the results if the carbon stars are
presumed to be oxygen-rich.
}
\end{figure}

\subsubsection{Dependency on pulsation period}

A dependence of the mass-loss rate on pulsation period is expected for several
reasons.
Firstly, more luminous stars are more able to drive off matter through
radiation pressure, and more luminous evolved stars are larger and therefore
pulsate with longer periods. Secondly, stars experiencing the most intense
mass loss are likely to already have lost a significant fraction of their
mass, to which they respond by bloating, thereby increasing the pulsation
period. And thirdly,
longer pulsation periods give the shocked atmosphere more time to grow grains,
onto which the radiation field can impart momentum to drive a wind
-- deemed important for carbon stars (e.g., Winters et al.\ 2000; Wachter
et al.\ 2002; Nowotny et al.\ 2010, 2011) as well as M-type stars (Woitke
2006; H\"ofner 2008; Norris et al.\ 2012; Bladh et al.\ 2015; Ohnaka, Weigelt
\& Hofmann 2016; H\"ofner \& Olofsson 2018).

 Periods for 1847 Mira candidates in M\,33 were derived independently by
Yuan et al.\ (2017), based on I-band images. We found 570 stars in common with
our survey.
These are likely biased towards the
redder
sources in the Yuan et al.\ catalogue, and against the
reddest
sources in our survey.


The mass-loss rate
increases with increasing period (Fig.\ 13). While not immediately obvious due
to the large scatter at short periods, it becomes clear when considering also
the (much fewer) stars with the longest periods ($>600$ d). However, the
latter show a much weaker trend with period, suggesting a ``saturation'' of
the mass loss, as the period increase as a result of said intense mass loss.
This behaviour had been already seen for Miras (Vassiliadis \& Wood 1993;
Groenewegen et al.\ 2009) or when the data are limited to carbon stars
(Whitelock et al.\ 2006).

\begin{figure}
\centerline{\psfig{figure=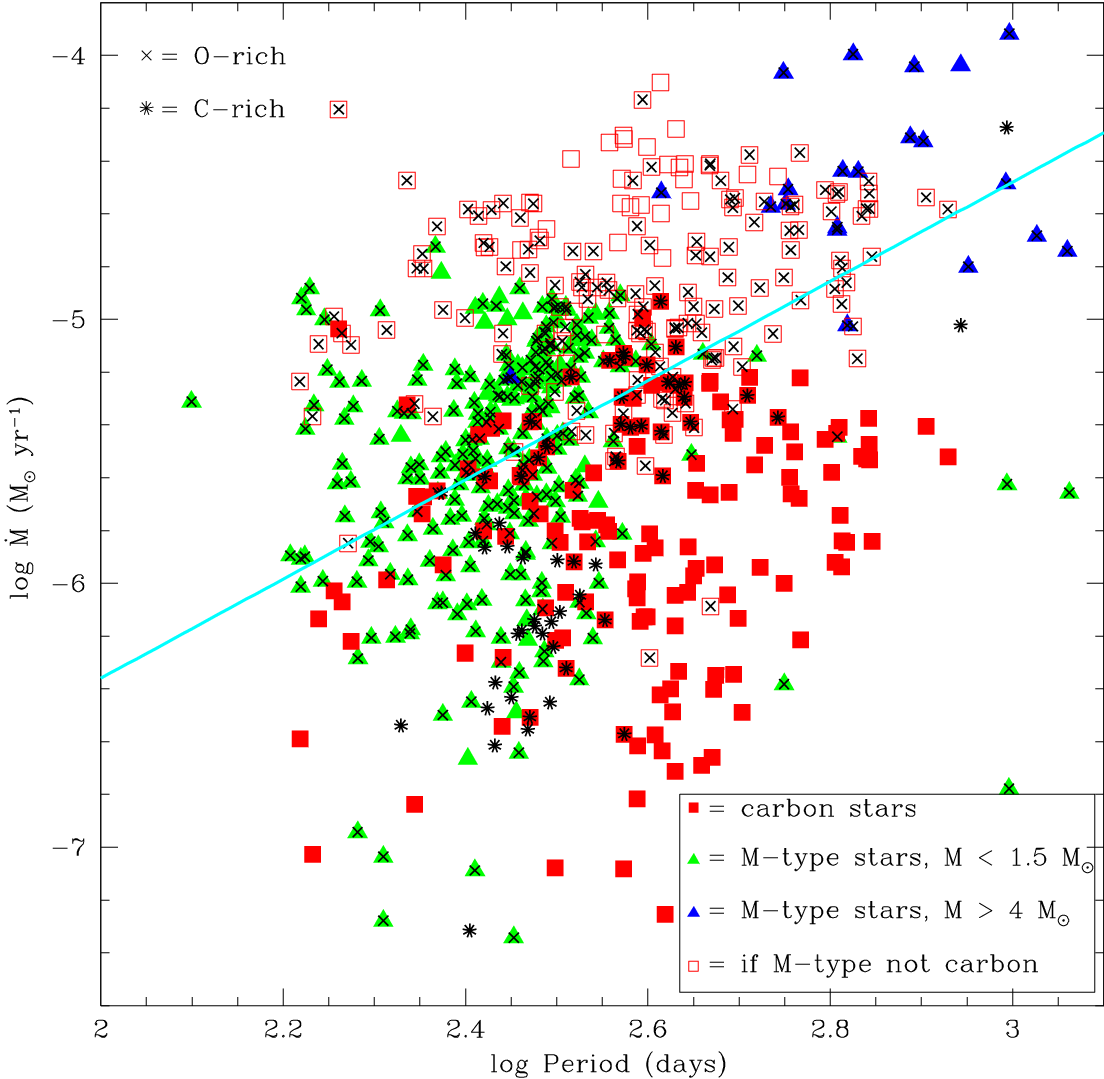,width=84mm}}
\caption[]{Mass-loss rate vs.\ pulsation period, for low-mass AGB stars (green
triangles), intermediate-mass carbon stars (red squares) and massive AGB stars
and RSGs (blue triangles). The open red squares show the results if the carbon
stars are presumed to be oxygen-rich. The crosses and star symbols show O-rich
and carbon stars based on the classification by Yuan et al.\ (2017).
The cyan line is the relation given by Eq.\ (2) that we derived.
}
\end{figure}

Using the Theil--Sen estimator to fit the mass-loss rate as a function of
period for the M-type stars (as classified
here)
results in the following
(see Fig.\ 13):
\begin{equation}
\log\left(\frac{\dot{M}}{{\rm M}_\odot\ {\rm yr}^{-1}}\right)=
(1.88\pm0.43)\times\log\left(\frac{P}{\rm d}\right)-10.12\pm0.06
\end{equation}
This suggests a very steep dependence on the pulsation period.
%
%
This
is somewhat misleading, as there is a clear, well-documented relation between
luminosity and pulsation period
and thus part of the above dependence of mass-loss rate on period is in
actual fact due to the already identified dependence on luminosity.

In figure 14 we show our
luminosities and the periods from Yuan et al.,
and the Period--Luminosity (PL) relationships from Menzies et al.\ (2008) and
Guandalini \& Busso (2008). There is good agreement between our data and the
PL relation from Guandalini \& Busso for $P<400$ d ($\log P<2.6$). At longer
periods, stars depart from the PL relation: low-mass stars probably
experiencing the effects of the reduced mass as a result of their sustained
heavy mass loss
(Wood 2000),
whilst the most massive AGB stars may be more luminous due to the effects of
HBB
(Whitelock et al.\ 2003).
Even more massive RSGs perhaps do not obey a similar PL relation as the AGB
stars do; Yang \& Jiang (2012) suggest RSGs pulsate in the first overtone. An
offset of $\sim0.4$ dex in $\log L$ is seen between our results and the PL
relation for carbon stars from Menzies et al. They assumed a relatively near
value for the distance modulus of the LMC, whereas we adopt a relatively
distant value for M\,33. This could probably explain half a magnitude in
distance modulus, or $\sim0.2$ dex in $\log L$.

It is thus clear that, while parameterisations of mass loss with single
stellar parameters is instructive with regard to the physical mechanisms, it
has limited value in offering recipes for deriving mass-loss rates from
observations or for incorporating mass-loss recipes into stellar evolution
models. What is really needed is the parameterisation in terms of all the
independent stellar parameters (or a combination of inter-dependent stellar
parameters that account for all independent parameters).

\begin{figure}
\centerline{\psfig{figure=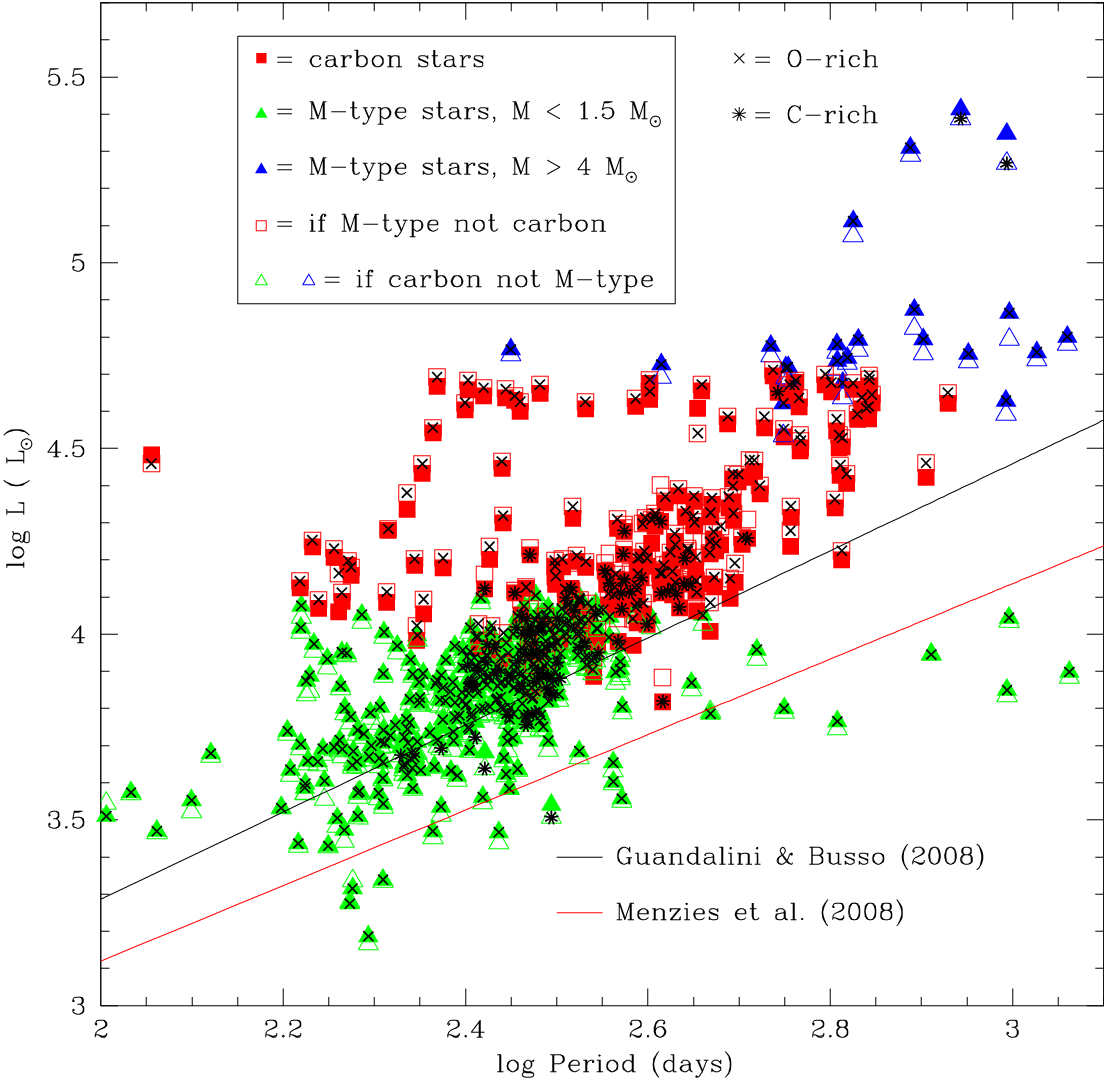,width=84mm}}
\caption[]{Luminosity vs.\ pulsation period. The symbols are as in figure 13;
the red and blue lines represent the Period--Luminosity relationships derived
by Menzies et al.\ (2008) and Guandalini \& Busso (2008), respectively.}
\end{figure}

\subsubsection{A multi-parameter description}

The pulsation period may exhibit a stronger development than luminosity during
the final stages of AGB and RSG evolution. Therefore, we use a multi-linear
bisquare regression to fit the mass-loss rates of the M-type stars as a
function of luminosity and period:
\begin{eqnarray}
\log\left(\frac{\dot{M}}{{\rm M}_\odot\ {\rm yr}^{-1}}\right)=
(0.88\pm0.12)\times\log\left(\frac{L}{{\rm L}_\odot}\right)+ \nonumber \\
(0.34\pm0.23)\times\log\left(\frac{P}{\rm d}\right)-9.72\pm0.38
\end{eqnarray}
The data are challenging such parameterisation, resulting in larger errors in
the period dependence and zero point, but it does show a
correlation with period, at the (only slight) cost of the dependence on
luminosity.

If we use the amplitude, instead of period, then the
proportionality with
luminosity (for the same sample of M-type stars with periods) is preserved to
a much better degree:
\begin{eqnarray}
\log\left(\frac{\dot{M}}{{\rm M}_\odot\ {\rm yr}^{-1}}\right)=
(1.05\pm0.08)\times\log\left(\frac{L}{{\rm L}_\odot}\right)+ \nonumber \\
(0.14\pm0.09)\times A-9.66\pm0.36
\end{eqnarray}
Interestingly, the errors on all coefficients are a little smaller than for the
paramaterisation with luminosity and period. This might mean that the period is
not as important as the amplitude is.

If the amplitude is not entirely dependent on period and luminosity, then we
expect a parameterisation of the mass-loss rate in terms of luminosity, period
{\it and} amplitude to yield the most accurate formula:
\begin{eqnarray}
\log\left(\frac{\dot{M}}{{\rm M}_\odot\ {\rm yr}^{-1}}\right)=
(0.94\pm0.13)\times\log\left(\frac{L}{{\rm L}_\odot}\right)+ \nonumber \\
(0.28\pm0.23)\times\log\left(\frac{P}{\rm d}\right)+ \nonumber \\
(0.12\pm0.10)\times A-9.90\pm0.40
\end{eqnarray}
While this does show a
correlation with
all three parameters, the errors on the coefficients have become slightly
larger again. This could be a sign that the amplitude {\it does} depend
strongly on luminosity and period, in which case the mass-loss rate should be
expressed in terms of either the period or amplitude, but not both.
Not surprisingly, given that the errors on the zero points are very
similar, the correlation coefficient is 0.61 for each of equations (3)--(5).

\subsection{Feedback into the interstellar medium}

To determine the total budget of mass returned to the ISM, we must consider
both the extreme, dusty stars which may contribute a disproportionately large
amount, but also the vastly more numerous stars with low mass-loss rates, but
which together may still contribute a significant fraction to the total. The
latter will be critically affected by interstellar reddening, which could
mimic mass loss, and photometric uncertainties, which could exaggerate or
annihilate the effects of circumstellar reddening (cf.\ McDonald et al.\ 2009;
McDonald, Zijlstra \& Boyer 2012). The interstellar reddening towards M\,33 is
modest, especially at IR wavelengths, but even such small values may matter.
In figure 2 we had assumed that stars with $(J-K_{\rm s})<1$ mag have $\tau=0$
and thus $\dot{M}=0$ M$_\odot$ yr$^{-1}$. Since photometric scatter has a
purely statistical nature, we can rectify this bias by considering also the
distribution of stars over negative values for the optical depth (see Paper
III).

Eventually, we obtain the binned and cumulative mass-loss rate distributions 
presented in figure 15. If the reddenings of all non-variable stars are taken
as due to circumstellar dust, then the low-mass red giants would contribute a
few times as much feedback as all other stars combined (Fig.\ 15, bottom
right). It is highly unlikely that there is such hike in the mass loss at the
lowest masses, because those stars lose less than half their mass whilst the
more massive AGB stars can lose up to 80 per cent of their (already higher)
mass. However, if also the presumed carbon stars are in fact all oxygen-rich,
then they would contribute about as much as those low-mass red giants. We thus
take that scenario as the extreme upper limit to the feedback: $<0.44$
M$_\odot$ yr$^{-1}$ ($<0.185$ if they are genuine carbon stars but the reddened
low-mass red giants do all have dusty winds).

Among the variable stars, the mass return rate is relatively constant over
mass from about 3 M$_\odot$ upwards. Below it, it drops, but this may be a
selection bias against the low-mass variables as we had also inferred a
rather modest star formation rate at ancient times ($10^{10}$ yr; Paper V).
The truth is probably in between, reinforcing a mass insensitivity of the mass
return rate for stellar populations as a whole, within a factor $\sim2$. The
mass return is dominated by AGB stars, but RSGs do contribute up to a third to
the total
(Fig.\ 15, top right: at $\log M/{\rm M}_\odot=0.9$--1.0 the cumulative
mass-loss rate reaches 0.010--0.013 M$_\odot$ yr$^{-1}$, compared to 0.034
M$_\odot$ yr$^{-1}$ in total);
Jura \& Kleinmann (1989, 1990) found a smaller contribution from
RSGs in the Solar Neighbourhood
 -- viz.\ 1--$3\times10^{-5}$ M$_\odot$ kpc$^{-2}$ yr$^{-1}$ compared to the
contribution from AGB stars of 3--$6\times10^{-4}$ M$_\odot$ kpc$^{-2}$ yr$^{-1}$
-- which is similar to that found by Boyer et al.\ (2012) in the metal-poor
Small Magellanic Cloud (SMC).
The mass return rate based on the WFCAM variables alone amounts to 0.034
M$_\odot$ yr$^{-1}$, or 0.09 M$_\odot$ yr$^{-1}$ if all presumed carbon stars are
in fact M-type stars. We know from cross-identification with known sources
that the latter is not the case, so again that number would be an extreme
upper limit.

However, we must correct for our survey incompleteness. In Paper V and Section
2.2 we estimated that the completeness in detecting LPVs is $\sim0.3$--0.5.
Taken all the above into consideration, therefore, we estimate a total mass
return rate of 0.1 M$_\odot$ yr$^{-1}$, give or take a factor two. Carbon stars
contribute about a quarter to the total. Not all of the presumed carbon stars
may be carbon stars -- we do not expect many presumed M-type stars to be
carbon stars as we have been generous in the carbon star classification
(see Appendix, section A1.3).
But as pointed out already, we may have missed a few extreme carbon stars
because they would have become too faint even at near-IR wavelengths. The
conclusion remains the same as that which we had reached in Paper III for the
central square kpc, namely that even in the slightly sub-solar metallicity
disc of M\,33 the dust input is dominated by silicates, with only a minor
carbonaceous component.

\begin{figure*}
\centerline{\vbox{
\hbox{\psfig{figure=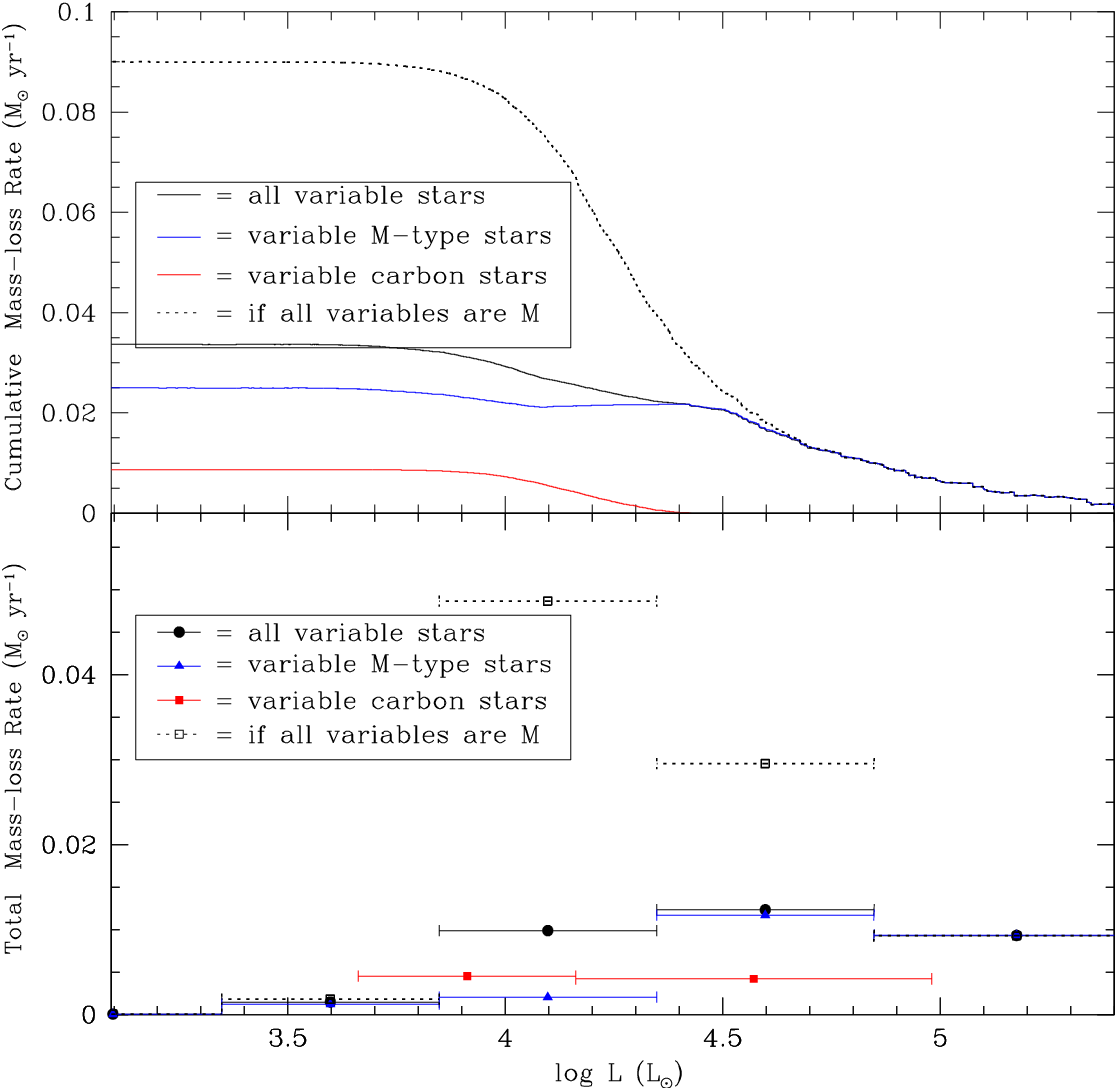,width=88mm}
\psfig{figure=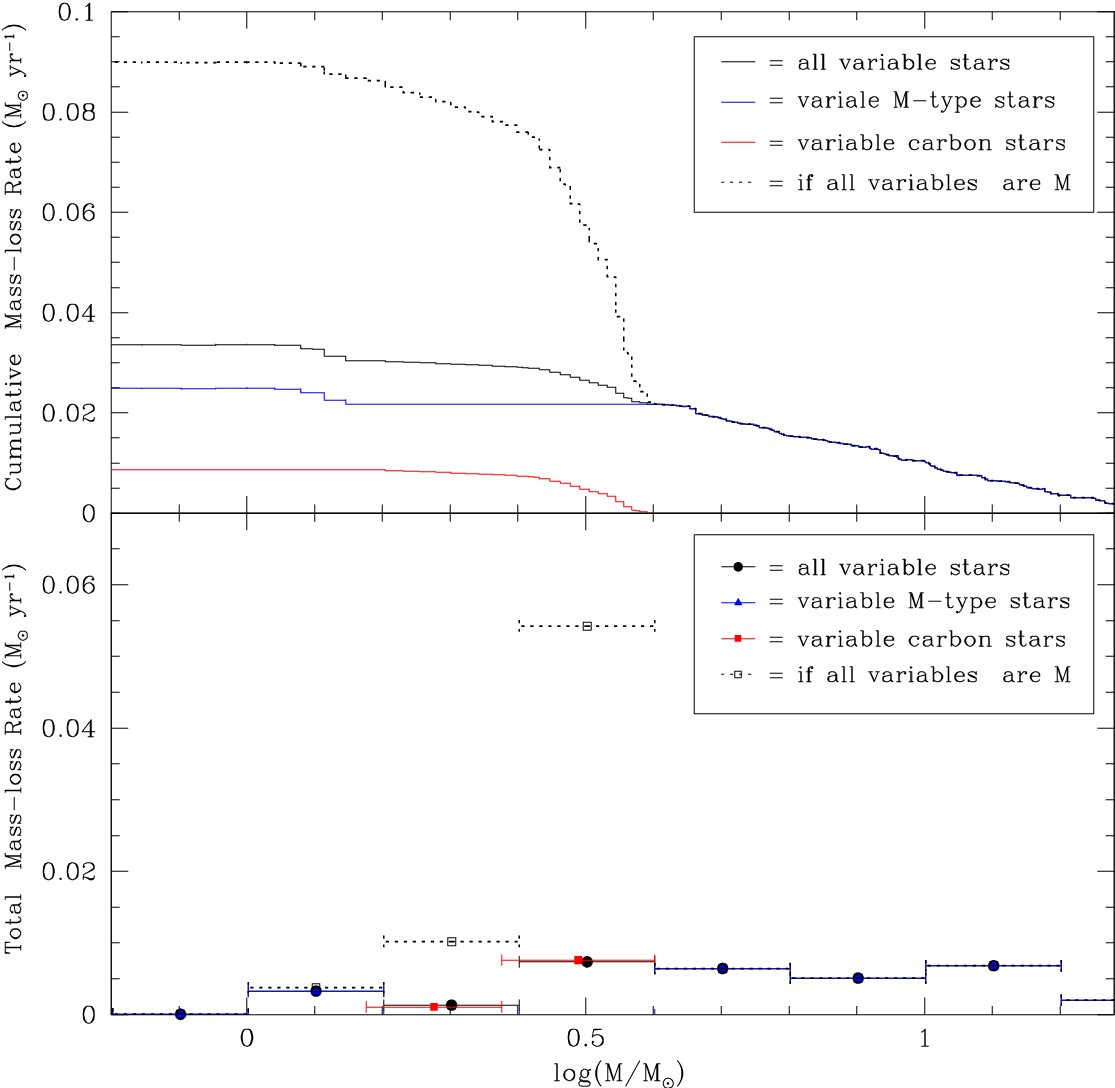,width=88mm}}
\hbox{\psfig{figure=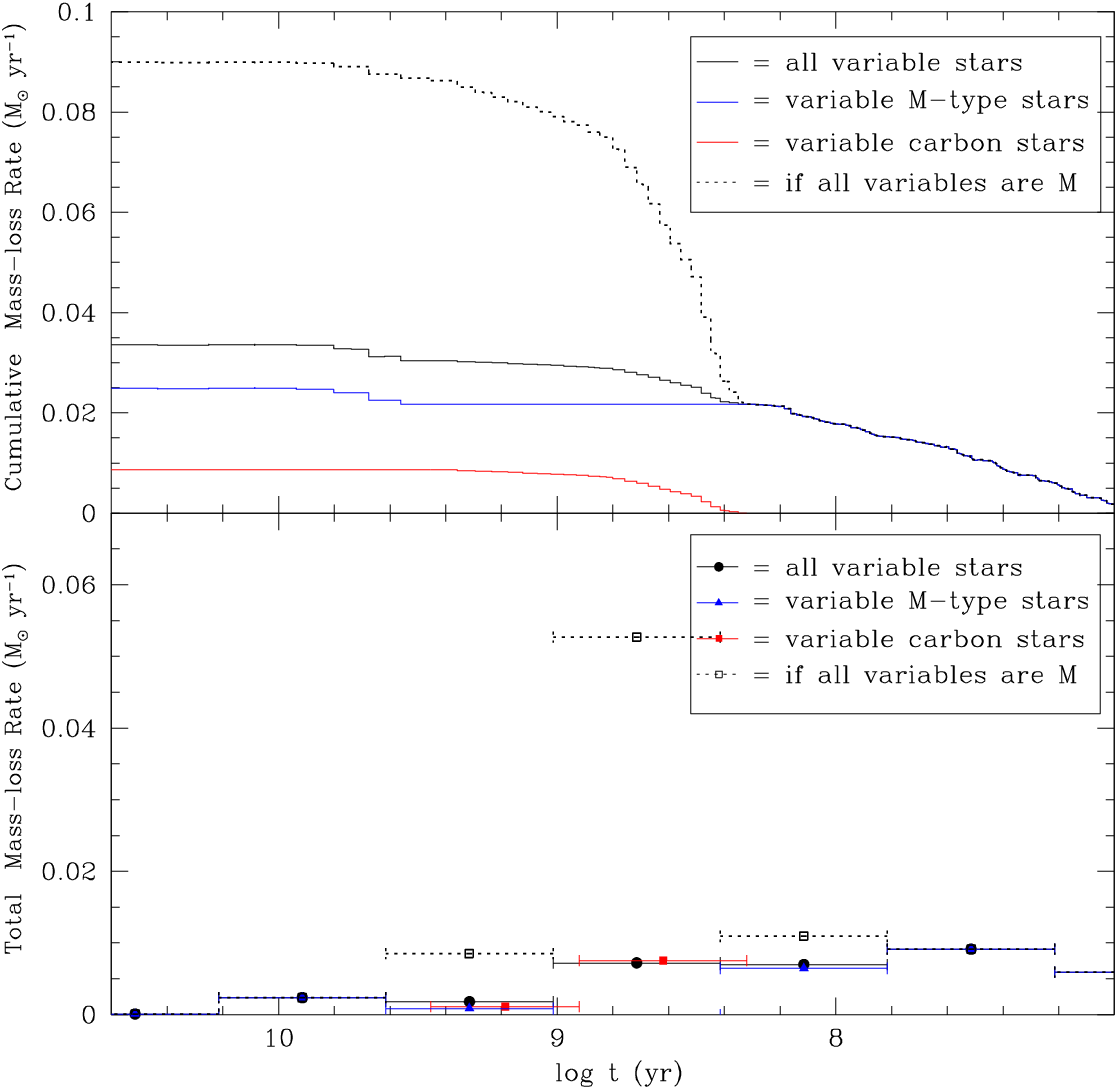,width=88mm}
\psfig{figure=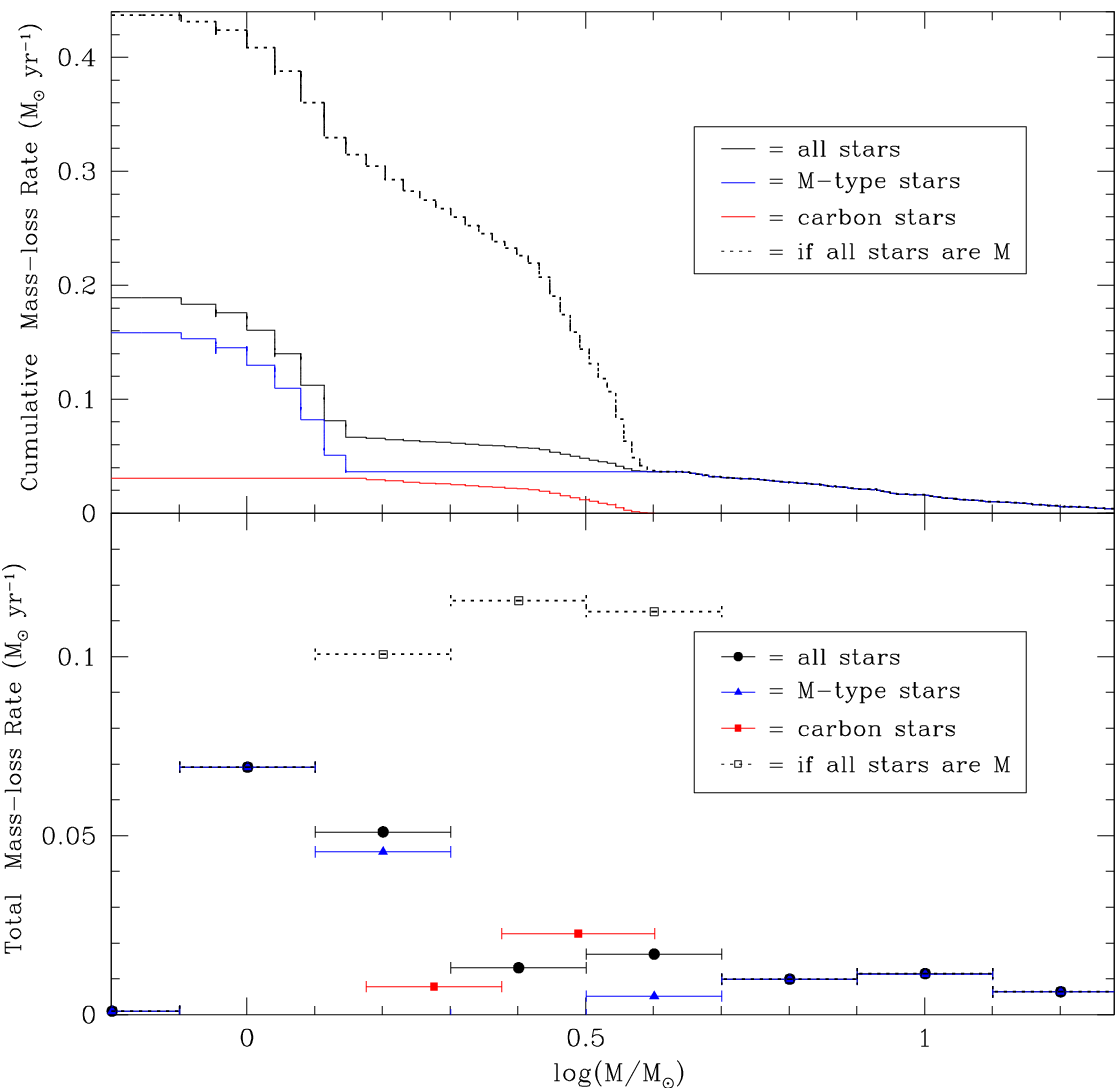,width=88mm}}
}}
\caption[]{Mass-loss rate vs.\ luminosity (top left), age (bottom left) and
birth mass (top right) for the WFCAM variable stars, and vs.\ birth mass for
all stars including non-variable stars (bottom right). The top panels show the
cumulative distribution (integrated from the right); the bottom panels show
binned contributions. Distinction is made between likely carbon stars (red)
and M-type stars (blue).}
\end{figure*}

The radial profile of mass-return-rate surface density, deprojected onto the
galaxy plane
(assuming an inclination angle of $i=56^\circ$ and a position angle of
$PA=23^\circ$ -- see Zaritsky et al.\ 1989),
is shown in figure 16. For comparison with the UIST results we
had derived in Paper III, it is shown in more detail for the central part of
M\,33 in the bottom panel. The WFCAM results broadly agree with the UIST
results, with a radial decline in mass return except for a relatively flat
profile within the inner few hundred pc. The radial profile over $\sim10$ kpc
is very similar for the carbon stars and M-type stars, with an exponential
profile characterised by a similar scalelength of $R=2.7$ kpc. They are offset
by about a factor four, as discussed above. Jura \& Kleinmann (1989)
determined a mass return rate of $\sim3$--$6\times10^{-4}$ M$_\odot$ yr$^{-1}$
kpc$^{-2}$ in the Solar Neighbourhood; this is similar to the (completeness
corrected) rate we find about a kpc from the centre of M\,33.

\begin{figure}
\vbox{
\centerline{\psfig{figure=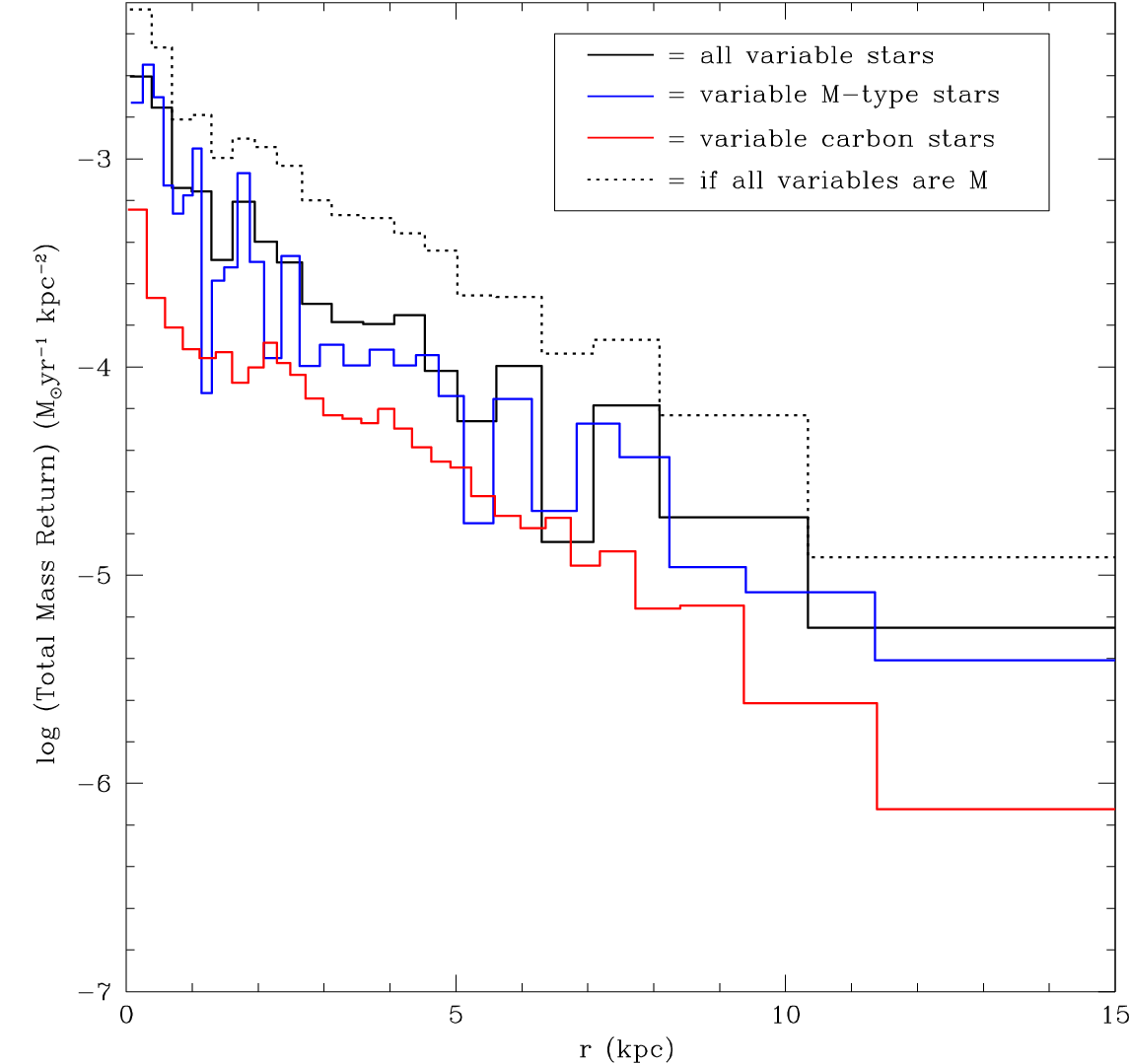,width=84mm}}
\centerline{\psfig{figure=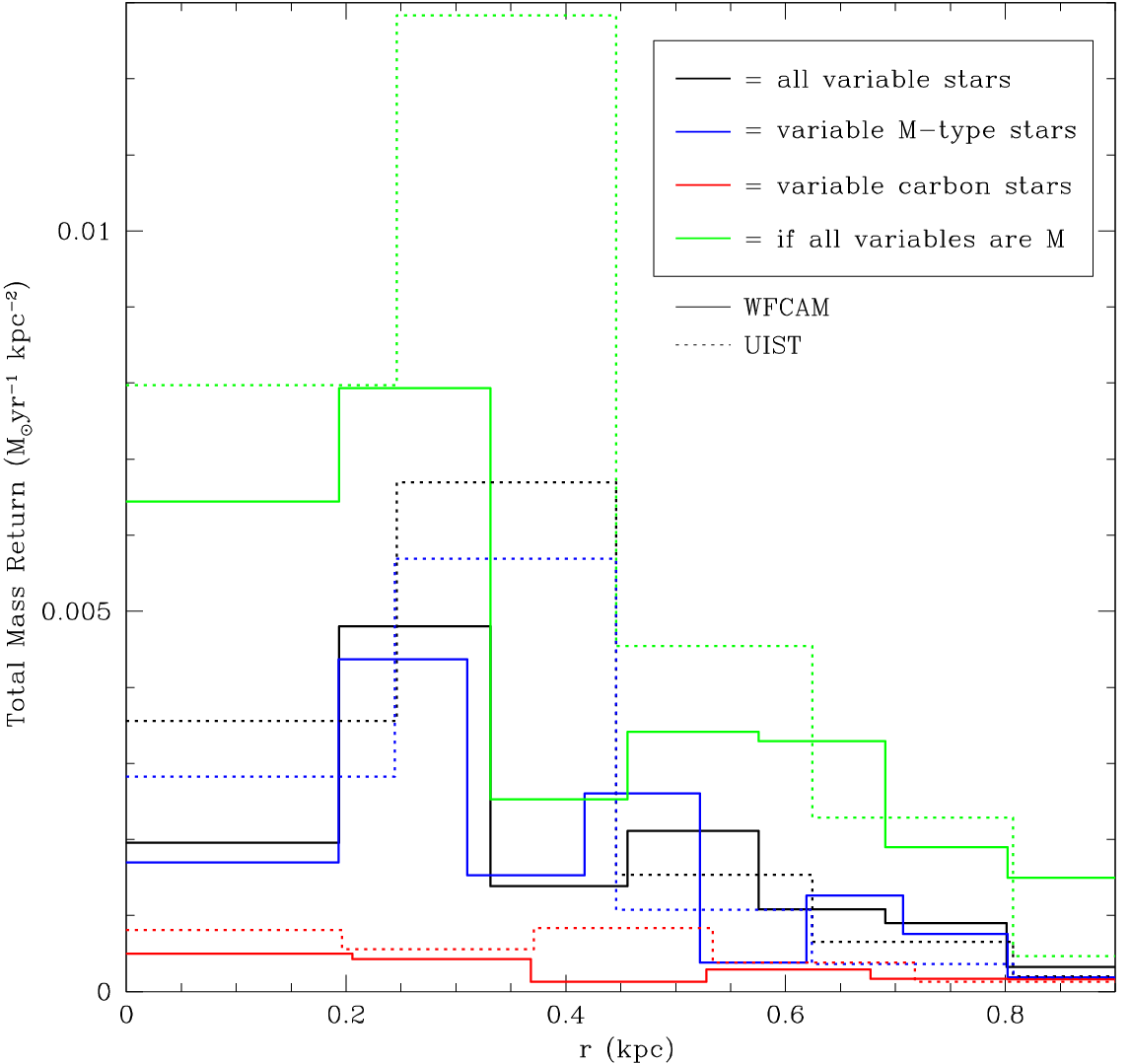,width=84mm}}}
\caption[]{Top: radial profile of mass-return-rate surface density across a
large part of M\,33 deprojected onto the galactic plane; bottom: the same, but
for the central square kiloparsec of M\,33.}
\end{figure}

\section{Discussion and conclusions}

Here we discuss the implications of our results for the evolution of stars and
the ISM.

\subsection{Mass loss from AGB stars and RSGs}

\subsubsection{Mass-loss mechanism}

We have studied how the mass-loss rate of AGB stars and RSGs depends on
luminosity, pulsation period and pulsation amplitude.
Overall, the distribution in the $\dot{\rm M}$--L diagram confirms that
variable AGB stars evolve in mass-loss rate, and that RSGs may do the same or
face phases of enhanced mass loss which they may or may not experience. Both,
AGB stars and RSGs exhibit maximum mass-loss rates which are higher for
more luminous and -- presumably, at least initially --
more massive stars. There is no indication that super-AGB stars do not reach
the rates expected for their mass;
while we have no means of confirming individual super-AGB stars, in the
mass-loss rate vs.\ luminosity plane (Figs.\ 9 \& 10) there is a continuous
``band'' of stars losing mass above the single-scattering limit ($\dot M_{\rm
classic}$) which crosses the classical AGB limit around $\dot M\sim 10^{-4}$
M$_\odot$ yr$^{-1}$, and likewise there is a continuous band of RSGs losing mass
around $\dot M\sim 10^{-5}$ M$_\odot$ yr$^{-1}$ that continues down to the same
classical AGB limit.
It is less clear how the mass loss from carbon stars compares to that
from M-type stars,
because their dust has different properties and the properties of both
carbonaceous dust and oxygenous dust (silicates) are uncertain. Also, we may
have missed the most extreme carbon stars, which become very red and may have
evaded our survey.

The
proportionality between
the mass-loss rate and luminosity is similar to the Reimers law for a typical
temperature of 3750 K as derived by Mauron \& Josselin (2011), with an
exponent of 1.09 compared to the 0.98 we determined. However, their baseline
$\log\dot{M}=-5.59$ for $L=10^5$ L$_\odot$ is more than an order of magnitude
smaller than our $-4.45$. Indeed, most of the RSGs studied by Mauron \&
Josselin lose mass at modest rates, akin to the sample studied by Verhoelst et
al.\ (2009) and corresponding to the lower ``branch''. We are more sensitive
to the RSGs with higher rates of mass loss, and clearly that phase cannot be
neglected for the population as a whole. Mauron \& Josselin also presented the
relation by Salasnich, Bressan \& Chiosi (1999), which has a steeper
dependence on luminosity (exponent 1.385) but a baseline $\log\dot{M}=-4.68$
which is much more similar to ours. Salasnich et al.\ also presented a version
with a gas-to-dust ratio which increases with luminosity, to account for cases
such as Betelgeuse that have much less dust than could potentially form (van
Loon et al.\ 2005a), giving rise to an even steeper luminosity dependence but
much larger constant of $-4$. Finally, Mauron \& Josselin presented the
relation by Vanbeveren, De Loore \& Van Rensbergen (1998), the only one with a
shallower luminosity dependence than ours (an exponent of 0.8) but a similar
baseline $\log\dot{M}=-4.7$. The best overall agreement, however, is found
between our relation derived in M\,33 and the van Loon et al.\ (2005a) formula
for a temperature of 3500 K, with a luminosity exponent of 1.05 and a baseline
$\log\dot{M}=-4.6$ ($-4.79$ for $T=3750$ K).

Goldman et al.\ (2017) developed a new formula for the mass-loss rate of the
most evolved OH/IR stars.
Their dependence on luminosity is very similar to that which we derive for the
cool evolved stars in M\,33, though their dependence on period is stronger.
They found a negligible dependence on the gas-to-dust ratio ($\psi$).
We remind the reader that we accounted for a factor two increase in $\psi$ as
a result of the metallicity gradient across the disc of M\,33, and so the
mass-loss rates we derived should not bear any hidden metallicity dependence.
For M-type stars
we can divide our measured mass-loss rate by the mass-loss rate that is
predicted on the basis of the formula from Goldman et al.; we thus derive
$\dot{M}_{\rm here}/\dot{M}_{\rm Goldman}\sim0.1$--1. Because the Goldman et al.\
sample is heavily biased towards the most extreme stars in the LMC and the
Centre and Bulge of the Milky Way, our sample exhibits somewhat lower rates on
average
(e.g., compare our Eq.\ (1) with theirs in Fig.\ 9),
but importantly the most extreme stars in our sample do reach similar high
rates. The results we derive in M\,33 are thus in good agreement with those
derived in more nearby systems of both lower (LMC) and higher (central Milky
Way) metallicity. For carbon stars this ratio would be $\dot{M}_{\rm
here}/\dot{M}_{\rm Goldman}\sim0.01$--0.3, though we note that the Goldman et
al.\ formula is valid only for oxygen-rich stars. However, while we do not
find any dependence of the ratio for either M-type or carbon stars on
luminosity or pulsation period (Fig.\ 17), the bulk of carbon stars would have
{\it higher} mass-loss rates than the Goldman et al.\ formalism if they are in
fact M-type stars. This suggests that those stars are indeed carbon stars; the
most luminous carbon stars (with the longest periods) do not reach such high
mass-loss rates, which could mean that at least some of them are in fact
massive, M-type AGB stars instead. This would explain some of the discrepancy
between our classifications and spectroscopically confirmed stars (which
otherwise was in fair agreement).


\begin{figure}\vbox{
\centerline{\psfig{figure=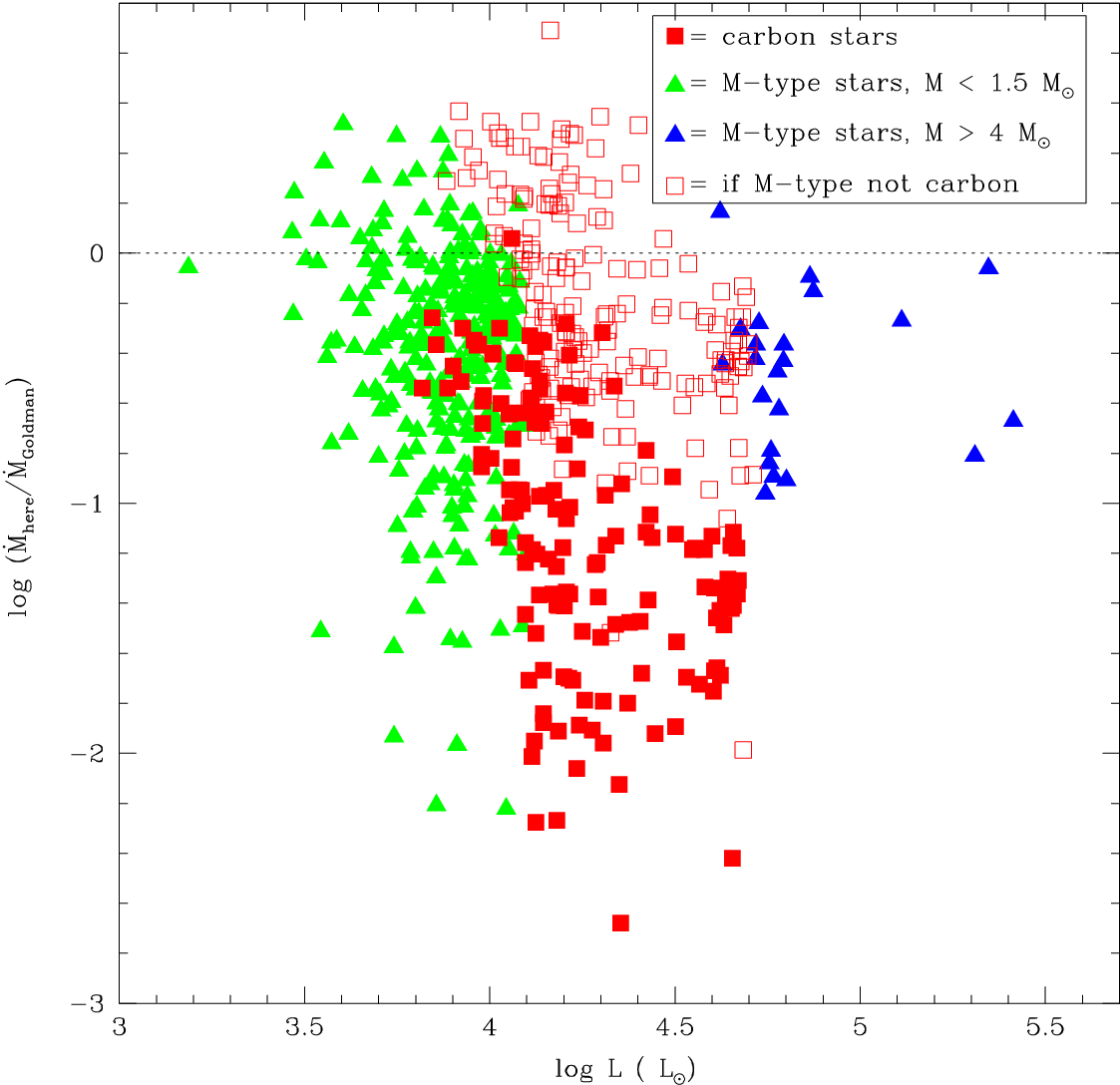,width=84mm}}
\centerline{\psfig{figure=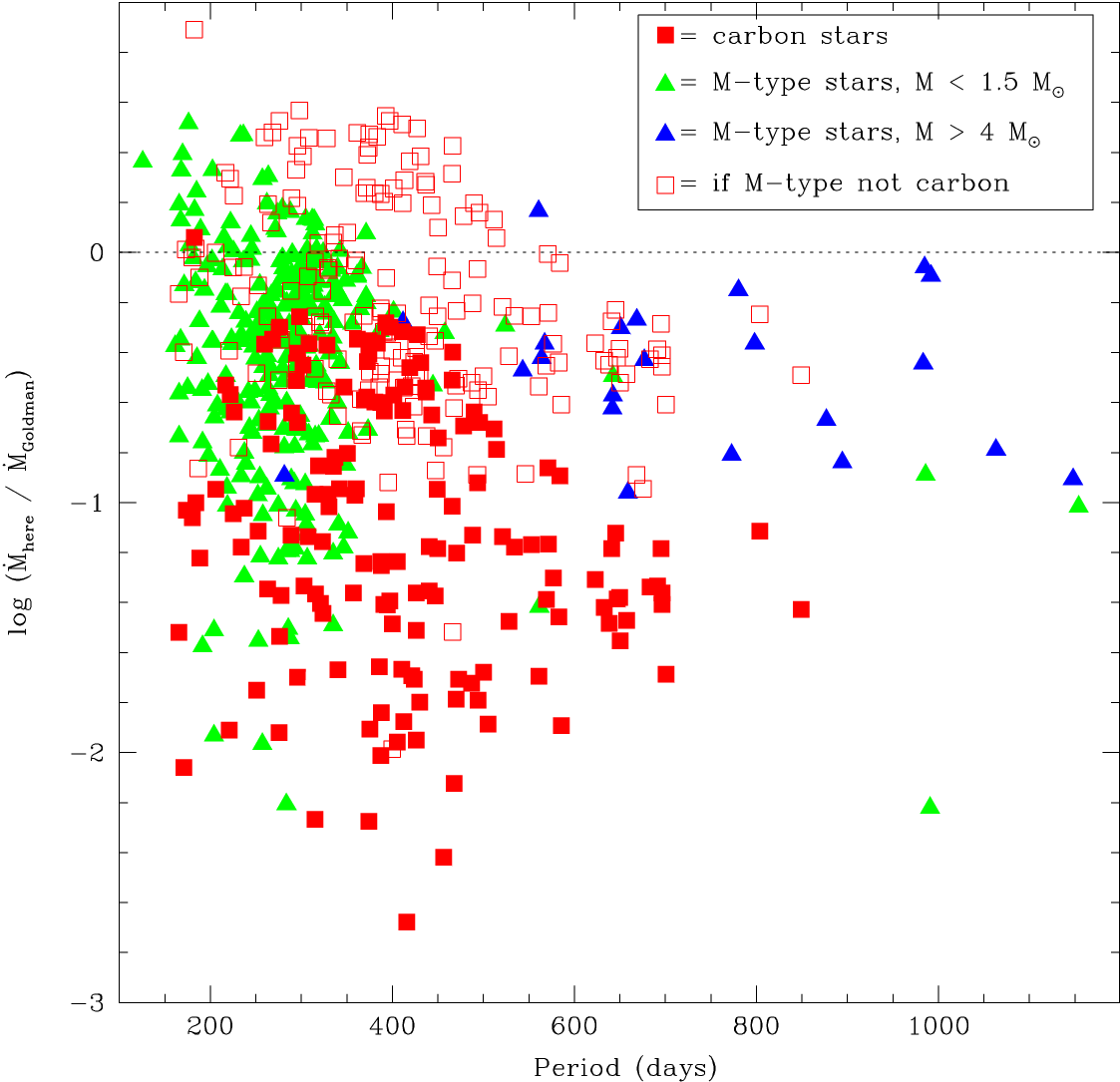,width=84mm}}}
\caption[]{Ratio of mass-loss rate estimated here, to that derived using the
recipe from Goldman et al.\ (2017), vs.\ luminosity (top) and pulsation period
(bottom).}
\end{figure}

The parameterisation with luminosity alone reflects
both
the dependence of the mass-loss rate on stellar (birth) mass,
as well as on evolution.
An additional parameter needs to be invoked
to separate these two effects.
Naturally this would be the stellar effective temperature (van Loon et al.\
2005a), but this is not available for most of the stars in our M\,33 sample.
It explains why Beasor \& Davies (2018) claim that the parameterisations
derived by Goldman et al.\ (2017) and van Loon et al.\ (2005a) grossly
overestimate the mass-loss rates of stars in their clusters, and that the
mass-loss rate increases more rapidly with increasing luminosity than our
approximate
proportionality.
At their own admittance, those star clusters represent stars with a single
birth mass $\sim16$ M$_\odot$, and therefore solely reflect the evolution of
mass loss in time, not the highest rates that are attained in the short phase
that will easily have been missed in just a few star clusters (van Loon,
Marshall \& Zijlstra 2005). The Beasor \& Davies work and ours are not at all
in contradiction with one another, but rather complement eachother.

Correlations with pulsation amplitude can be rather muddled by the fact that
the magnitude scale is a relative scale, and more luminous stars naturally
tend to pulsate with smaller amplitudes expressed this way. Following the
procedure in van Loon et al.\ (2006;
their Eq.\ (1) in \S\,4.5)
(cf.\ van Loon et al.\ 2008), to express the amplitude in terms of the
luminosity (energy) variations, we find a much clearer growth of mass-loss
rate with pulsation amplitude (Fig.\ 18):
\begin{equation}
\log\left(\frac{\dot{M}}{{\rm M}_\odot\ {\rm yr}^{-1}}\right)=1.7\times
\log\left(\frac{A_L}{{\rm L}_\odot}\right)-12.0
\end{equation}
This suggests that even the pulsation mechanism -- by temporarily storing it
in the stellar mantle -- really employs the energy produced through nuclear
fusion in order to drive the mass loss. The remaining scatter in the
correlation is likely to include additional dependencies such as on the
(effective) gravity. We also note the lack of deviation of the carbon stars
from this trend (Fig.\ 18) -- a further indication that most of the carbon
stars have been identified correctly and that their mass-loss rates have not
been severely under-estimated.

\begin{figure}
\centerline{\psfig{figure=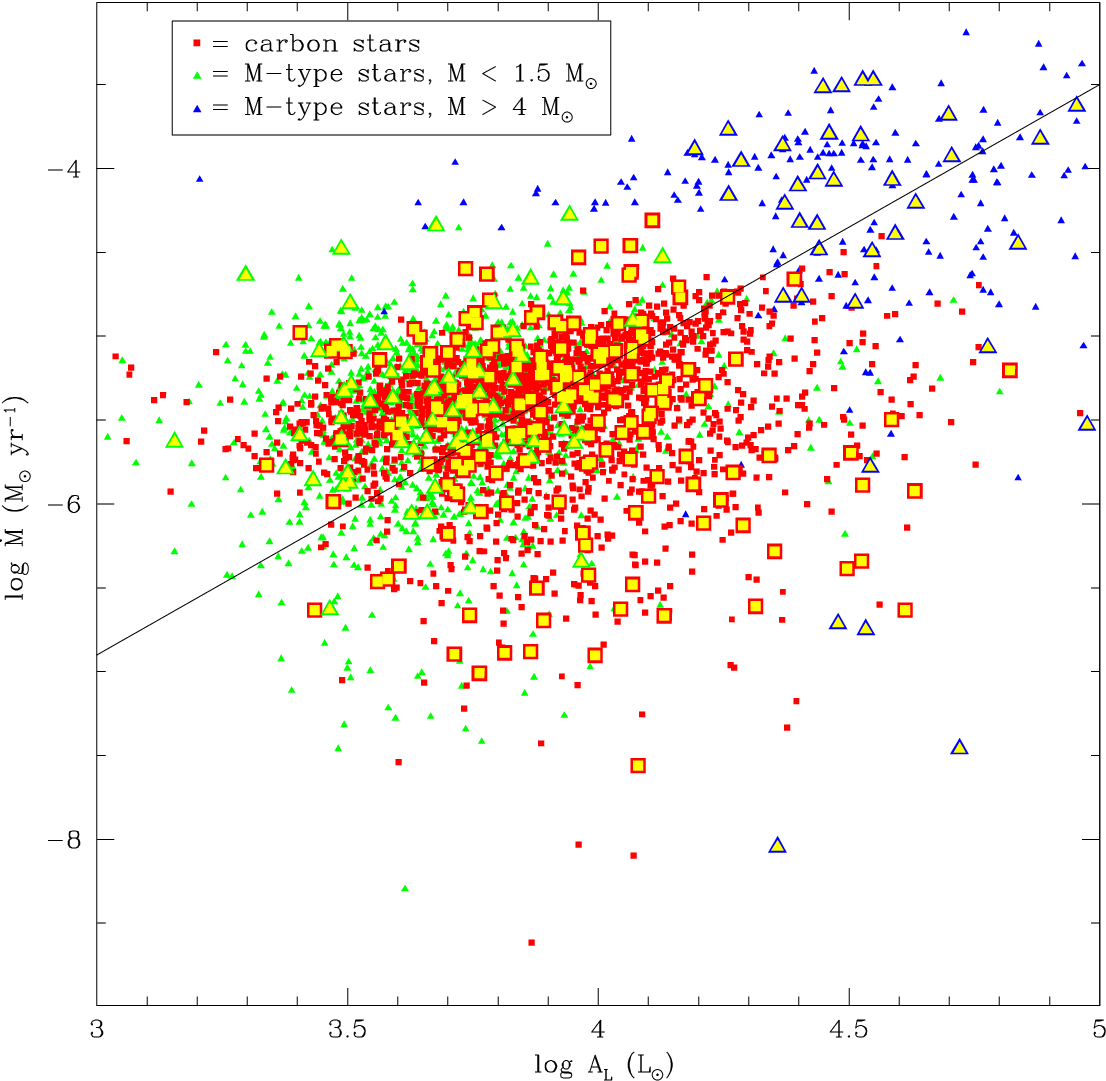,width=84mm}}
\caption[]{Mass-loss rate vs.\ luminosity amplitude. The symbols have the same
meaning as in figure 10.}
\end{figure}

\subsubsection{Mass-loss timescales}

In Papers II and V, comparison between our derived SFRs and those derived by
other, independent means suggested our SFRs needed to be corrected by a factor
seven (WFCAM) to ten (UIST). We also found that, when integrated over the
pulsation phase duration, stars appeared to lose more mass than they were born
with, again requiring an order of magnitude correction. This seemed to confirm
the correction needed to the SFRs, if it is due solely to overestimation of
the pulsation duration ($\delta t$).
The reasons for such mismatch might be related to the treatment of the
mass loss and/or pulsational stability in the Padova models.
However, we have also since found that the survey completeness contributes to
the correction factor to the SFRs. But this would not affect the birth/final
mass assessment, thus leading to tension between the measured mass-loss rates
and pulsation duration. We here revisit the issue, with the aim to resolve it.
We calculate the ratio of the integrated mass loss and the birth mass, as a
function of birth mass (Paper III):
\begin{equation}
\eta = \frac{\sum_{i=1}^{N}{\left(\dot{M}_i\times(\delta t)_i\right)}}
            {\sum_{i=1}^{N}{M_i}},
\end{equation}
and also compare this to the initial--final mass relation determined by
Williams, Bolte \& Koester (2009). The results are presented in figure 19.

\begin{figure}
\vbox{
\centerline{\psfig{figure=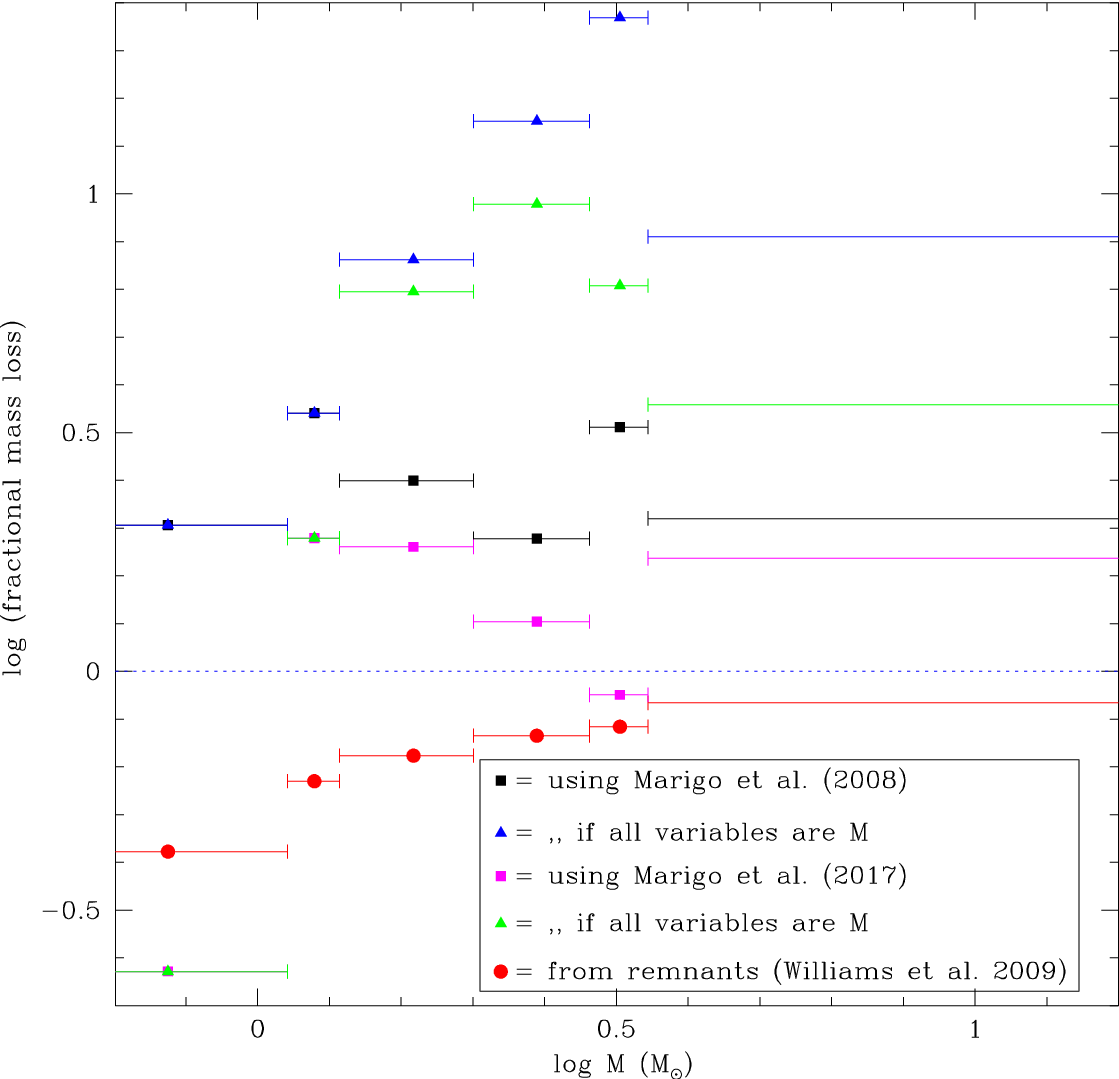,width=84mm}}
\centerline{\psfig{figure=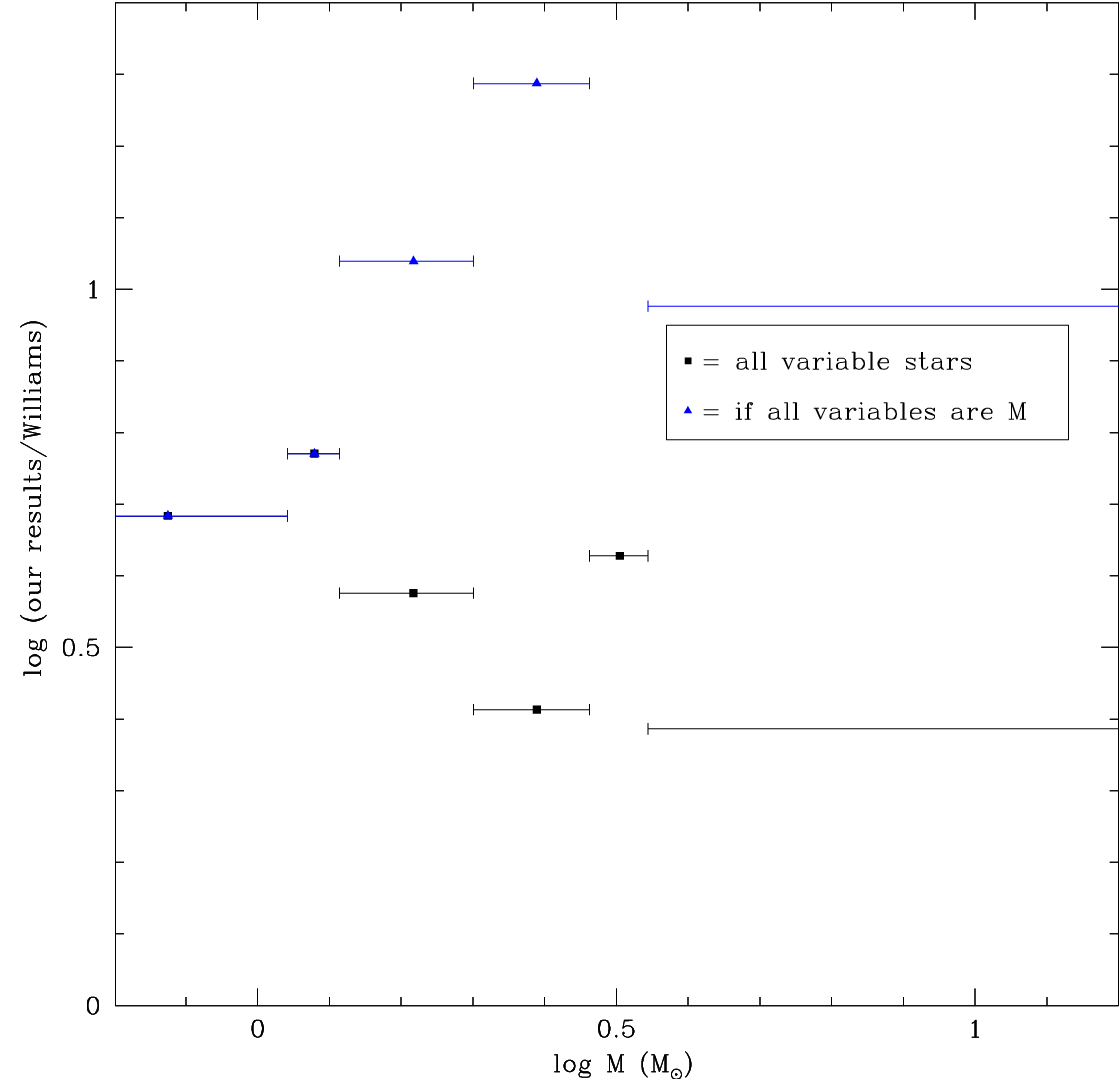,width=84mm}}
}
\caption[]{Top: ratio of the mass lost during the LPV phase, and the birth
mass. The horizontal dotted lines indicate the situation where the star expels
all of its mass. For comparison we also plot the same fractional mass loss
derived from the initial--final mass relation determined by Williams et al.\
(2009). Bottom: ratio of our fractional mass loss, and that of Williams et al.
Clearly, the stars appear to lose more mass than they were born with, and a
correction must be identified and applied.}
\end{figure}

The situation is somewhat less severe as previously noted, with corrections
needed of 0.4--0.7 dex, or factors of about 2.5--5.
We attribute this to the need to shorten the pulsation phase duration by
this factor. The survey incompleteness would then contribute an additional
factor 2--3 correction to the SFRs (but not affect the mass discrepancy).
Since the models of Marigo et al.\
(2008), their carbon star lifetimes were recalibrated by Rosenfield et al.\
(2016). Previous determinations of the duration of the Thermally-Pulsing AGB
(TP-AGB) by Girardi et al.\ (2010) and Rosenfield et al.\ (2014) had been
restricted to initial masses 0.8--2.5 M$_\odot$ and metallicities $-1.54<$
[Fe/H] $<-0.58$. Using ten galaxies with recent star formation from the
AGB--SNAP sample (Dalcanton et al.\ 2012), Rosenfield et al.\ (2016) extended
the analysis to initial masses as high as 4 M$_\odot$, though at nearly the
same, low, metallicities. They used $N_{\rm TP-AGB}/{N_{\rm RGB}}$ as a proxy
for the mean TP-AGB lifetime. Marigo et al.\ (2017) presented updated models
that included revision of the TP-AGB lifetimes. The pulsation duration in
those new models is indeed reduced by about a factor two (Fig.\ 20), though no
new information is presented for stars above 4 M$_\odot$.

We thus suggest that the LPV phase must be reduced by another factor two, and
that it must also be reduced by a similar factor for the most massive AGB
stars and RSGs. This is not surprising, as the pulsation duration in the
Padova models is nearly that of the entire TP-AGB lifetimes (cf.\ Ventura et
al.\ 2018). The latter must be an upper limit to the duration of the more
extreme phase of LPV and heavy mass loss and it is therefore no surprise that
we would infer a shorter duration of that catastrophic phase.

Likewise, if we had wrongly identified stars as carbon stars most of the time,
then the discrepancy worsens. Hence it is unlikely that most of the presumed
carbon stars are in fact oxygen rich; only some of the most luminous examples
may at times have been misclassified because of photometric uncertainties et
cetera.

\begin{figure}
\centerline{\psfig{figure=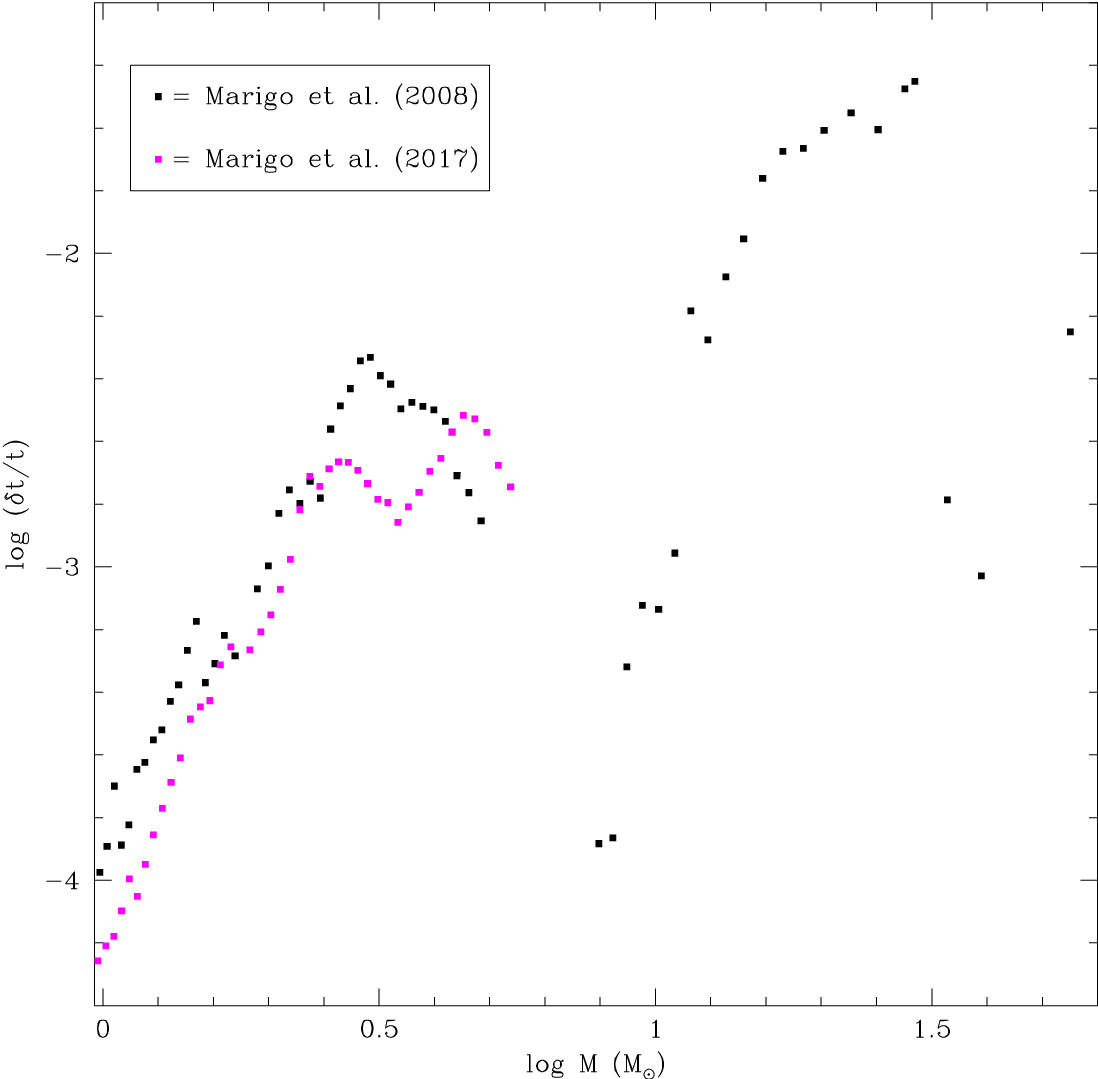,width=84mm}}
\caption[]{Ratio of pulsation duration and age, as tabulated in the Marigo et
al.\ (2008) isochrones (black) and Marigo et al.\ (2017) isochrones (magenta),
vs.\ stellar birth mass.}
\end{figure}

In figure 21 we plot the radial variation in the integrated mass-loss rate,
normalised to the stellar mass (both for the variable stars only). This is a
measure of the duration of the dominant mass-loss phase -- or rather the
inverse of it. In fact, it sets an upper limit to the duration, as the stars
do not completely vanish but leave remnants
(white dwarves, neutron stars or black holes).
The timescale is rather uniform across the disc, and suggests $\delta
t<3\times10^5$ yr for the population as a whole. This is a few times shorter
than the
radial pulsation-phase
timescales of $\sim10^6$ yr as predicted by the models
(for a quick derivation, combine a crude approximation to figure 20:
$\log(\delta t/t)\sim-4+2\log M({\rm M}_\odot)$, with a similar approximation
to Fig.\ 6 in Paper II: $\log t\sim10-2\log M({\rm M}_\odot)$); the latter
are more typical of the TP-AGB duration rather than that of the LPV phase and
the accompanied heavy mass loss. The above estimated population-average
timescale of mass loss is only a little shorter than the duration of the RSG
phase (Jura \& Kleinmann 1990).

\begin{figure}
\centerline{\psfig{figure=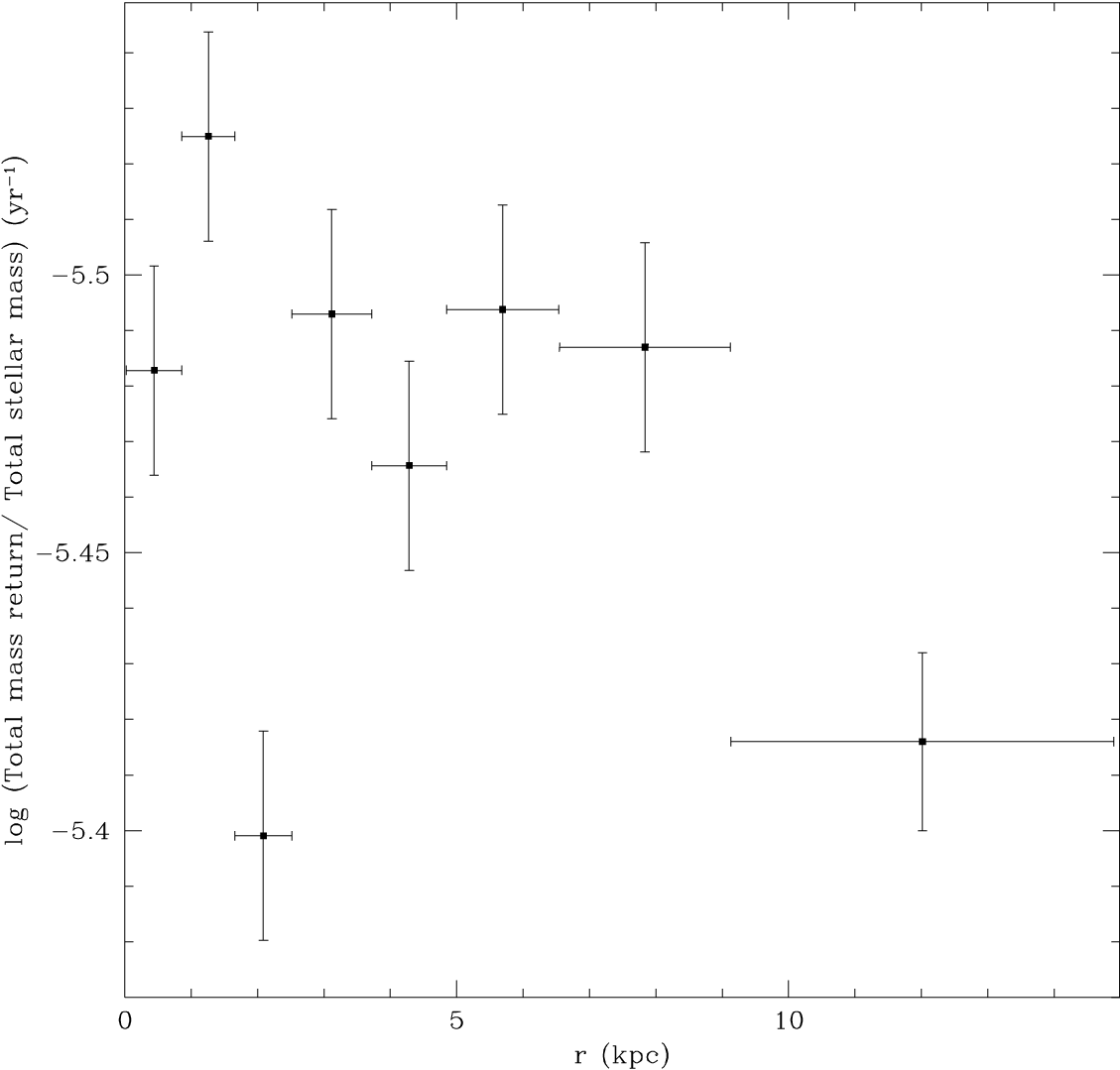,width=84mm}}
\caption[]{Ratio of total mass returned by variable stars, and the integrated
stellar mass of the variables, plotted vs.\ radial distance to the centre of
M\,33.}
\end{figure}

Motivated by this, we have constructed pseudo-evolutionary tracks of the
mass-loss rate and the accumulated mass loss. Assuming that the mass-loss rate
increases monotonically
in time,
we order stars within a given mass range in terms of
their mass-loss rate. This would be an uncalibrated timeline. In order to
calibrate these timelines relatively between different mass ranges, we assume
a constant SFR over the past 10 Gyr and a Salpeter IMF (the relative numbers
within the 1--1.5, 1.5--4, 4--8 and $>8$ M$_\odot$ mass ranges are then
expected to be 0.4215 : 0.4246 : 0.0935 : 0.0604). We assume that this
LPV
phase lasts $2\times10^5$ yr for carbon stars
 (with birth masses in the range 1.5--4 M$_\odot$)
and shall see how reasonable this is. Qualitatively, we see that in all mass
ranges the mass-loss rate steadily increases before shooting up by almost an
order of magnitude in the final $\sim10^4$ yr (Fig.\ 22, top). Until that
final ``super-wind'' phase, the mass-loss rate scales approximately
proportional
with time, $\dot{M}\propto t$. This is very similar across all AGB mass
ranges, and is only -- possibly -- marginally steeper for RSGs overall; but
note again the three different modes of mass loss for RSGs, with jumps around
$\log t\sim3.5$ and $\sim4.1$.

The accumulated mass loss shows an even more strikingly similar evolution
(Fig.\ 22, bottom), which extends into the super-wind phase because of its
short duration: $\Delta M\propto t^2$. With the chosen calibration we find
that the durations of the LPV phase and the accompanied mass loss are
$8\times10^4$ yr for the low-mass AGB stars, $2\times10^5$ yr for the carbon
stars (assumed), and $6\times10^4$ yr for both the massive AGB stars and the
RSGs. The total mass lost in this phase is then in fair agreement with the
initial--final mass relation for low-mass AGB stars (1--1.5 M$_\odot$), which
would lose on average 0.5 M$_\odot$, carbon stars (1.5--4 M$_\odot$), which
would lose on average 1.0 M$_\odot$, and massive AGB stars (4--8 M$_\odot$),
which would lose on average 4.4 M$_\odot$. The amount of mass lost on average
by RSGs ($>8$ M$_\odot$) is with 11 M$_\odot$ rather high; while this population
probably includes stars $>20$ M$_\odot$ most stars in this category will have
birth masses in the 8--20 M$_\odot$ range and besides leaving behind a neutron
star also lose mass in the supernova explosion. On the other hand, Beasor \&
Davies (2018) find a much lower mass lost by RSGs of $\sim16$ M$_\odot$, of
$\Delta M\sim0.6$ M$_\odot$. They conclude that this is well below the
predictions from stellar evolution models, and we would argue that we find
evidence that indeed the amount of mass lost by RSGs exceeds a solar mass,
possibly by several times. The reason for our higher estimate is that we
analysed a much larger population that included sufficient examples of the
brief episode of the highest mass-loss rates that were missed in their star
cluster sample.

\begin{figure}
\centerline{\psfig{figure=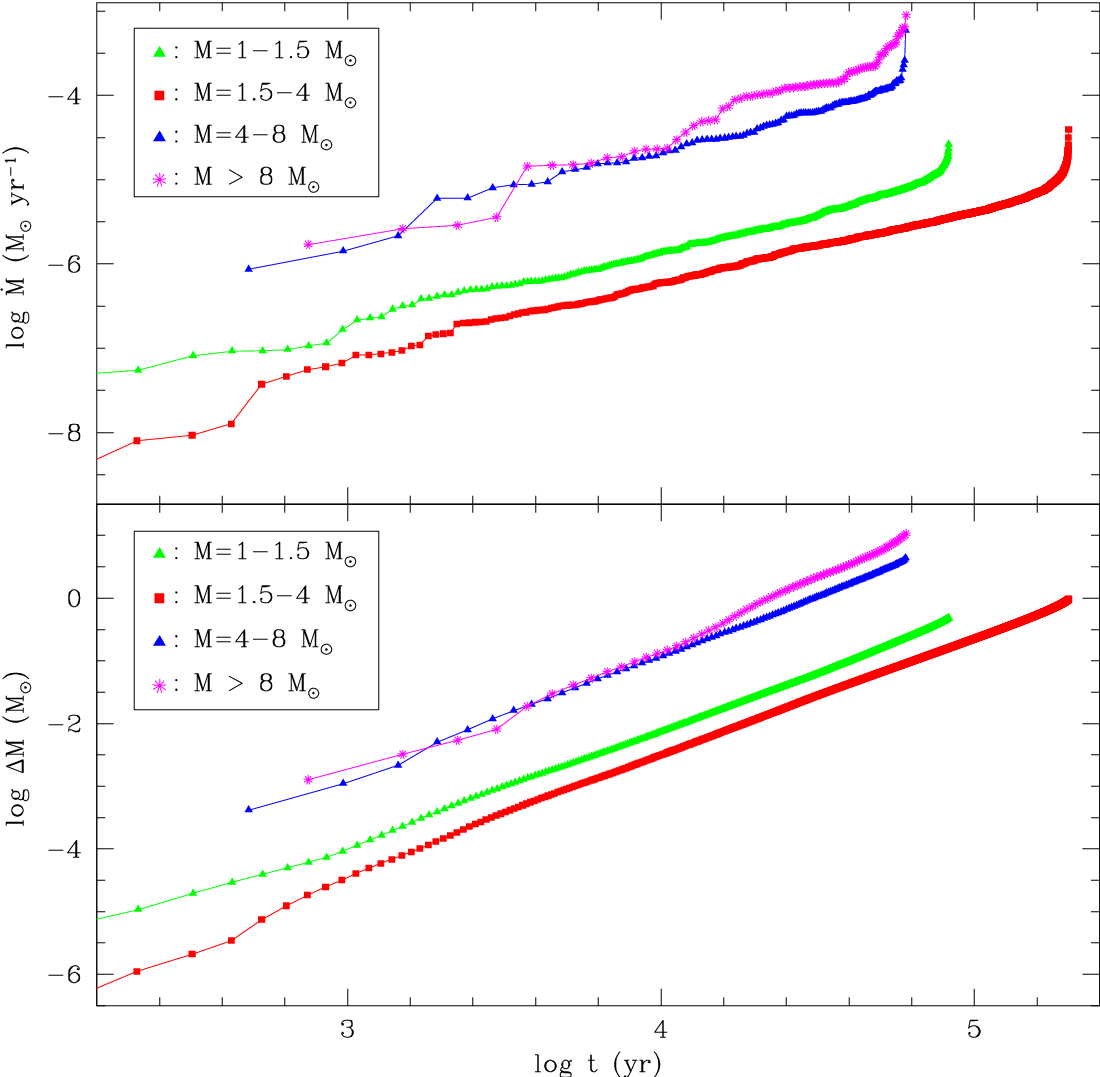,width=84mm}}
\caption[]{Pseudo-evolutionary tracks of the mass-loss rate and accumulated
mass loss, for populations of low-mass AGB stars, intermediate-mass (carbon
star) AGB stars, massive AGB stars and RSGs. This is based on a constant SFR
over the past 10 Gyr, Salpeter IMF and monotonically increasing mass-loss
rate.}
\end{figure}

The good consistency across all AGB stars
in terms of the amount of mass they lose over their (LPV) lifetime
suggests that the mass loss from carbon stars has been estimated accurately.
We may have missed a few extreme carbon stars which could increase $\Delta M$
a little. Our analysis certainly does not support the suggestion that carbon
stars could produce much larger amounts of dust than O-rich AGB stars because
they produce carbon. It would either increase the dust:gas ratio in the
outflow, thus reducing the inferred mass-loss rate from the IR SED and leading
to a shortfall in the mass carbon stars must lose in order to leave behind a
white dwarf. Or the dust:gas ratio would remain constant, leading to a much
higher mass-loss rate which is then in tension with the birth mass and the
requirement for the carbon stars to leave behind a white dwarf.

If one wonders about the shorter duration for the low-mass AGB stars, this is
explained by them sooner losing enough mass to expose a white dwarf interior.

\subsection{Replenishment of the ISM}

A comprehensive picture of the replenishment and enrichment of the ISM is
obtained from a two-dimensional map of the mass return rate surface density
(Fig.\ 27; not deprojected). Apart from a general radial decline in feedback,
and the outline of the inclined disc, localised features are discernible
related to star forming regions and/or spiral arms. We confirm the three
local enhancements surrounding the remarkably quiescent nucleus found already
in Paper III. We also detect enhancements in the IS spiral arm just South from
the centre (see figure 12 in Paper V; see also Gratier et al.\ 2010 and
Rela\~no et al.\ 2018) around RA $\sim23\rlap{.}^\circ35$--$23\rlap{.}^\circ45$,
Dec $\sim30\rlap{.}^\circ5$--$30\rlap{.}^\circ55$, and possibly the onset of the
Northern spiral arms IN, IIN and IIN starting at RA $\sim23\rlap{.}^\circ45$,
Dec $\sim30\rlap{.}^\circ75$. On the other hand, the giant H\,{\sc ii} region
NGC\,604, at RA $\sim23\rlap{.}^\circ64$, Dec $\sim30\rlap{.}^\circ78$, is
striking for its total absence from the mass return map; possibly it is too
young, for the dusty feedback to have kicked in just yet.

\begin{figure*}
\hbox{{\psfig{figure=fig23a.ps,width=89mm}}
{\psfig{figure=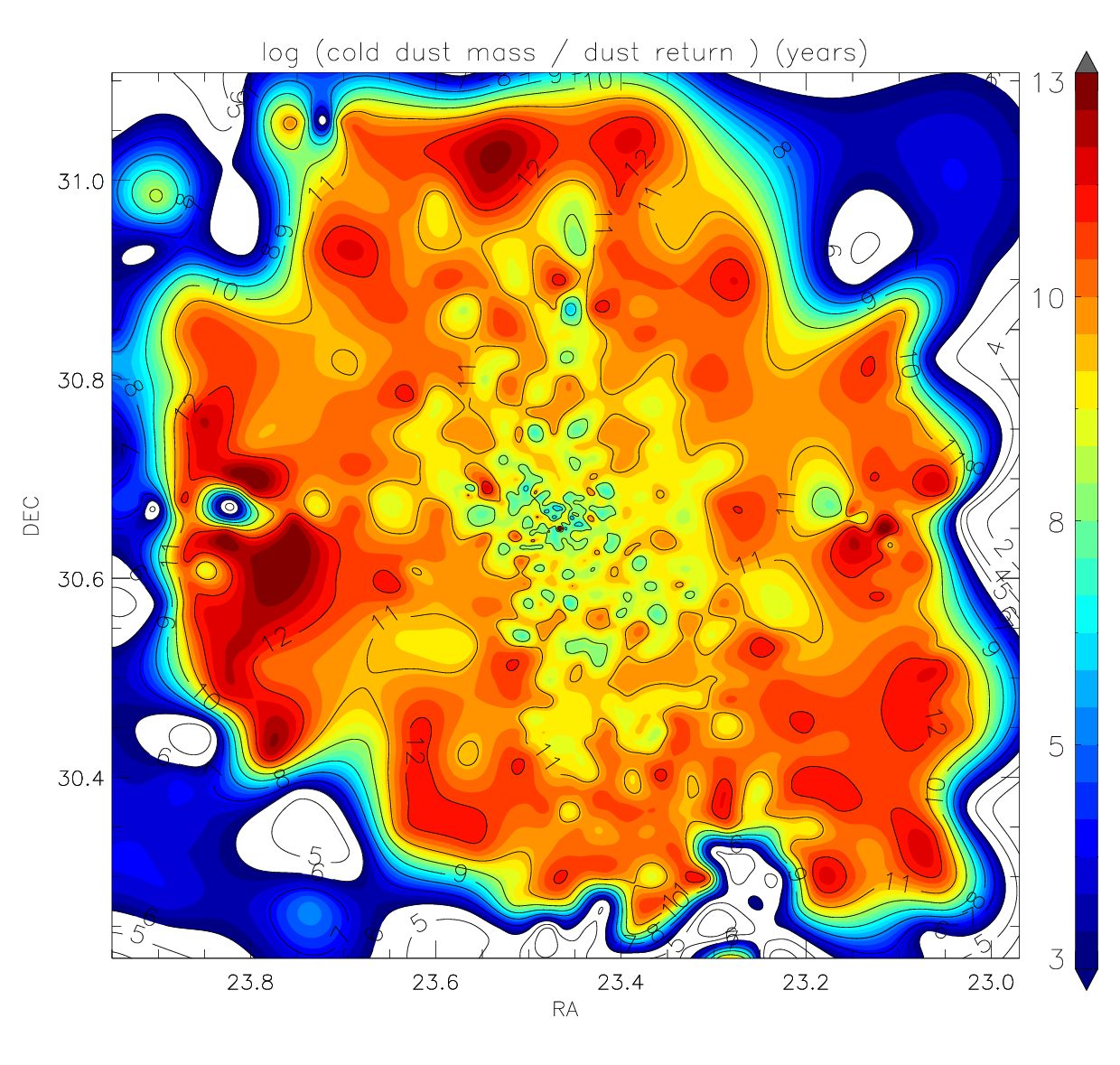,width=89mm}}}
\caption{Left: map of mass-return-rate surface density over the disc of M\,33;
Right: ratio of the cold dust mass and dust-mass return-rate surface density.}
\end{figure*}

Compared to the cold interstellar dust map from Tabatabaei et al.\ (2014),
there appears to be a decline in the ability for feedback to replenish the ISM
dust (Fig.\ 27, right panel), with timescales for full replacement of the
current ISM dust mass ranging from $\sim10^8$ yr in the central parts, to a
Hubble time in the outskirts. However, this also depends on the rate at which
the ISM is consumed in star formation and/or removed by other processes. Kang
et al.\ (2012) estimated the depletion timescale to vary from $\sim0.3$ Gyr in
the centre of M\,33 to $\sim2$ Gyr in the outskirts. This would mean that the
lower feedback in the outskirts is partially offset by the longer depletion
timescale, resulting in a more -- but not totally -- uniform replenishment of
the ISM.

Verley et al.\ (2009) estimated the recent star formation rate across M\,33 to
be $\xi=0.45\pm0.10$ M$_\odot$ yr$^{-1}$. Our estimated mass return rate of
$\zeta=0.1$ M$_\odot$ yr$^{-1}$ thus falls short by a factor $\xi/\zeta=4.5\pm1$
of sustaining the ISM at its current level, i.e.\ at more than 3 $\sigma$
significance.
Matsuura et al.\ (2009) found a similar mass deficit in the LMC, and
Boyer et al.\ (2012) and Matsuura, Woods \& Owen (2013) found the same in the
SMC.

We
note that it is hard to see how the mass-loss rates could have been
underestimated, as this would make it even more problematic that stars
appeared to lose more mass than they were born with. The total size of the
evolved star population also sets a firm limit on the possible incompleteness
level of our survey; we recall that we detected more than half of the heavily
reddened evolved stars as variables.
While very red sources can be missed these are often faint and not
necessarily the most prolific mass-losers (van Loon et al.\ 1997) or so rare
they do not dominate unless by stochastic effects in small systems or locally
(van Loon et al.\ 2005b; Gonz\'alez-L\'opezlira 2018; cf.\ Jones et al.\ 2017
\& Goldman et al.\ 2018). Srinivasan et al.\ (2009) reach similar values for
the total mass return as compared to Matsuura et al.\ (2009), but they did
this without including extreme carbon stars but by including more
low-mass-loss-rate stars. Looking at figures 2, 10, and 15, the sources we did
not include in our sample are at most about twice as red in the near- and/or
mid-IR (those would be the equivalent of the most extreme sources known in the
aforementioned LMC studies). There are fewer of those than there are of the
reddest sources that we did include, and not all of those will be mass-losing
evolved stars. It is therefore unlikely that we have underestimated the mass
return rate by more than a factor two, and our value already errs on the high
side for reasons of incompleteness.
Hence, there is little room for our estimated mass return rate to be elevated
above $\zeta=0.1$ M$_\odot$ yr$^{-1}$.

However, additional mass is also returned by supernov{\ae}, LBV eruptions and
hot massive-star winds. These contributions come from massive stars, which
contribute about one fifth to the mass returned through dusty stellar winds
(see Fig.\ 19, top right panel). Assuming a standard Salpeter initial mass
function, the total mass in stars more massive than $M_0$ compared to that in
stars between 1 and $M_0\gg 1$ M$_\odot$ is $\sim M_0^{-1.35}$, or about 6 per
cent for $M_0=8$ M$_\odot$.
Lower-mass stars leave degenerate remnants that are more massive compared
to their birth mass than massive stars do. However, while a solar mass star
has indeed $M_{\rm birth}/M_{\rm remnant}\sim2$ this decreases to $\sim7$ already
before the realm of the massive stars is reached, and most massive stars
tjhemsleves will leave behind a remnant of $\sim1.4$ M$_\odot$ or more.
Accounting for this will therefore not change the contribution from massive
stars by more than a few per cent.
While the low-mass stars that formed more recently will not have started to
contribute to the feedback, for a constant SFR this would only make a factor
two difference at most. Thus, unaccounted-for mass return by massive stars
cannot elevate the estimated total mass return rate by more than about ten per
cent.

For star formation to continue beyond the next Gyr or so, gas must flow into
the disc of M\,33, via cooling flows from the circum-galactic medium and/or by
inward migration from gas reservoirs in the outskirts of the disc (cf.\
Putman, Peek \& Joung 2012). Indeed, Putman et al.\ (2009) see evidence for
gas that had been tidally disrupted from M\,33 by M\,31, to be falling back
onto M\,33, although they estimate that M\,33 will exhaust this gas supply
within the next few Gyrs.

The star formation rate peaks within the central $r<0.2$ kpc (Papers II \& V),
and the molecular gas density peaks at $r\sim0.1$ kpc from the centre (Tosaki
et al.\ 2011). Figure 20, on the other hand, suggests that the replenishment
rate is highest around $r\sim0.3$ kpc. Therefore, the very centre of M\,33,
while characterised by high levels of gas depletion is not where the ISM is
replenished at the highest rate. This suggests a mechanism exists by which gas
moves inwards from where it was injected, to where stars form from it (cf.\
Nelson et al.\ 2018). This would imply some delay between feedback and re-use
of gas and dust.

We see something similar on larger scales (Fig.\ 28), when we express the mass
return rate in relation to the local, recent star formation rate. This ratio
approaches (but does not quite reach) unity around $r=3$--4 kpc, whereas the
mass return rate falls short of sustaining star formation by a few times in
the outskirts of the disc, and by as much as an order of magnitude in the
central kpc. This was not obvious from inspecting the radial profile of the
mass return rate (Fig.\ 20) and SFR (Paper V) separately, as the rather
subtle change in slope around 3 kpc from a shallower radial profile in mass
return rate, to a shallower radial profile in SFR causes a peak around that
pivotal point. One may thus wonder what happens in the galaxy around 3 kpc,
that causes this. Has it got to do with the galactic structure, for instance
a dynamical resonance, or is it the result of the delay (and positional
displacement) between where stars form and where they return matter?

\begin{figure}
\centerline{\psfig{figure=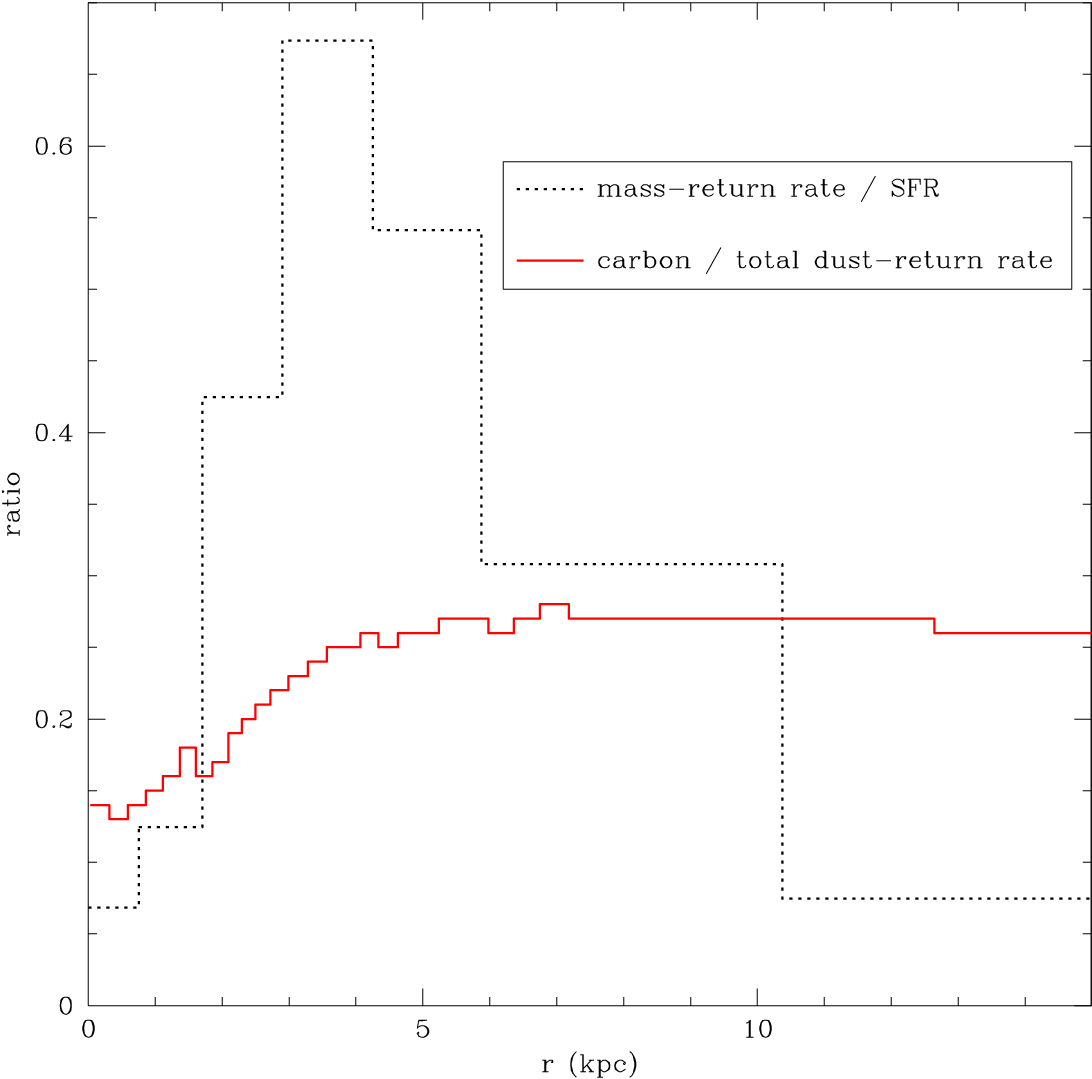,width=84mm}}
\caption[]{Ratio of total mass returned by variable stars, and the recent SFR
(black), and the mass fraction of carbonaceous dust returned to the ISM (red),
both plotted vs.\ radial distance to the centre of M\,33.}
\end{figure}

Finally, in figure 29 we show the mass returned by just the low-mass AGB stars
(1--1.5 M$_\odot$) and carbon stars (1.5--4 M$_\odot$); the high-mass map is
too stochastic to be meaningful. It is very clear that the feedback from the
low-mass stars occurs predominantly in the nuclear region whereas the feedback
from the carbon stars occurs much more uniformly across the disc. The reason
for this is twofold: the ancient star formation epoch -- which is responsible
for the low-mass AGB stars we see today -- was concentrated in the centre, and
the lower metallicity in the disc has favoured the formation of carbon stars.
This is also visible in figure 28, in which we plot the mass ratio of the dust
that is being returned to the ISM in the form of carbonaceous grains compared
to the total. This ratio is just over a quarter in the outer disc, but drops
to about a seventh in the nuclear region where the solar metallicity is not
(as) conducive to the formation of carbon stars.
At lower metallicity, and in generally older stellar systems, carbon
stars are more plentiful and will make a relatively larger contribution to
interstellar dust -- see for comparison the SMC (Boyer et al.\ 2012) and outer
disc of the Milky Way (Ishihara et al.\ 2011).

\begin{figure}
\vbox{{\psfig{figure=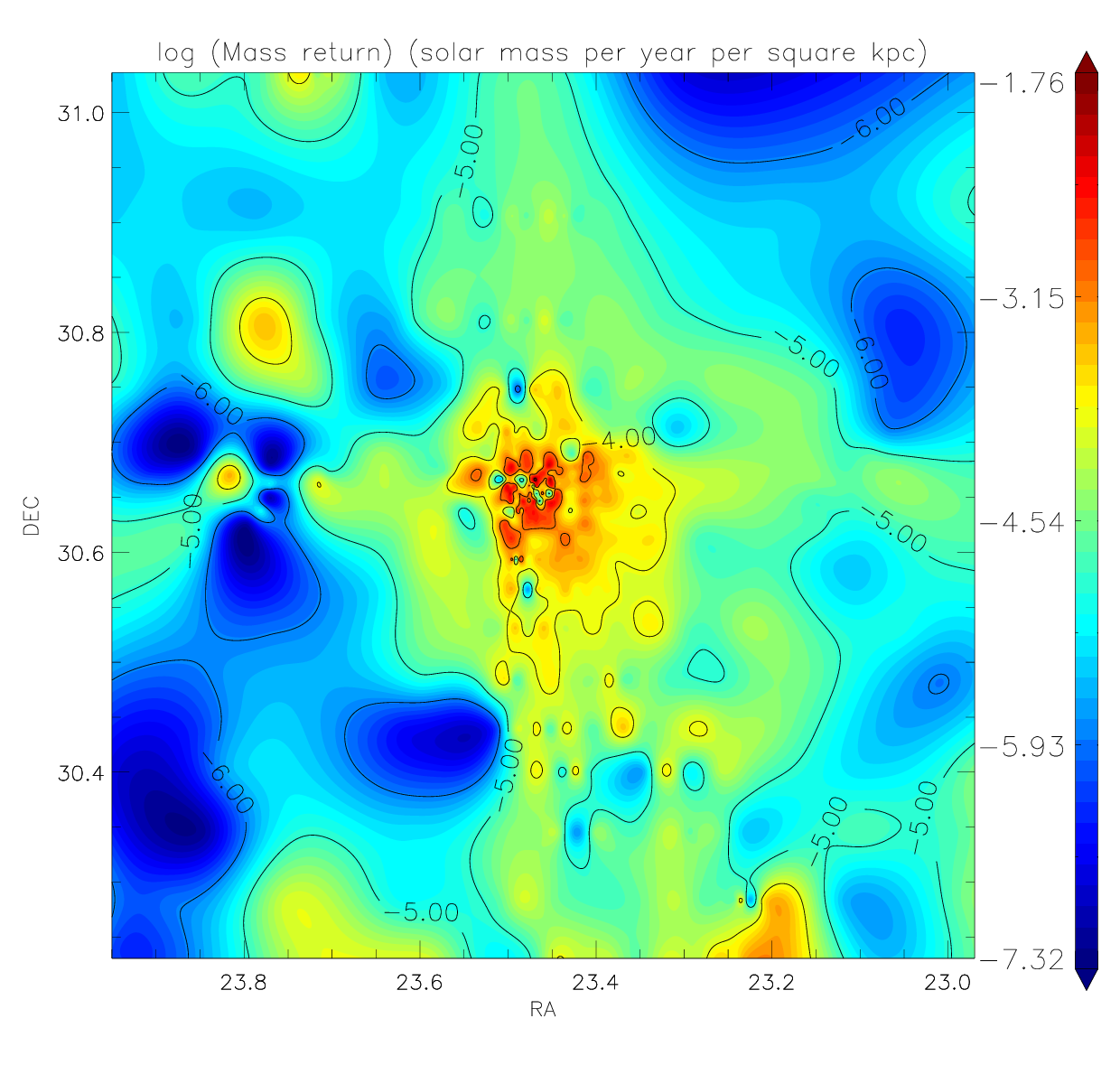,width=84mm}}
{\psfig{figure=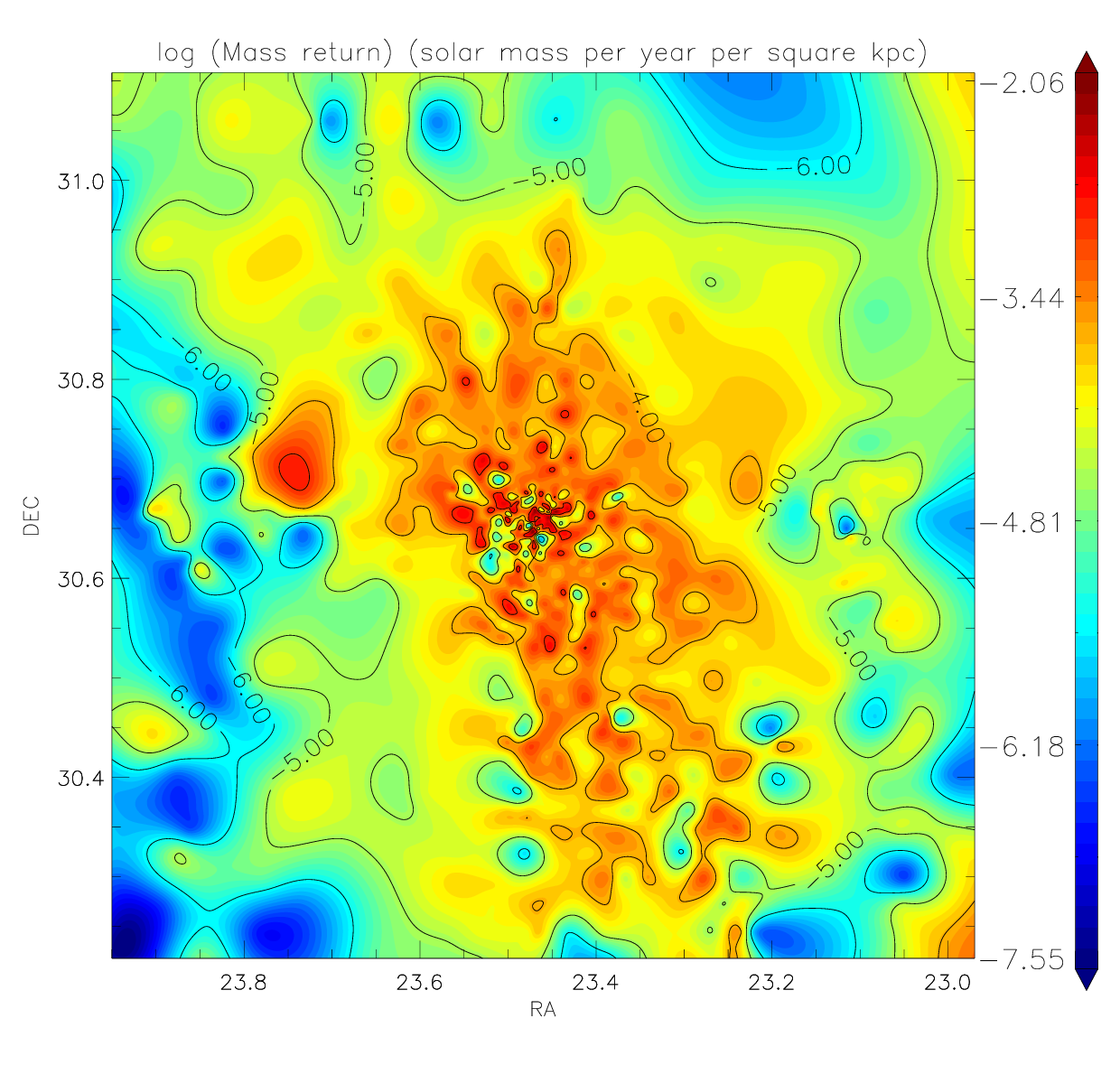,width=84mm}}}
\caption{Top: map of mass-return-rate surface density for low-mass stars
(1--1.5 M$_\odot$); Bottom: for carbon stars (1.5--4 M$_\odot$).}
\end{figure}

\section{Summary of conclusions}

This paper is the culmination of our long-term monitoring programme of
infrared variable star in M\,33, presenting the measurement of the mass-loss
rates and their analysis in terms of stellar evolution and galaxy evolution.
This has led to a better understanding of the mass-loss and dust formation
process, the timescales on which this mass loss happens, and the rate at which
the ISM is replenished and what this means for global galactic processes such
as star formation and gas recycling and the migration of stars and gas. The
pertinent results can be summarised as follows:
\begin{itemize}
\item[$\bullet$]{We have shown that the mass-loss rates from AGB stars and
RSGs increase with birth mass, reflected in an approximate
{\bf proportionality with}
luminosity, but for stars of a given mass there is also an evolutionary term
that needs to be quantified. We have shown that the luminosity amplitude
accomplishes this to a better degree than the magnitude amplitude or period
do.}
\item[$\bullet$]{Carbon stars do not lose mass at higher rates than O-rich
stars of similar luminosity, and super-AGB stars lose mass at rates as high as
expected for their luminosities. The mass loss from pulsating RSGs shows
discrete modes: below the nuclear burning rate ($\sim10^{-6}$ M$_\odot$
yr$^{-1}$), around it ($\sim10^{-5}$ M$_\odot$ yr$^{-1}$), and well above it
($>10^{-4}$ M$_\odot$ yr$^{-1}$).}
\item[$\bullet$]{To comply with the initial--final mass relation, we have
shown that the timescale of dominant mass loss must be a few times shorter
than the TP-AGB or RSG duration, from $\sim2\times10^5$ yr for carbon stars
down to $\sim6\times10^4$ yr for massive AGB stars and RSGs, with low-mass AGB
stars spending $\sim8\times10^4$ yr in this phase. Throughout this phase, for
both AGB stars and RSGs the mass-loss rate and accumulated mass lost evolve
with time as, respectively, $\dot{M}\propto t$ (until a brief final superwind)
and $\Delta M\propto t^2$ (the superwind only makes a modest contribution).}
\item[$\bullet$]{The replenishment of the ISM by mass loss, accounting for
some mass lost in other end stages of stellar evolution and survey
incompleteness, at $\dot{M}\sim0.1$ M$_\odot$ yr$^{-1}$ falls short by about a
factor four to sustain star formation at the current rate, thus requiring
external sources of gas supply. The discrepancy is largest in the centre and
outskirts, and no more than a factor two around 3--5 kpc.}
\item[$\bullet$]{The dust being returned to the ISM is mostly oxygenous, with
the mass fraction of dust being carbonaceous ranging from 1/7 in the centre to
1/4 beyond 5 kpc.}
\end{itemize}

\section*{Acknowledgments}
We thank the staff at UKIRT for their excellent support of this programme.
We are also grateful to the referee for their thorough review, which
helped improve the clarity of the presentation.
JvL thanks the School of Astronomy at IPM, Tehran, for their hospitality
during his visits. We are grateful for financial support by The Leverhulme
Trust under grant No.\ RF/4/RFG/2007/0297, by the Royal Astronomical Society,
and by the Royal Society under grant No.\ IE130487.


\appendix

\section{Known sources in M\,33}

Below we describe several sub-samples of interest. This helps to assess the
completeness and reliability of our work and that of others, and enables us to
make informed decisions to exclude certain sources from further analysis. It
also highlights the results for certain types of objects. Interestingly, the
contamination of our sample by variable young stellar objects seems to be
negligible, in contrast to the situation in the Milky Way plane (Contreras
Pe\~na et al.\ 2017).

\subsection{Stars apparently more massive than 50 M$_\odot$}

Our WFCAM survey identified a small number of variables with masses derived
from their K$_{\rm s}$-band magnitude in excess of 50 M$_\odot$ (Table A1). Stars
that massive are not expected to reach the RSG branch. Below we describe
attempts to identify them with known objects, and in figure A1 we present their
light-curves.

Combination of near-IR colours and radial velocities suggests that many of
these sources are Galactic foreground stars. Hence we classified \#138, \#535,
\#300, \#320, \#436, \#2926, \#462, \#379 and \#93347 as Galactic foreground
stars. None of these sources have been identified at 24 $\mu$m (Montiel et
al.\ 2015; see also Paper IV). Drout et al.\ (2012) listed source \#58111
($\log L/{\rm L}_\odot=6.42$, $\log\dot{M}=-2.25$) as a foreground dwarf star.

Source \#323 is listed as a possible yellow supergiant by Drout et al.\
(2012). They determined $\log T_{\rm eff}=3.711$ and $\log L/{\rm L}_\odot=6.26$
if in M\,33). There is no {\it Spitzer} detection. Furthermore, there is an
offset between the estimated radial velocity of this source and its expected
velocity based on the position in M\,33, which makes it just as likely to be a
foreground Halo star.

Sources \#150838 ($\log L/{\rm L}_\odot=5.20$, $\log\dot{M}=-3.25$), \#301407
($\log L/{\rm L}_\odot=5.59$, $\log\dot{M}=-3.02$) and \#301327
($\log L/{\rm L}_\odot=5.21$, $\log\dot{M}=-3.39$) are located very near to
the edge of the image and therefore their photometry are unreliable. They are
not detected with {\it Spitzer} -- which would have been expected if their
luminosities and mass-loss rates were correct -- and thus we reject them being
mass-losing evolved stars.

Source \#335 ($\log L/{\rm L}_\odot=5.71$, undetectable mass loss) is
identified as a globular cluster in M\,33, and the light-curve looks messy.

Source \#404 is in the middle of a massive OB association next to a dark
cloud. Supergiant star [HS80]4C is located a mere half an arcsecond from this
source. The region is severely crowded, therefore the {\it Spitzer} photometry
might be affected by blending. It certainly is not a RSG, but possibly an
embedded young stellar object or a blue supergiant illuminating dust
surrounding the OB association.

Source \#640 is a stellar object located in a compact cluster within a small
H\,{\sc ii} region. Within $0\rlap{.}^{\prime\prime}7$ lies a supernova
remnant. Obviously, the {\it Spitzer} photometry is affected by blending with
the remnant, so it is difficult to know what the near-IR object really is.
There is no 8-$\mu$m measurement. Our near-IR data appears to be affected by
blending as well. This is likely an evolved massive star, but its estimated
mass-loss rate is unreliable.

Source \# 636 ($\log L/{\rm L}_\odot=5.43$, $\log\dot{M}=-4.42$) is listed as
an M1\,Ia red supergiant in Drout et al.\ (2012). The radial velocity
$v=-103.2$ km s$^{-1}$ is very close to what is expected. It is also in a star
cluster, so likely to be in M\,33. Its photometry is affected by crowding, and
this is why it does not have {\it Spitzer} photometry. While this is a RSG in
M\,33, the estimated mass-loss rate should be considered inaccurate.

Indeed, there is good evidence for some of the sources to be dusty evolved
stars. Source \#262487 ($\log L/{\rm L}_\odot=4.25$, $\log\dot{M}=-4.58$) has
no J-band detection but a very red H--K$_{\rm s}$ colour, and the light-curve
also suggests this is possibly a very dusty evolved star. Sources \#246352 and
\#9025 are very red both in the near- and mid-IR. The light-curves of these
stars are consistent with heavily dust-enshrouded OH/IR stars with periods
around three years.

Furthermore, source \#176091 ($\log L/{\rm L}_\odot=4.96$, $\log\dot{M}=-3.47$)
is very red at near-IR wavelengths -- there is no optical counterpart. It has
definitely mid-IR excess, so it is not just red because of interstellar
extinction. The luminosity is very reasonable and the light-curve shows a long
decline, so perhaps this is a Luminous Blue Variable (LBV). Source \#447 is
bright but not particularly red, and its 8-$\mu$m flux density is in excess of
those at shorter wavelengths; it could be a post-RSG for which the luminosity
estimated by us is correct and only the mass is overestimated.

\begin{table*}
\caption{Variables with $M>50$ M$_\odot$ in M\,33, with UKIRT ID No.\ (Paper
IV) and photometry in magnitudes.}
\begin{tabular}{rllccccccl}
\hline \hline
ID & RA(J2000) & DEC(J2000) & $K_{\rm s}$ & $(J-K_{\rm s})$ & $(H-K_{\rm s})$ & [3.6] & [4.5] & [8] & included? \\
\hline
   320 & 01:31:46.074 & +30:28:16.37 & 12.655 &           0.382     &           0.572 &        &        &        & no  \\
150838 & 01:32:04.789 & +30:53:46.03 & 16.656 &                     &           2.484 &        &        &        & no  \\
   300 & 01:32:10.867 & +30:41:12.75 & 12.999 &           0.602     &           0.550 &        &        &        & no  \\
   535 & 01:32:15.377 & +30:26:33.24 & 12.888 & \llap{$-$}0.012     &           0.349 & 12.711 & 12.583 & 12.402 & no  \\
   335 & 01:32:24.027 & +30:12:43.23 & 12.974 &           0.557     &           0.065 & 12.644 & 12.671 & 12.539 & no  \\
   323 & 01:32:50.662 & +30:45:10.37 & 13.099 &           1.334     &           0.647 &        &        &        & no  \\
   138 & 01:32:58.638 & +30:52:52.40 & 12.187 &           0.302     & \llap{$-$}0.193 & 12.013 & 12.028 & 11.809 & no  \\
   447 & 01:33:05.445 & +30:31:38.10 & 13.355 &           1.114     &           0.333 & 12.843 & 12.624 & 11.848 & no  \\
262487 & 01:33:12.128 & +30:14:39.15 & 17.315 &                     &           2.837 & 15.477 & 14.875 & 13.366 & yes \\
246352 & 01:33:15.455 & +30:35:04.36 & 16.525 &                     &           2.676 & 14.699 & 14.176 & 11.844 & yes \\
  9025 & 01:33:38.667 & +30:21:04.45 & 15.156 &           4.495     &           1.976 & 13.876 & 13.231 & 11.897 & yes \\
   636 & 01:33:39.341 & +30:31:18.91 & 13.459 &           1.170     &           0.580 &        &        &        & no  \\
301838 & 01:33:44.553 & +30:32:48.02 & 16.628 &           1.333     & \llap{$-$}0.079 &        &        &        & no  \\
   436 & 01:33:53.573 & +30:37:05.02 & 13.166 &           0.825     &           0.249 & 12.475 & 12.549 & 12.819 & no  \\
  2926 & 01:33:53.834 & +30:34:21.04 & 13.437 & \llap{$-$}0.868     &           1.341 & 12.242 & 12.234 & 12.085 & no  \\
   404 & 01:33:58.510 & +30:34:19.60 & 13.222 &           0.960     &           0.283 & 12.398 & 12.521 & 10.768 & no  \\ 
176091 & 01:34:11.738 & +30:53:58.61 & 16.470 &                     &           2.197 & 15.196 & 15.111 & 14.141 & no  \\
   640 & 01:34:16.510 & +30:51:55.26 & 13.400 &           1.549     &           0.670 & 12.025 & 11.626 &        & no  \\
   462 & 01:34:48.178 & +31:06:27.35 & 12.154 &           0.680     &           0.669 & 11.315 & 11.446 & 11.257 & no  \\
 58111 & 01:35:02.727 & +30:26:29.73 & 15.057 &                     &           2.928 &        &        &        & no  \\
 93347 & 01:35:04.638 & +30:26:33.44 & 15.806 &                     &           2.766 &        &        &        & no  \\
   379 & 01:35:56.452 & +30:16:44.97 & 13.127 &           0.350     &           0.284 &        &        &        & no  \\ 
301327 & 01:35:56.782 & +30:26:31.17 & 15.308 &           4.044     &                 &        &        &        & no  \\
\hline
\end{tabular}
\end{table*}

\begin{figure}
\centerline{\psfig{figure=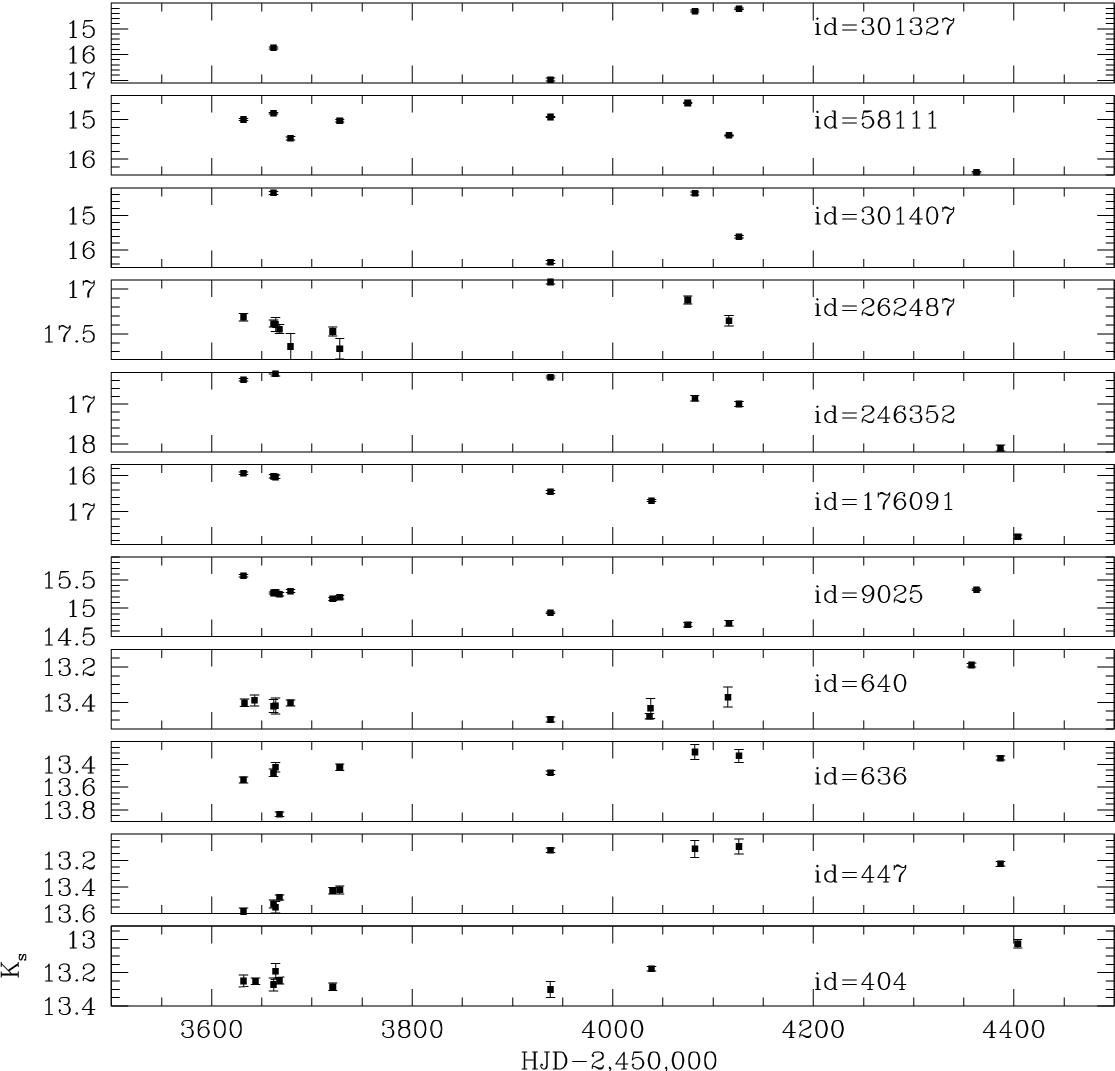,width=84mm}}
\caption[]{Light-curves for stars with estimated $M>50$ M$_\odot$.}
\end{figure}

\subsection{Spitzer 24-$\mu$m variables}

Montiel et al.\ (2015) detected a small number of variables among the point
sources detected with {\it Spitzer} at 24 $\mu$m. We had shown, in Paper IV,
that only eleven sources are credible dusty evolved stars. Among these, eight
are WFCAM variables and two probably as well\footnote{We have since found that
the counterpart of VC\,17 is a different, certain WFCAM variable rather than a
probable variable.}, and another one (VC\,14) is an optically known LPV RSG
(Drout et al.\ 2012) of which we had not detected near-IR variability. In
table A2 we list the results from our SED modelling; figure A2 shows the SEDs
and the fits to the SEDs.

On the basis of the shape of the SED, the success with which we can fit it,
and the variability we assess the likely nature of these sources. We thus
confirm that all WFCAM variables and the known RSG are cool evolved stars;
including the 24-$\mu$m flux density suggests that VC\,17, too, is a RSG.
VC\,14 is detected because it is so luminous, not because it is so dusty.
VC\,13 has a rather flat SED, so we are not hundred per cent convinced it is
an AGB star; likewise, there is some uncertainty regarding the counterpart of
VC\,7, hence the question mark. We interpret VC\,22 as a YSO; the SED is not
consistent with an obscured star, and if it were an AGB star it would have an
unprecedented mass-loss rate for such modest luminosity. VC\,2 and VC\,4, too,
are more consistent with a YSO; we note the similarity between VC\,2 and
IRAS\,05346$-$6949 in the LMC (van Loon et al.\ 2001b; Jones et al.\ 2017).

\begin{table*}
\caption{List of 24-$\mu$m variables in M\,33, with UKIRT ID No.\ (Paper IV).
The luminosities and mass-loss rates are derived from SED fits, which should
be appropriate for AGB stars and RSGs but not YSOs.}
\begin{tabular}{rrccccll}
\hline\hline
name & ID & RA(2000) & DEC (2000) & $\log L/{\rm L}_\odot$ & $\log\dot{M}$ (M$_\odot$ yr$^{-1}$) & variable? & type \\
\hline
 2 & 311369 & 01:34:22.85 & +30:34:09.9 & 5.17 & \llap{$-$}2.79 & no       & YSO? \\
 4 &  39836 & 01:33:32.64 & +30:36:55.5 & 5.06 & \llap{$-$}3.37 & no       & YSO? \\
 6 & 304069 & 01:33:29.70 & +30:24:08.6 & 4.61 & \llap{$-$}3.56 & yes      & AGB  \\
 7 &  17077 & 01:34:12.95 & +30:29:38.5 & 4.81 & \llap{$-$}3.55 & no       & AGB? \\
   & 160486 & 01:34:12.87 & +30:29:40.1 &      &                & no       & AGB? \\
 8 & 305279 & 01:33:28.38 & +30:36:47.9 & 4.52 & \llap{$-$}3.75 & yes      & AGB  \\
 9 & 304597 & 01:34:27.85 & +30:43:40.0 & 4.66 & \llap{$-$}3.65 & yes      & AGB  \\
10 & 252686 & 01:33:50.06 & +30:16:31.7 & 4.41 & \llap{$-$}3.66 & probably & AGB  \\
13 &  16033 & 01:33:26.65 & +30:57:14.4 & 4.76 & \llap{$-$}3.80 & yes      & AGB? \\
14 &    453 & 01:34:12.25 & +30:53:14.1 & 5.46 & \llap{$-$}4.60 & no       & RSG  \\
16 &   8656 & 01:33:47.34 & +30:16:32.4 & 4.52 & \llap{$-$}4.09 & yes      & AGB  \\
17 &  24352 & 01:33:49.86 & +30:52:41.3 & 5.06 & \llap{$-$}3.55 & yes      & RSG? \\
20 & 249854 & 01:33:19.68 & +30:31:05.1 & 4.26 & \llap{$-$}4.04 & probably & AGB  \\
21 &  53418 & 01:33:41.54 & +30:14:12.7 & 4.44 & \llap{$-$}4.03 & yes      & AGB  \\
22 & 182878 & 01:33:37.43 & +30:55:50.4 & 3.96 & \llap{$-$}3.46 & no       & YSO  \\
23 &  43590 & 01:34:09.40 & +30:55:18.2 & 4.41 & \llap{$-$}4.09 & yes      & AGB  \\
\hline
\end{tabular}
\end{table*}

\begin{figure}
\centerline{\psfig{figure=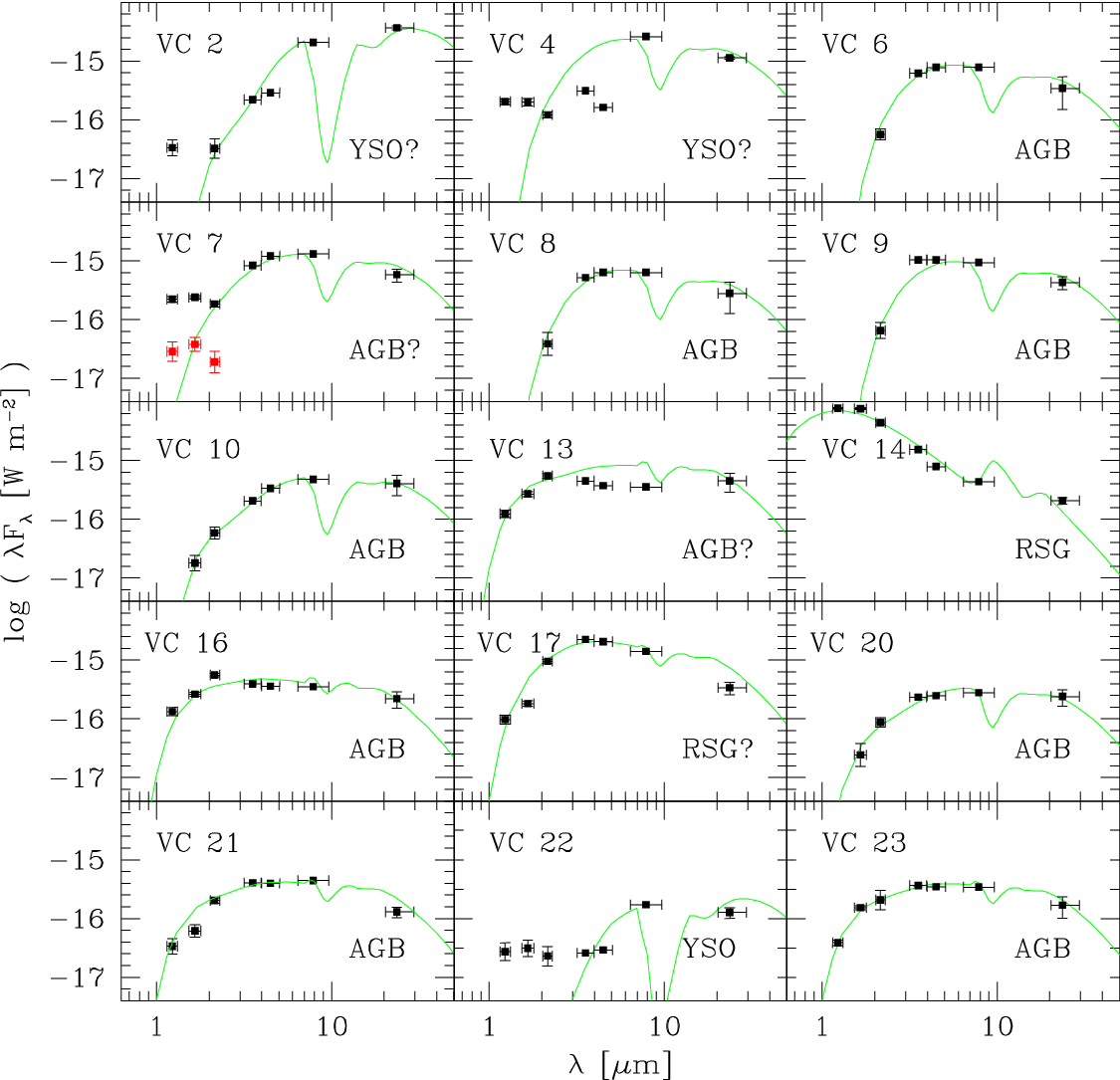,width=84mm}}
\caption[]{SEDs for stars with 24-$\mu$m variability. For VC\,7, also the
near-IR data for another, nearby source are displayed (in red).}
\end{figure}

Overall, the luminosities and mass-loss rates obtained from SED fitting
including the 24-$\mu$m datum are higher than those obtained from applying the
{\bf near-IR}
scaling relations. This should not come as a surprise, as the 24-$\mu$m
sources are at the extreme end of the spectrum. They were already included in
the census of mass loss in M\,33, and their revised combined mass-loss rate --
almost a factor two higher -- does not noticeably change this (it would add a
mere $10^{-3}$ M$_\odot$ yr$^{-1}$ to the total of $10^{-1}$ M$_\odot$ yr$^{-1}$
-- see below).

\subsection{Spectroscopically confirmed AGB stars}

Spectra of seven carbon star candidates located in the northern plume of M\,33
were obtained with the 10-m {\it Keck} telescope (Block et al.\ 2007).
{\bf Strong absorption bands of CN between 0.7--0.9 $\mu$m}
were seen in all seven spectra, confirming their carbon star nature. We have
identified all of these stars in our WFCAM survey as (massive) carbon stars,
but only three were classified as LPVs (Table A3).

Another object worth noting is the first confirmed Mira star in M\,33
(Barsukova et al.\ 2011). A luminous red variable, it was originally detected
as a possible nova. Spectroscopy carried out with the Russian 6-m telescope
revealed an M2--3 spectral type with emission lines characteristic of
dissipative pulsation shocks. The photometric variability is typical for
Mira-type pulsation, with a long period of 665 days and an amplitude exceeding
seven magnitudes in the R-band. It is \#2878 in our WFCAM catalogue, with
$K_{\rm s}=15.39$ mag and an estimated amplitude of $A_{\rm Ks}=0.99$ mag. The
estimated mass for this star based on its K-band magnitude is $M=3.8$
M$_\odot$. While strictly speaking this classifies it as a carbon star in our
classification scheme, this comes with considerable uncertainty and our
estimated mass compares well with the 4 M$_\odot$ reported in Barsukova et
al.\ (2011). The colour of $(J-K)=0.94$ mag suggests that this star has a low
mass-loss rate. We derive a bolometric magnitude of $M_{\rm bol}=-6.97$ mag,
which is in fair agreement with the $M_{\rm bol}=-6.2$ mag reported in
Barsukova et al.\ (2011).

\begin{table*}
\caption{List of spectroscopically confirmed AGB stars in M\,33, with UKIRT ID
No.\ (Paper IV).}
\begin{tabular}{rllclccl}
\hline\hline
ID & RA(2000) & DEC (2000) & $M$ (M$_\odot$) & spectroscopic type & $\log L/{\rm L}_\odot$ & $\log\dot{M}$ (M$_\odot$ yr$^{-1}$) & variable? \\
\hline
11381 & 01:34:09.78 & +31:55:52.4 & 3.6 & carbon star *  & 4.05 & \llap{$-$}5.27 & yes \\
18181 & 01:34:15.44 & +30:54:17.9 & 3.4 & carbon star *  & 4.22 & \llap{$-$}5.32 & no  \\
13626 & 01:34:16.97 & +30:52:33.8 & 3.2 & carbon star *  & 4.22 & \llap{$-$}5.40 & no  \\
12163 & 01:34:17.52 & +30:52:16.9 & 3.5 & carbon star *  & 4.31 & \llap{$-$}5.28 & yes \\
23388 & 01:34:19.50 & +30:54:32.6 & 3.3 & carbon star *  & 4.19 & \llap{$-$}5.30 & no  \\
19223 & 01:34:19.91 & +30:53:22.8 & 3.3 & carbon star *  & 4.21 & \llap{$-$}5.35 & no  \\
 2878 & 01:34:27.13 & +30:58:42.7 & 3.8 & M-type star ** & 4.69 &           0.00 & yes \\
89880 & 01:34:30.66 & +30:53:09.9 & 3.6 & carbon star *  & 4.25 & \llap{$-$}4.94 & yes \\  
\hline
\multicolumn{8}{l}{* Block et al.\ (2007)} \\
\multicolumn{8}{l}{** Barsukova et al.\ (2011)} \\
\end{tabular}
\end{table*}

The division between carbon stars and more massive, oxygen-rich AGB stars
around 3.7 M$_\odot$ is in between our adopted boundary of 4 M$_\odot$, and the
slightly lower value of 3.5 M$_\odot$ suggested by recent models presented by
Ventura et al.\ (2018) for solar metallicity.

\subsection{Spectroscopically confirmed RSGs}

Drout et al.\ (2012) identified RSGs (and yellow supergiants) in M\,33 using
the Hectospec multi-fiber spectrograph on the 6.5-m Multiple Mirror Telescope
in two observing campaigns, in 2009 and 2010. We cross-correlated our WFCAM
survey with their catalogue in paper IV. Our WFCAM survey detected 381 (93\%)
of the red stars in the survey by Drout et al., of which 186 (98\%) rank-1
stars and 13 rank-2 stars. Of the rank-1 stars, 14 were found by us to be
variable, as was one rank-2 star. The stars marked with rank-1 were selected
both photometrically and kinematically, while the rank-2 stars were confirmed
with just one method. We list their properties in Table A4.

Based on the birth mass of these variable stars estimated in paper V, seven or
eight are indeed expected to be RSGs, and another one or two appear to be
massive AGB stars. However, six have relatively low estimated masses,
comparable to (massive) carbon stars. This indicates that there remains
significant uncertainty regarding the (massive) carbon star population without
spectroscopic confirmation.

\begin{table*}
\caption{Variable stars listed as RSGs in M\,33 by Drout et al.\ (2012), with
UKIRT ID No.\ (Paper IV).}
\begin{tabular}{rcccccl}
\hline\hline
ID & $M$ (M$_\odot$) & \multicolumn{2}{c}{$\log L/{\rm L}_\odot$} & \multicolumn{2}{c}{$\log\dot{M}$ (M$_\odot$ yr$^{-1}$)} & type \\
 & & {\it if M-type star} & {\it if carbon star} & {\it if M-type star} & {\it if carbon star} & \\
\hline
2751 &         3.6 & 4.53 & 4.53 & $-5.58$ & $-6.63$ & carbon \\
5107 &         3.6 & 4.49 & 4.46 & $-4.31$ & $-5.14$ & carbon \\
7703 &         2.7 & 4.27 & 4.25 & $-4.95$ & $-5.88$ & carbon \\
1168 &         7.0 & 4.99 & 4.98 & $-4.81$ & $-6.14$ & M-type \\
1003 &         9.2 & 5.09 & 5.08 & $-5.58$ &         & M-type \\
3606 &         3.6 & 4.59 & 4.60 &         &         & carbon \\
 539 & \llap{1}1.2 & 5.39 & 5.29 & $-6.45$ & $-7.03$ & M-type \\
5108 &         3.2 & 4.28 & 4.24 & $-4.65$ & $-5.49$ & carbon \\
 587 & \llap{2}1.9 & 5.39 & 5.38 &         &         & M-type \\
1647 & \llap{1}9.3 & 5.41 & 5.36 & $-3.92$ & $-4.84$ & M-type \\
 991 & \llap{1}2.5 & 5.18 & 5.17 & $-4.71$ &         & M-type \\
2329 &         3.6 & 4.71 & 4.71 & $-5.18$ & $-6.28$ & carbon \\
1368 &         4.8 & 4.92 & 4.94 &         &         & M-type \\
 916 & \llap{1}1.4 & 5.17 & 5.17 &         &         & M-type \\
 682 & \llap{2}9.7 & 5.39 & 5.37 & $-4.53$ &         & M-type \\
\hline
\end{tabular}
\end{table*}

\subsection{Optically luminous variable stars}

An ongoing photometric survey is being conducted by Martin \& Humphreys (2017)
to investigate the variability of luminous evolved stars in M\,33 (and M\,31).
Using the 0.51-m telescope at the University of Illinois they obtained
photometry for 199 stars over a period of four years, down to $V\sim19$ mag,
including classical LBVs, LBV candidates, post-RSG A/F-type hypergiants, and
B[e] supergiants. From those within M\,33, 20 show substantial variability. We
have identified all of them in our WFCAM survey, but only four stars are
identified as variable stars by us (Var\,B, Var\,C, UIT\,003 and M33\,C-4119).
Their properties are listed in Table A5 -- most mass estimates (especially
those between parentheses) are suspect because the assumption that the stars
are on the AGB or RSG branches is no longer valid. Some stars have a nearby
bright star, making it difficult to decide which one is the true counterpart.
In such cases the properties of both stars are mentioned.

Regarding Var\,C (Humphreys et al.\ 2014), several eruptions have been
reported since its discovery (Hubble \& Sandage 1953; Rosino \& Bianchini
1973; Humphreys et al.\ 1988; Szeifert et al.\ 1996). But from 1998 onward it
had returned to a minimum state (Burggraf et al.\ 2015). Shorter episodes of
brightening were seen from 2001 until 2005 (Viotti et al.\ 2006; Clark et al.\
2012). At the moment it is in a hot, quiescent phase (Humphreys et al.\ 2014).
With $K_{\rm s}=15.75$ mag and $(J-K_{\rm s})=0.57$ mag it is one of the
confirmed blue variables in our list. We estimated $\log L/{\rm L}_\odot=4.61$
and $M=3.5$ M$_\odot$, but these must have been underestimated as the star is
not a RSG (or AGB star). The light-curve, which was presented in Paper IV,
shows that its brightness has steadily diminished, however by 2007 it seems to
have stabilised.

\begin{table*}
\caption{Luminous variables in M\,33 from Martin \& Humphreys (2017), with
UKIRT ID No.\ (Paper IV) and photometry in magnitudes.}
\begin{tabular}{lrllccccc}
\hline\hline
name & ID & RA(2000) & DEC(2000) & $K_{\rm s}$ & $(J-K_{\rm s})$ & \multicolumn{2}{c}{$\log L/{\rm L}_\odot$} & $M$ (M$_\odot$) \\
     &    &          &           &           &                & {\it if M-type star} & {\it if carbon star} &                 \\
\hline
{\it OB supergiants} &&&&&&&&\\
M\,33\,C-23048  & 39484 & 01:33:09.14 & +30:49:54.5 & 17.665 & 0.175 & 3.92 & 3.95 &        (1.0) \\
M\,33\,C-4119   & 44270 & 01:33:12.81 & +30:30:12.6 & 17.200 & 0.615 & 4.01 & 4.02 &        (1.2) \\
M\,33\,C-19725  & 11894 & 01:33:39.52 & +30:45:40.5 & 17.058 & 0.066 & 4.20 & 4.23 &        (1.3) \\
V-139873        &  5168 & 01:34:37.25 & +30:38:17.7 & 16.348 & 0.298 & 4.42 & 4.45 &        (2.6) \\
\hline
{\it Warm Hypergiants} &&&&&&&&\\
V-093351        &  2671 & 01:33:52.42 & +30:39:09.6 & 15.153 & 1.048 & 4.76 & 4.75 &        (4.2) \\
V-125093        &  1394 & 01:34:15.42 & +30:28:16.4 & 14.139 & 1.324 & 5.15 & 5.13 & \llap{1}1.7  \\
\hline
{\it Yellow Supergiants} &&&&&&&&\\
M\,33-013303.40 &  3095 & 01:33:03.4  & +30:30:51.5 & 15.687 & 0.490 & 4.64 & 4.65 &        (3.6) \\
M\,33-013303.60 &  2094 & 01:33:03.6  & +30:29:03.4 & 15.275 & 0.496 & 4.80 & 4.82 &        (3.9) \\
M\,33\,C-22178  & 66321 & 01:33:55.78 & +30:48:31.3 & 18.160 & 0.251 & 3.71 & 3.73 &        (0.8) \\
                & 32740 &             &             & 17.180 & 1.223 & 3.94 & 3.92 &        (1.2) \\
V-104139        &  2172 & 01:33:58.96 & +30:41:39.5 & 15.382 & 0.356 & 4.79 & 4.81 &        (3.8) \\
V-104958        &  7136 & 01:33:59.37 & +30:23:10.9 & 16.43  & 0.156 & 4.33 & 4.37 &        (1.7) \\
M\,33\,C-14120  & 42128 & 01:34:20.65 & +30:39:42.6 & 17.839 & 0.027 & 3.90 & 3.94 &        (1.0) \\
                & 77403 &             &             & 18.265 & 1.020 & 3.52 & 3.51 &        (0.8) \\
\hline
{\it Classical LBVs} &&&&&&&&\\
Var\,C          &  4896 & 01:33:35.14 & +30:36:00.4 & 15.750 & 0.570 & 4.60 & 4.61 &        (3.5) \\
Var\,B          &  8197 & 01:33:49.23 & +30:38:09.1 & 16.416 & 0.825 & 4.29 & 4.29 &        (2.4) \\
Var\,83         &  2161 & 01:34:10.93 & +30:34:37.6 & 15.295 & 0.448 & 4.81 & 4.82 &        (3.9) \\
\hline
{\it  B[e]sg} &&&&&&&&\\
M\,33-013242.26 &   733 & 01:32:42.26 & +30:21:14.1 & 13.856 & 1.010 & 5.29 & 5.28 & \llap{1}6.6  \\
\hline
{\it Candidate LBVs} &&&&&&&&\\
M\,33\,C-4174   & 19026 & 01:32:35.25 & +30:30:17.6 & 16.919 & 0.587 & 4.12 & 4.14 &        (1.3) \\
M\,33\,C-21192  &  4298 & 01:34:32.74 & +30:47:09.6 & 16.191 & 0.099 & 4.53 & 4.57 &        (2.9) \\
\hline
{\it Unknown Class} &&&&&&&&\\
UIT\,003        & 24329 & 01:32:37.72 & +30:40:05.6 & 16.944 & 0.372 & 4.16 & 4.19 &        (1.3) \\
M\,33\,C-12559  &  5000 & 01:33:51.00 & +30:38:18.8 & 16.223 & 0.282 & 4.47 & 4.50 &        (2.8) \\
\hline
\end{tabular}
\end{table*}

\subsection{Infrared eruptive stars}

Using {\it Spitzer} IRAC images, Khan et al.\ (2013) listed nine candidate
eruptive stars of the $\eta$\,Carin{\ae} ($\eta$\,Car) kind in M\,33. Later,
in Khan et al.\ (2015) using data from the optical through to the far-IR, they
characterised these candidates. They showed that none of the candidates are
true analogues of $\eta$\,Car. Some of these objects are dusty stars with
$25<M_{\rm ZAMS}/{\rm M}_\odot<60$ undergoing, once or twice, an obscured phase
at most lasting a few thousand years. Five of the candidates (M\,33-2, -5, -6,
-8 and -9) have SEDs rising nearly monotonically from the optical to 24
$\mu$m, which suggests the presence of cold dust associated with young star
clusters and/or H\,{\sc ii} regions. This inference is vindicated by the
recent detection, by Koch et al.\ (2018), of the first, single-peaked OH
(main-line) maser in M\,33 at the position of M\,33-8. That mid-IR source had
already been associated with an H\,{\sc ii} region and radio continuum source
(Buckalew et al.\ 2006), and a water maser had been detected by Greenhill et
al.\ (1990).

Except two (M\,33-2 and -6), we found counterparts in our WFCAM catalogue
(Table A6). Regarding M\,33-2, while there is no WFCAM source on its exact
location, a luminous, massive star is found at a distance of $2^{\prime\prime}$
and we therefore include this star in our analysis. Khan et al.\ (2015)
estimated a luminosity of $\log L/{\rm L}_\odot=5.68$ for M\,33-1 (Object\,X --
see below), i.e.\ similar to our estimate. However, the even higher
luminosities estimated by Khan et al.\ for M\,33-3, -4 and -7 are not
corroborated by our photometry. On the exact location of M\,33-4, we find
WFCAM source \#18510, which is neither a luminous nor massive star; star \#741
at a distance $<3^{\prime\prime}$ {\it is} luminous and massive, though, and we
therefore include it. We detect variability in M\,33-8.

\begin{table*}
\caption{$\eta$\,Carin{\ae} candidates in M\,33 and their near-IR properties,
with UKIRT ID No.\ (Paper IV) and photometry in magnitudes.}
\begin{tabular}{lrllcccccl}
\hline\hline
name & ID & RA(2000) & DEC(2000) & $K_{\rm s}$ & $(J-K_{\rm s})$ & $M$ (M$_\odot$) & $\log L/{\rm L}_\odot$ & $\log\dot{M}$ (M$_\odot$ yr$^{-1}$) & included? \\
\hline
M\,33-1 &   1961 & 01:33:24.04 & +30:25:34.54 & 13.66 & 3.40 & \llap{2}1.1 & 5.62 & $-3.15$ & yes \\
M\,33-2 &   1241 & 01:33:34.10 & +30:36:26.56 & 14.19 & 1.75 & \llap{1}1.0 & 5.12 & $-3.92$ & no  \\
M\,33-3 &  23552 & 01:33:45.45 & +30:36:48.85 & 16.24 & 1.67 &         3.1 & 4.30 & $-5.46$ & yes \\
M\,33-4 &  18510 & 01:34:13.56 & +30:33:42.30 & 16.18 & 1.65 &         3.2 & 4.28 & $-5.46$ & yes \\  
        &    741 &             &              & 13.85 & 1.29 & \llap{1}6.6 & 5.27 & $-4.27$ & yes \\
M\,33-5 &   2087 & 01:33:16.54 & +30:52:49.94 & 14.36 & 1.47 &         8.9 & 5.06 & $-4.31$ & no \\
M\,33-7 &  15675 & 01:33:35.50 & +30:39:28.98 & 16.03 & 1.75 &         3.5 & 4.34 & $-5.33$ & yes \\
M\,33-8 &   2105 & 01:34:00.21 & +30:40:47.53 & 14.42 & 1.90 &         8.3 & 5.04 & $-3.90$ & no  \\
M\,33-9 & 301623 & 01:33:29.03 & +30:40:21.94 & 16.43 & ...  & \llap{1}3.0 &  ... & ...     & yes \\
\hline
\end{tabular}
\end{table*}

Object\,X (M\,33-1 = \#1961) is of special interest, having been identified
as the brightest mid-IR source in M\,33 (Khan et al.\ 2011). It listed as 
\#M\,33--1 in Khan et al.\ (2013). Its bolometric luminosity is
$L\sim5\times10^5$ L$_\odot$; it is optically variable on short time scales
(tens of days) and slightly variable in the mid-IR. Its properties suggest
that this star has a complex dusty circumstellar structure resulting from
episodic mass-loss events over at least half a century. There are suggestions
that Object\,X is an analogue to Var\,A (Humphreys et al.\ 2006) or the
Galactic post-RSG IRC\,+10$^\circ$420 (Jones et al.\ 1993) and over the next
few decades will become visible in the optical. Our WFCAM catalogue lists this
source as \#1961. While we did not detect variability, with $K_{\rm s}=13.66$
mag and $(J-K_{\rm s})=3.40$ mag it is very red, and the mass-loss rate of
$\dot{M}>10^{-3}$ M$_\odot$ yr$^{-1}$ is extreme.

\subsection{Symbiotic binaries}

Symbiotic stars (SySt) are binary systems with an evolved giant (normal a red
giant star or a Mira surrounded by an opaque dust shell) in which the evolved
giant transfers mass onto a hot, luminous and compact white dwarf companion.
Among twelve symbiotic stars identified in M\,33 by Miko{\l}ajewska et al.\
(2017) we recovered five (\#12118, \#47715, \#42739, \#13728 and \#66184). Of
these, two (\#42739 and \#66184) show variability and both are detected with
{\it Spitzer}. We list their properties in Table A7. Interestingly, the
spectral types of these stars are in excellent agreement with what we would
have suggested based on their masses as derived from their K-band magnitudes.

\begin{table*}
\caption{Symbiotic stars in M\,33 that are identified in our WFCAM survey,
with UKIRT ID No.\ (Paper IV).}
\begin{tabular}{rllllccl}
\hline\hline
   ID &  RA(2000) & DEC(2000) & $M$ (M$_\odot$) &  spectroscopic type & $\log L/{\rm L}_\odot$ &$\log\dot{M}$ (M$_\odot$ yr$^{-1}$) & variable? \\
\hline
12118 & 01:33:03.27 & +30:35:28.3 & 3.1 & carbon star & 4.23 & $-5.48$ & no  \\
47715 & 01:33:11.10 & +30:15:18.2 & 1.0 & M-type star & 3.73 & $-5.93$ & no  \\
42739 & 01:34:35.17 & +30:34:09.4 & 1.5 & carbon star & 3.98 & $-5.88$ & yes \\
13728 & 01:34:49.5  & +30:47:36.9 & 1.7 & carbon star & 4.11 & $-6.63$ & no  \\
66184 & 01:34:57.79 & +31:00:54.2 & 1.0 & M-type star & 3.71 & $-6.24$ & yes \\
\hline
\end{tabular}
\end{table*}

\label{lastpage}
\end{document}